\documentclass[12pt,preprint]{aastex}
\usepackage{psfig,natbib}
\newcommand{\h}     {$h^{-1}_{70}\,$~kpc}

\newcommand{\flux}      {ergs cm$^{-2}$ s$^{-1}$ \AA$^{-1}$}

\shortauthors{JENKINS ET AL.}
\shorttitle{LYMAN LIMIT ABSORBER}
\begin{document}
\title{A Near-Solar Metallicity, Nitrogen-Deficient Lyman Limit Absorber
Associated with two S0 Galaxies\footnote{Based on observations from
(1)the NASA/ESA {\it Hubble Space Telescope\/} obtained at the Space
Telescope Science Institute, which is operated by the Association of
Universities for Research in Astronomy, Inc., under NASA contract
NAS5-26555, and (2) the NASA-CNES-CSA {\it Far Ultraviolet Spectroscopic
Explorer (FUSE)} mission operated by Johns Hopkins University, supported
by NASA contract NAS5-32985.}}
\author{Edward B. Jenkins, David V. Bowen}
\affil{Princeton University Observatory,
Princeton, NJ 08544-1001}
\email{ebj@astro.princeton.edu, dvb@astro.princeton.edu}
\author{Todd M. Tripp}
\affil{Department of Astronomy, University of Massachusetts\\
Amherst, MA 01003-9305}
\email{tripp@fcrao1.astro.umass.edu}
\author{Kenneth R. Sembach}
\affil{Space Telescope Science Institute\\
3700 San Martin Dr., Baltimore, MD 21218}
\email{sembach@stsci.edu}

\begin{abstract}

The UV spectrum of the bright quasar PHL~1811 at $z_{\rm em}=0.192$
reveals a foreground gas system at $z=0.080923$ with $\log N({\rm
H~I})=17.98\pm 0.05$.  We have determined the abundances of various
atomic species in this system from a spectrum covering the wavelength
range 1160 to 1730$\,$\AA\ recorded at $7\,{\rm km~s}^{-1}$ resolution
by the E140M grating of STIS on the {\it Hubble Space Telescope} ({\it
HST}), supplemented by coverage at shorter wavelengths by the {\it Far
Ultraviolet Spectroscopic Explorer}.  The abundances of C~II, Si~II,
S~II and Fe~II compared to that of O~I indicate that a considerable
fraction of the gas is in locations where the hydrogen is ionized.  An
oxygen abundance $[{\rm O/H}]=-0.19\pm 0.08$ in the H~I-bearing gas
indicates that the chemical enrichment of the gas is unusually high for
an extragalactic QSO absorption system.  However, this same material has
an unusually low abundance of nitrogen, $[{\rm N/O}] < -0.59$,
indicating that there may not have been enough time during this
enrichment for secondary nitrogen to arise from low and intermediate
mass stars.  From the convergence of high Lyman series lines we can
determine the velocity width of H~I, and after correcting for turbulent
broadening shown by the O~I absorption feature, we derive a temperature
$T=7070^{+3860}_{-4680}\,$K.  We determine a lower bound for the
electron density $n(e)>10^{-3}\,{\rm cm}^{-3}$ by modeling the
ionization by the intergalactic radiation field and an upper bound
$n(e)<0.07\,{\rm cm}^{-3}$ from the absence of much C~II in an excited
fine-structure level.  The thermal pressure in the range
$4<p/k<140\,{\rm cm}^{-3}\,$K could be confined by a warm-hot
intergalactic medium (WHIM) structure with $\rho/\bar\rho\sim 20$ that
might accompany a wall of galaxies at the same redshift, seen in data
from the {\it Sloan Digital Sky Survey}.  An $r$-band image of the field
surrounding PHL~1811 recorded by the ACS instrument on {\it HST} shows
that two galaxies at the same redshift as the gas are S0 galaxies,
separated by only 34 and 87\h\ from the line of sight.  One or both of
these galaxies may be the source of the material in the Lyman limit
system, which may have been expelled from them in a fast wind, by tidal
stripping, or by ram-pressure stripping.  Subtraction of the ACS
point-spread function from the image of the QSO reveals the presence of
a face-on spiral galaxy under the glare of the quasar; while it is
possible that this galaxy may be responsible for the Lyman limit
absorption, the exact alignment of the QSO with the center of the galaxy
suggests that the spiral is the quasar host.

\end{abstract}
\keywords{galaxies: abundances --- galaxies: halos ---
 intergalactic medium --- quasars:
absorption lines --- quasars: individual (PHL~1811)}

\section{Introduction}\label{intro}

The absorption lines detected in the spectra of background quasars allow
us to examine in detail the nucleosynthetic products released to the
gaseous media within and between galaxies  (Lauroesch et al. 1996;
P\'eroux et al. 2003; Pettini 2003; Prochaska 2004). This field of
research has flourished over the last few years due to the availability
of high-resolution spectrographs coupled to 8$-$10$\,$m class
ground-based telescopes. Elemental abundance patterns of absorbing gas
clouds provide insights into the character and maturity of stellar
systems associated with such absorbers, either through direct
comparisons with abundances measured for old stars in our own Galaxy 
(Prochaska 2003) or theoretical models for chemical evolution 
(Matteucci, Molaro, \& Vladilo 1997; Calura et al. 2004; Dessauges-Zavadsky et al. 2004).  However, there remains a large scatter in
abundances within the Damped Lyman Alpha (DLA) population at any given
redshift, an outcome which has hampered efforts to synthesize a coherent
picture of how heavy elements build up in the universe.  For instance,
recent studies have shown that low-redshift QSO absorbers have
metallicities that range from substantially sub-solar abundances  (Tripp
et al. 2002, 2004) to nearly solar values  (Prochaska et al. 2004). 
Some systems show significant nitrogen underabundances  (Tripp et al.
2004) while other absorbers appear to have normal (solar) nitrogen
abundances with respect to other metals  (Savage et al. 2002). 
Nevertheless, there has been some progress in noting a general trend: by
accumulating a large database of metallicity measurements from DLAs and
averaging these values over broad redshift intervals, Prochaska, et al 
(2003) have found evidence for an increase in elemental abundances with
increasing cosmic time.  This finding is consistent with our expectation
that the universe becomes more metal-rich as it evolves, but the wide
range in observed abundances still indicates that absorption systems
probably arise from gas residing within, or cast off from, a very
heterogeneous population of galaxy types and ages. Clearly, more studies
are needed to understand the chemical enrichment of low-z QSO absorbers,
which will ultimately lead to a better understanding of chemical
evolution over the past 10 Gyr in parts of the universe that have baryon
densities that are well above average.  

A considerable amount of effort has gone into trying to understand the
nature of galaxies that may be associated with QSO absorption systems,
but for high-$z$ systems whose abundances are obtained from ground-based
telescopes, detailed investigations into the origin of the absorption
are hampered by the great distances. That is, without the ability to
study well resolved images of galaxies near quasar lines of sight,
investigating the exact causes of the abundance fluctuations will
inevitably prove extremely difficult. For example, variations could
arise from QSO sightlines intercepting similar galaxies at different
galactocentric radii; alternatively, differences may be a direct results
of galaxy age, morphological type, star formation history, and/or gas
dynamical processes.

Taken together, these arguments offer a compelling case for studying
absorption systems at {\it low} redshift, where both individual galaxies
and galactic large scale structure can be studied in detail. An
additional need arises from the present undersampling of low-$z$
systems, which are more difficult to survey since the absorption lines
required for abundance measurements lie in the ultraviolet (UV).
Unfortunately, present-day UV spectrographs and space telescopes can
observe only the very brightest quasars. As a result, an extrapolation
to $z=0$ of the $N$(H~I)-weighted metallicity trend at high-$z$ shown by
Prochaska et al  (2003) is highly uncertain because there are not enough
observations of systems with $z < 1.5$.

In this paper we report abundance measurements for one such low-redshift
system, the $z=0.08$ Lyman limit absorber towards PHL~1811. We also
assess the probable contributors (galaxies) that may be the origin of
the absorption. The discovery of a gas system that produced continuous,
nearly total absorption shortward of the Lyman limit in the spectrum of
PHL~1811 arose from an initial, brief observation of the QSO using the
{\it Far Ultraviolet Spectroscopic Explorer\/} ({\it FUSE\/}). In
\S\ref{FUSE_survey} we give some background on the discovery of the
system and the limited findings we obtained from the early dataset, and
in \S\ref{nearby_galaxies} we recount our initial investigations of the
galaxies in the immediate vicinity of the quasar sightline that we found
to have redshifts similar to that of the absorption system.

The traditional definition of a ``Lyman Limit System'' is one that has
an H~I column density high enough to cause absorption beyond the Lyman
limit, but that is too low to show damping wings in the Ly$\alpha$
transition.  A DLA was originally defined by Wolfe et al  (Wolfe et al.
1986) to have $\log N$(H~I)$\geq 20.3$, based on their ability to detect
damped absorption in low resolution spectra with an equivalent width
$\geq 10$~\AA . In the case of the absorption towards PHL~1811 though,
the high resolution used to record the Ly$\alpha$ transition enables us
to detect a damping wing in the line (\S\ref{H_I}) that leads to an
accurate determination of $N$(H~I), slightly less than $10^{18}$
cm$^{-2}$. To be consistent with the accepted terminology for a system
with this much neutral hydrogen, however, we will continue to describe
this absorption system as a Lyman limit system.

Our abundance measurements came from an {\it HST} program, which used
the medium resolution echelle grating (E140M) and {\it Space Telescope
Imaging Spectrograph} (STIS) (\S\ref{stis}). We have supplemented these
data with more extensive observations from  {\it FUSE\/} (\S\ref{fuse}),
which provide data at improved S/N at the shortest wavelengths. We
derive abundances of various species using different methods appropriate
for each case at hand in the subsections of \S\ref{coldens}.  From the
convergence of Lyman series lines at the shortest wavelengths in the
{\it FUSE\/} data, coupled with observations of O~I in the STIS data, we
can determine permissible temperatures for the H~I-bearing material
(\S\ref{temp}). This information, coupled with a range for the electron
density,  [with an upper bound defined by the population of C~II in an
excited fine-structure level (\S\ref{elec_dens}) and a lower bound from
a model for the photoionization of the material (\S\ref{uniform_slab})]
help to define the physical characteristics of the system.

From our overview of our fundamental measurements of various atoms and
ions in \S\ref{general_patterns}, we deduce that we are observing an
absorber in which ionized gas dominates over neutral gas. For this
reason, we have considered whether our elemental abundances have been
distorted by our inability to see all stages of ionization.  We
investigate various ionization scenarios, including collisional
ionization (\S\ref{collisional}), photoionization from the intergalactic
radiation field for two values of the ionization parameter $U$
(\S\ref{uniform_slab}), and a special case where the apparent abundances
of N and O could in principle be altered by an intense stellar radiation
field (\S\ref{more_detailed}) if stars that are invisible to us are
indeed present inside this system.  We caution that the level of
technical detail within the last of these three subsections
(\S\ref{more_detailed} and supporting equilibrium equations given in the
Appendix) is greater than that of the others, and the arguments are
intended to defend our conclusions against the effects arising from a
speculative extreme in ionization conditions; some readers may wish to
skip this discussion.

As noted above, we are also interested in determining the origin of the
Lyman limit system towards PHL~1811, and associating the results from
our absorption line measurements with the galaxies and galactic large
scale structure in the immediate vicinity of the quasar sightline. A
small part of our {\it HST} observing program (one orbit) was devoted to
obtaining an $r$-band image of the field around PHL~1811 using the {\it
Advanced Camera for Surveys} (ACS) (\S\ref{basic_reduction}); our goal
was to apply standard surface photometry metrics (\S\ref{surface_phot})
to determine the morphology of the two galaxies close to the sightline
that have redshifts similar to that of the Lyman limit system. We also
wished to search for bars, spiral-arm patterns, or indicators of
interesting dynamical phenomena in or near the galaxies, by enhancing
the high-frequency image information in the data
(\S\ref{unsharp_masking}). The results from these analyses for each
galaxy are given in \S\ref{s_g158} and \S\ref{s_g169}. Even though we
did not use the coronagraphic mode of STIS to block light from the
quasar, we were nevertheless able to subtract off a substantial part of
the image flare from the QSO (\S\ref{psf_subtraction}) to reveal a
loosely wound spiral arm pattern exactly centered at the position of the
quasar (\S\ref{s_qso}).  However we are uncertain if this is the galaxy
that is responsible for the Lyman limit absorption or is simply the host
galaxy of the quasar. 

Beyond the immediate area around the quasar sightline,  we present in
\S\ref{LSS} evidence from the {\it Sloan Digital Sky Survey} ({\it
SDSS\/}) that the Lyman limit system is within (or near the end of) a
large-scale filament (or sheet) of galaxies at a similar redshift.  We
close with a discussion on how the abundance pattern in the absorber
might be related to the chemical evolution of the gas (\S\ref{chemev}),
speculate on the possible origins of the material (\S\ref{source}) and
present a general summary of our results (\S\ref{summary}).

\section{Previous Findings and the Motivation for New
Observations}\label{overview}

\subsection{Survey of Absorption Systems using {\it
FUSE}}\label{FUSE_survey}

The research discussed here was triggered when Jenkins et al.  (2003;
hereafter Paper~I) used {\it FUSE\/}  (Moos et al. 2000; Sahnow et al.
2000) to perform an exploratory study of the Galactic and intergalactic
absorption lines appearing in the spectrum of the extraordinarily bright
quasar PHL~1811 at $z_{\rm em} = 0.192$  (Leighly et al. 2001).  We
identified 7 extragalactic gas systems, one of which was a Lyman limit
system at $z=0.08093$.\footnote{The value $z=0.08093$ reported in
Paper~I is revised in this paper to $z=0.080923$ on the basis of the
more accurate wavelengths that are available in the STIS E140M
spectrum.}  Three of the systems have redshifts that differ by less than
$2500\,{\rm km~s}^{-1}$ from that of the Lyman limit system.  In Paper~1
we reported on the unusually favorable opportunity for research on
intergalactic systems at low redshift: this quasar, the second brightest
in the sky, shows twice the expected number of absorption systems over a
path $\Delta z = 0.192$ in the local universe.  Moreover, there was only
a 6\% chance of seeing a Lyman limit system.  Incidental information
about PHL~1811 is given in Table~1 of Paper~I.

The {\it FUSE\/} spectrum was of great value in providing a general
picture of what absorption systems were present, but it had several
drawbacks that made it difficult to derive quantitative results: (1) A
large part of the spectrum contained many Galactic H$_2$ features, and
chance overlaps with them, together with atomic features from the Galaxy
and the other redshifted systems, blocked many lines of interest, (2)
the wavelength resolving power of only $\lambda/\Delta\lambda\lesssim
20,000$ and the modest signal-to-noise ratio (maximum S/N was 22 per
resolution element) generally led to the detection of only the
strongest, most saturated lines and (3) various useful Lyman series
lines (Ly$\beta$ and higher) in the Lyman limit system were on the flat
portion of the curve of growth, which, along with a limit for the
transmission below the Lyman limit, resulted in 2 orders of magnitude
uncertainty in $N$(H~I).  Hence, only rudimentary information on element
abundances and degrees of ionization could be obtained.  Clearly, better
spectra were needed to reach meaningful conclusions on the nature of gas
within the Lyman limit system.  A subsequent observing program using
STIS in a moderate resolution echelle mode (E140M) provided such a
spectrum. These data, combined with new, longer exposure FUSE
observations intended to improve upon the S/N of the original spectra,
now enable us to derive more precise measurements of absorption lines in
the spectrum of PHL~1811.

\subsection{Locations of Nearby Galaxies}\label{nearby_galaxies} 

In addition to using {\it FUSE} to observe the far-UV spectrum of
PHL~1811, we presented in Paper~I an $R$-band image of the $7\farcm 6
\times 7\farcm 6$ field surrounding the quasar.  We also measured the
redshifts of seven of the galaxies in the field.  Two of the galaxies,
which we designated as G158 and G169\footnote{We will retain these
simple designations for discussions later in this paper.  Catalog names
listed in the 2MASS point-source catalog are given in
Table~\protect\ref{tab_acs}.}, had redshifts close to that of the Lyman
limit system.  Their separations on the sky were $23\arcsec$ and
$59\arcsec$ from PHL~1811, respectively, corresponding to transverse
distances $\rho$ of 34 and 87\h.\footnote{$h_{70}\:=\:H_0/(70\,{\rm
km~s}^{-1}{\rm Mpc}^{-1})$, where $H_0$ is the Hubble constant.
$\Omega_m$=0.3 and $q_0\:=\:0$ is assumed throughout this paper.}  The
images of the galaxies were not good enough to obtain definitive
morphological classifications.  Since we felt that it was important to
learn more about these galaxies, we used the ACS on board {\it HST\/} to
register images of the galaxies in greater detail.  These observations
are discussed in \S\ref{ACS}.

\section{New Spectroscopic Observations}\label{spec_obs}

\subsection{STIS}\label{stis}

PHL~1811 was observed for 33.9$\,$ks (13~{\it HST\/} orbits) with STIS 
(Kimble et al. 1998; Woodgate et al. 1998) using the E140M grating and
the $0.2\times0.06$~arcsec entrance aperture\footnote{While the
$0.2\times0.06$~arcsec aperture admits less light than the $0.2\times
0.2$~arcsec one, it gives a profile that does not have undesirable,
broad shoulders under the main profile of the line-spread function; see
Figure~13.91 in the STIS Instrument Handbook  (Kim Quijano et al.
2003).} under a Cycle~11 {\it HST\/} observing program (ID~=~9418).  The
observing time was distributed between two sessions; one was held on
2002 October 7 and 9\footnote{Archive dataset names O8D902010,
O8D902020, O8D902030, O8D904010, O8D904020, O8D904030, O8D904040}, while
the other was on 2003 July 8 and 9\footnote{Archive dataset names
O8D901010, O8D901020, O8D901030, O8D903010, O8D901030, O8D903030}.  The
resolving power of the spectrum was $\lambda/\Delta\lambda=46,000$
[$\Delta v=7\,{\rm km~s}^{-1}$  (Kim Quijano et al. 2003)], and we
obtained continuous coverage from about 1160$\,$\AA\ to 1730$\,$\AA,
except for only 5 small gaps longward of 1634$\,$\AA\ caused by
incomplete coverage of the echelle grating's free spectral range by the
MAMA detector.  The signal-to-noise ratio in each resolution element (2
pixels) increased in an approximately linear fashion from 10 to 16 over
the interval 1160$\,$\AA\ to 1250$\,$\AA, remained at 16 from
1250$\,$\AA\ to 1470$\,$\AA, and then it decreased linearly to 7 from
1470$\,$\AA\ to 1730$\,$\AA.  The spectra were reduced and combined in
the manner described by Tripp et al.  (2001).

\subsection{{\it FUSE}}\label{fuse}

In addition to the original observations reported in Paper~I, new
observations totaling 65.8$\,$ks of integration time were made on 2 and
3 June 2003.  All {\it FUSE\/} spectra were reduced using {\tt CALFUSE}
Version 2.2.2 and combined in the manner described in Paper~I. The
higher S/N {\it FUSE} spectrum provided more accurate equivalent widths
for key lines belonging to the Lyman limit system.  Also, high members
of the Lyman series were recorded with better accuracy.

\section{Analysis of the Spectra}\label{spectra_analysis}
\subsection{Equivalent Widths and Column Densities}\label{coldens}

Table~\ref{eqw_coldens} lists the equivalent widths of absorption
features associated with the Lyman limit system.  The definitions of
continua and their uncertainties followed from the methods outlined in
the appendix of a paper by Sembach \& Savage  (1992).  Errors in the
equivalent widths represent the combined effects of noise in the lines
and uncertainties in the continuum levels (the two were combined in
quadrature to obtain the final error).  Some of the measurements give
results that are less than or about equal to their errors, but they are
listed since they establish interesting limits to the column densities. 
In Figure~\ref{vstack} we show selected features of various heavy
elements on a common velocity scale.  Lines for different species
warranted different methods of interpretation.  In the following
subsections, we outline the basic approaches for deriving the column
densities shown in Table~\ref{eqw_coldens} (see also endnote $e$ in the
table for brief statements on these methods).

\subsubsection{H I}\label{H_I}

\begin{figure}[h!]
\epsscale{0.5}
\plotone{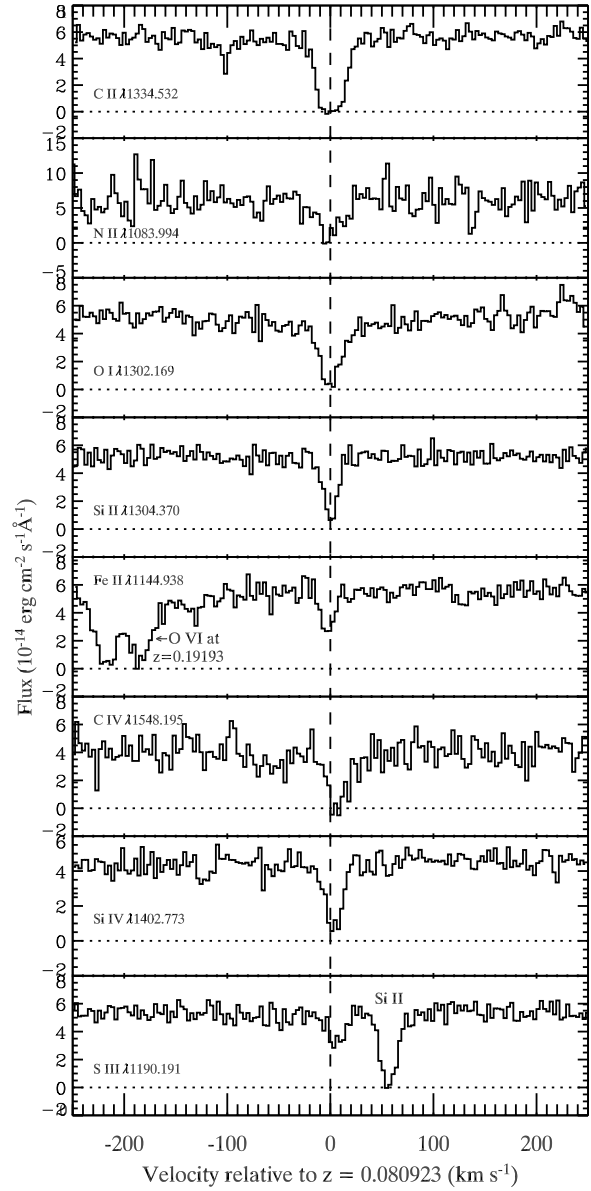}
\caption{Absorption features in the Lyman limit system at $z=0.080923$
in the STIS E140M spectrum of PHL~1811.  The feature at $-102\,{\rm
km~s}^{-1}$ for C~II may not be real; see endnote $f$ in
Table~\protect\ref{eqw_coldens}.}\label{vstack}
\end{figure}

\begin{figure}
\epsscale{1.0}
\plotone{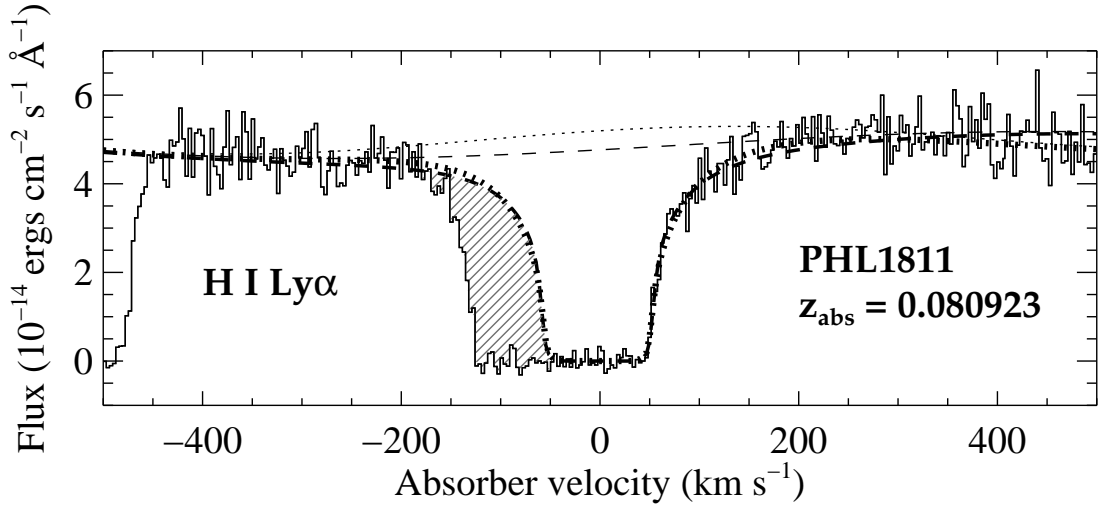}
\caption{Ly$\alpha$ absorption for the Lyman limit system at
$z=0.080923$ in the STIS E140M spectrum of PHL~1811.  The velocity zero
point is defined by the centroid of the apparent optical depth
$\tau_{\rm a}$ of the feature belonging to the 1302$\,$\AA\ transition
of O~I (see Fig.~\protect\ref{oi_profile}).  Two possible continua are
shown by thin dotted and dashed lines, and their best-fit Voigt profile
counterparts are shown with thick dotted and dashed lines.  The two
derived values of $N$(H~I) differ from each other by 0.035$\,$dex.  The
shaded portion shows the absorption by a small amount of extra hydrogen
at velocities well outside the extent of the absorptions by heavier
species shown in Fig.~\protect\ref{vstack}.\label{lalpha}}
\end{figure}

As discussed earlier in \S\ref{FUSE_survey}, the {\it FUSE\/} recordings
of the high members of the Lyman series and the Lyman limit absorption
gave very poor constraints on $N({\rm H~I})$.  In Figure~\ref{lalpha} we
present the Ly$\alpha$ line at $z=0.080923$ from the new STIS spectrum. 
As expected, this line is strongly saturated.  However, the right-hand
side of the profile shows a well developed damping wing, from which we
can measure a precise H~I column density.  As we discuss later in
\S\ref{more_detailed}, we expect O~I to be the most reliable tracer of
the H~I-bearing material.  For this reason, we use the center of the
1302$\,$\AA\ O~I feature to give the velocity zero point for most of the
neutral hydrogen.  We also observe that the maximum extent of the O~I
feature visible in Fig.~\ref{vstack} is only $40\,{\rm km~s}^{-1}$,
which is small compared to the width of the Ly$\alpha$ feature.  From
this we conclude that most of the hydrogen is creating a damped profile
that matches either of the two smooth absorption curves (drawn for two
extreme choices of continua) shown in Fig.~\ref{lalpha}.  Some remaining
material, an amount too small to show up in the other atomic features,
arises over a velocity range $-170$ to $-50\,{\rm km~s}^{-1}$.

\placetable{eqw_coldens}
\placefigure{vstack}
\begin{deluxetable}{
c    
r    
r    
l    
r    
c    
c    
}
\tabletypesize{\footnotesize}
\tablecolumns{7}
\tablewidth{0pt}
\tablecaption{Equivalent Widths and Column Densities\label{eqw_coldens}}
\tablehead{
\colhead{} & \colhead{Observed} & \colhead{Transition} &
\colhead{} & \colhead{$W_{\rm r} \pm 1\sigma$ Error} &
\colhead{$\log N \pm 1\sigma$ Error} &
\colhead{Analysis}\\
\colhead{Species} & \colhead{$\lambda$ (\AA)\tablenotemark{a}} &
\colhead{$\lambda$ (\AA)} & \colhead{$\log f\lambda$\tablenotemark{b}} &
\colhead{(m\AA)\tablenotemark{c}} & \colhead{(${\rm cm}^{-2}$)\tablenotemark{d}} &
\colhead{Method\tablenotemark{e}}
}
\startdata
H I\dotfill &  1314.046 (S)&1215.670 &2.704& 883$\pm  14$&$17.98\pm 0.05$&1\\
C II\dotfill & 1120.200 (F)&1036.337 &2.088&  95.3$\pm   5.7$&$\gg 14.70$&2\\
& 1442.527 (S)&1334.532 &2.234&147.8$\pm   3.8$\\
& 1442.037\tablenotemark{f} (S)&&& 20.1$\pm   3.7$&$13.00^{+0.07}_{-0.09}$&3\\
C II$^*$\dotfill& 1120.937 (F)&1037.018 &2.088&$-9.9\pm   7.9$&$<12.37$&3,4,5\\
&1443.797\tablenotemark{g} (S)&1335.708 &2.188&$-7.4\pm   6.4$\\
C III\dotfill& 1056.083 (F)& 977.020 &2.869& 264.4$\pm   7.6$\tablenotemark{h}
&$\gg 13.50$&6\\
C IV\dotfill & 1673.480 (S)&1548.195 &2.468&121.8$\pm  10.9$&
$13.90^{+0.51}_{-0.20}$&7\\
& 1676.263 (S)&1550.770 &2.167& 90.8$\pm   9.9$\\
N I\dotfill &  1296.621 (S)&1199.550 &2.199& 3.4$\pm   5.8$&
$12.44^{+0.30}_{-\infty}$&3,4\\
&  1297.349 (S)&1200.223 &2.018& 5.4$\pm   5.7$\\
&  1297.875 (S)&1200.710 &1.715& 0.2$\pm   5.8$\\
N II\dotfill & 1171.714\tablenotemark{i} (S)&1083.994 &
2.079& 85.6$\pm  10.3$&
$13.92^{+0.49}_{-0.41}$&8\\
&\phm{1171.714 }(F)&&& 86.7$\pm 8.6$\\
O I\dotfill & 1050.351 (F)& 971.738\tablenotemark{j}&1.128\tablenotemark{j}&
21.9$\pm  13.1\tablenotemark{k}$&\nodata\\
&1055.443 (F)& 976.448 &0.509&0.7$\pm 6.9$&$< 14.54$&3,5\\
&1068.788 (F)& 988.773\tablenotemark{l}&1.773\tablenotemark{l} & 66.1$\pm  
9.8$\tablenotemark{k}&\nodata\\
&1123.328 (F)&1039.230 &0.974&  16.6$\pm   4.4$&
$14.32^{+0.12}_{-0.17}$\tablenotemark{m}&9\\
&1407.544 (S)&1302.169 &1.796&112.2$\pm   4.2$&
$14.52^{+0.06}_{-0.04}$\tablenotemark{m}&10\\
Si II\tablenotemark{n}\dotfill &1286.748 (S)&1190.416 &2.541&  79.0$\pm   3.1$&
$13.95^{+0.05}_{-0.03}$&11\\
&1289.854 (S)&1193.290 &2.842&  94.2$\pm   3.3$\\
&1409.924 (S)&1304.370 &2.052&  62.9$\pm   4.9$\\
&1650.252 (S)&1526.707 &2.307& 102.3$\pm  10.8$\\
Si IV\dotfill &1506.547 (S)&1393.760 &2.854& 108.1$\pm   5.9$&
$13.53^{+0.23}_{-0.37}$&7\\
&1516.289 (S)&1402.773 &2.552&  81.7$\pm   5.4$\\
S II\tablenotemark{o}\dotfill & 1355.263 (S)&1253.805 &1.136&  6.7$\pm   5.5$&
$13.70^{+0.16}_{-0.25}$&3,4\\
& 1361.438 (S)&1259.518 &1.320& 12.6$\pm   6.6$\\
S III\dotfill & 1094.436 (F)&1012.501 &1.647&  35.7$\pm   8.0$\tablenotemark{k}&
\nodata\\
&1286.505 (S)&1190.191 &1.449&  36.2$\pm   3.8$&
$14.19\pm 0.06$&12\\
Ar~I\dotfill & 1133.045 (F)&1048.220 &2.440&$-1.9\pm   7.8$&
$< 12.60$&3,5\\
Fe II\dotfill & 1235.736 (S)&1143.226 &1.342&  10.0$\pm   6.0$\tablenotemark{k}&
\nodata\\
&1237.586 (S)&1144.938 &1.978&  28.9$\pm   4.1$&
$13.59^{+0.14}_{-0.12}$&7\\
&2810.586 (S)&2600.172 &2.793&  172$\pm 24$\tablenotemark{p}\\
\enddata
\tablenotetext{a}{Not measured values, but computed using the laboratory
wavelength and our best general fit to the absorption system's redshift,
$z_{\rm abs}=0.080923$.  Notations following wavelength: (S) = observed
with STIS; (F) = observed with {\it FUSE}.}
\tablenotetext{b}{From Morton  (2003)}
\tablenotetext{c}{Equivalent width in the rest frame of the Lyman limit
system.}
\tablenotetext{d}{Blank entries indicate that the transition shown on
the row was used to help derive the column density shown on a preceding
row for the element, while ellipses indicate that the line was not
employed in the derivation but is generally supportive of our
conclusions.  Error estimates do not include possible systematic errors
that could arise from uncertainties in the adopted $f-$values for the
transitions.}
%
%
\tablenotetext{e}{Key to methods for deriving column densities (see
\S\protect\ref{coldens} for details): {\bf (1)} Voigt profile fit to
portions of the damping wings -- see Fig.~\protect\ref{lalpha}, {\bf
(2)} Line is saturated and stronger than Si~II $\lambda 1193$ -- see
\S\protect\ref{C_II} for details on how the lower limit was derived for
C~II, {\bf (3)} Line is weak and should be on the linear portion of the
curve of growth, so that $\log N = \log (W_\lambda/\lambda) - \log
f\lambda + 20.053$, {\bf (4)} Average of two or more lines, with weights
proportional to the respective $\sigma_i^{-2}$ and overall error given
by $(\sum \sigma_i^{-2})^{-0.5}$, {\bf (5)} Formal measurement yields a
negative equivalent width: a $1\sigma$ upper limit is assigned using the
method of Marshall  (1992), {\bf (6)} The line is probably very strongly
saturated, so we derived a lower limit based on $\tau_0=4$ and
$b=9\,{\rm km~s}^{-1}$, {\bf (7)} Curve of growth fit to two lines
(i.e., the ``doublet ratio method''), with the upper bound for $N$
defined from the combination of $W_{\rm r}({\rm weak})+1\sigma$ and 
$W_{\rm r}({\rm strong})-1\sigma$, while the lower bound for $N$ arises
from  $W_{\rm r}({\rm weak})-1\sigma$ and $W_{\rm r}({\rm
strong})+1\sigma$, {\bf (8)} After correcting for possible O~VI
absorption from another system, the N~II line's nominal strength is
about equal to that of Si~II $\lambda 1304$, so the saturation of the
latter was used as a model -- see \S\protect\ref{N_II} for details on
this calculation and the determinations of the error limits, {\bf(9)}
Appearance of the 1302$\,$\AA\ line was used as a model to apply a small
correction (0.04~dex) for saturation for the best value and the upper
limit, but no such saturation was assumed for the lower limit, {\bf(10)}
From an integration of the apparent optical depth, {\bf(11)} From an
integration of apparent optical depths with corrections for unresolved,
saturated features within the profile using the method of Jenkins 
(1996) applied to all 4 lines, and {\bf (12)} Both lines are probably
slightly saturated, so $N$ and its limits were derived from $W_{\rm r}$
of the 1190$\,$\AA\ line assuming $b=9\,{\rm km~s}^{-1}$, as was found
for C~IV and Si~IV.}
%
%
\tablenotetext{f}{This is an apparent component with a low column
density and displaced with respect to the main absorption by $-102\,{\rm
km~s}^{-1}$.  However, its reality is doubtful, as an expected
corresponding feature with $W_{\rm r}=11\,$m\AA\ for the 1036.34$\,$\AA\
transition of C~II should be visible above the $1\sigma$ noise of
5$\,$m\AA\ in the {\it FUSE\/} spectrum.  No such line is visible, so
the feature in the STIS spectrum may arise from a defect in the
detector.  (It was detected in both of the main observing sessions
discussed in \S\protect\ref{stis}.)}
\tablenotetext{g}{This line appears in a region of the spectrum where
two echelle orders are spliced together.  While there seem to be no
obvious artifacts in the spectrum at this location, some subtle
systematic effects might compromise the reliability of the equivalent
width to an extent beyond that indicated by the formal errors.}
\tablenotetext{h}{The equivalent width measurement for the C~III feature
avoided an unidentified interfering line on the short wavelength side of
the absorption.  Hence it may understate the true value of $W_{\rm r}$.}
\tablenotetext{i}{The position of this line coincides with the location
of a possible absorption by O~VI $\lambda 1032$ for the system at
$z_{\rm abs}=0.13541$ identified in Paper~I.  Thus, in order to estimate
the strength of the N~II absorption, we had to subtract off the
contribution arising from O~VI, using measurements of the $\lambda 1037$
feature to gauge its strength, as we describe in \S\protect\ref{N_II}.}
\tablenotetext{j}{Includes contributions from three different
transitions at about the same wavelength.}
\tablenotetext{k}{This line is of limited value in deriving a column
density, but its presence at about the correct strength supports our
determination of the column density from the other line(s).}
\tablenotetext{l}{Includes contributions from nearby lines at 988.578
and 988.655$\,$\AA.}
\tablenotetext{m}{This is the only case for which we list two values of
$\log N$.  The best compromise value is $\log N({\rm O~I})=14.47\pm
0.05$; see \S\protect\ref{O_I} for details.}
\tablenotetext{n}{The strongest line of Si~II at $\lambda_{\rm
lab}=1260.422\,$\AA\ ($\log f\lambda = 3.171$) suffers interference from
another strong line, which might be a Ly$\alpha$ forest line.  Hence
this line was not considered.} 
\tablenotetext{o}{The weakest line of S~II at $\lambda_{\rm
lab}=1250.578\,$\AA\ ($\log f\lambda = 0.832$) is near a transition from
one echelle order to the next, where weak lines are not recorded
reliably.  Hence this line was not measured.}
\tablenotetext{p}{From a measurement in a G230MB STIS spectrum of
PHL~1811 reported in Paper~I.}
\end{deluxetable}

\placefigure{lalpha}

An optimum Voigt profile fit to the right-hand portion of the Ly$\alpha$
profile was obtained using a fitting procedure created by Fitzpatrick \&
Spitzer  (1997).  Our best value for $\log N({\rm H~I})$ is 17.98.  A
significant source of uncertainty in the fit is the choice of the
continuum level.  The two choices shown in Fig.~\ref{lalpha} give values
of $N$(H~I) that indicate a (roughly 2$\sigma$) uncertainty of $\pm
0.035\,$dex.  This uncertainty combined with the formal uncertainty
found by the fitting program results in a net uncertainty of $\pm
0.05\,$dex at the 1$\sigma$ level.

\subsubsection{O I}\label{O_I}

The O~I feature recorded at the highest S/N and resolution was the
1302.17$\,$\AA\ transition observed at 1407.54$\,$\AA\ in the STIS E140M
spectrum.  The integral of the apparent optical depth $\tau_{\rm a}$
over the complete profile, $\int \ln[I_{\rm cont.}/I(v)]dv$, multiplied
by the factor $10^{14.576}/f\lambda$ yields $N({\rm O~I})=10^{14.52}{\rm
cm}^{-2}$.  This value is higher than another determination $N({\rm
O~I})=10^{14.32}{\rm cm}^{-2}$ based on a measurement of a rest-frame
equivalent width $W_{\rm r}=16.6\pm 4.4\,$m\AA\ for the 1039.23$\,$\AA\
transition.  The line is not resolved by {\it FUSE}, so we had to assume
that it has a shape for its optical depth that is similar to the one
shown at 1302$\,$\AA\ (see Fig.~\ref{oi_profile} in \S\ref{temp}).  The
most error-prone portion of the 1302$\,$\AA\ profile is the part where
the absorption is very deep, reaching a minimum intensity over a span of
3 pixels of about 0.06 times the local continuum (see
Fig.~\ref{vstack}).  If we reduce the assumed strength of the absorption
at this intensity trough enough to arrive at an intermediate $\log
N({\rm O~I})=14.47$ for the entire profile, the $\chi^2$ increases by
1.6 above the minimum value (for $\log N({\rm O~I})=14.52$).  For this
case, we find that a reconstruction of $\int \tau_{\rm a}(v)dv$ for the
1039$\,$\AA\ transition should give $W_{\rm r}=22.2\,$m\AA, which is
only $1.3\sigma$ above the measured value.  Thus $\log N({\rm
O~I})=14.47$ represents the best compromise between the two
measurements, and its uncertainty is about 0.05$\,$dex.

The measurement $W_{\rm r}/\lambda$ for the group of three lines near
988.8$\,$\AA\ is nearly the same as that for the 1302$\,$\AA\
transition, which has a value of $f\lambda$ that is about equal to the
sum of the $f\lambda$ values for the three lines.  Likewise, the upper
limit for the strength of the 976.4$\,$\AA\ line and the measurement of
three transitions at 971.738$\,$\AA\ yield results that are consistent with
our adopted column density (the error in $W_{\rm r}$ for the latter is
large because nearby strong lines prevent us from accurately measuring
the continuum level).  Generally, these lines support our conclusions
based on the 1039 and 1302$\,$\AA\ lines, but with lower accuracy.

\subsubsection{Si~II}\label{Si_II}

Of all the species for which we have absorption line data, the
information for Si~II is the most extensive.  Four features were
observed whose transition strengths span a factor of 6 and whose level
of saturation range from moderate to strong.  (The $\lambda 1260.422$
transition is apparently blended with another strong line that might be
a Ly$\alpha$ forest line.\footnote{We are unable to verify that this is
a Ly$\alpha$ feature because its Ly$\beta$ counterpart coincides with
the Fe~II $\lambda 1063.18$ absorption feature from the Lyman limit
system.  Higher Lyman series lines are probably too weak to detect.}) 
Under these circumstances, one can in principle derive a column density
by modeling the observed result in terms of an instrumentally smoothed
Voigt profile, with one or more velocity components.  A simpler
procedure that does not rely on a specific adopted model is to integrate
the apparent optical depth $\tau_{\rm a}(v)$ over velocity $v$, (as we
did for O~I in \S\ref{O_I} above), but correct for the possible
under-representation of the (smoothed) opacities caused by unresolved,
saturated structures within the profile using the method of Jenkins 
(1996).  A generalization of this method beyond the analysis of two
transitions involves finding at each velocity $v$ the best-fit curve of
growth to the four values of $\tau_{\rm a}(v)$ as a function of
$f\lambda$.\footnote{Tripp et al.  (2004) performed a similar analysis
of several Si~II transitions that had differing outcomes for $N_a\equiv
10^{14.576}\int \tau_{\rm a}(v)dv/f\lambda$ and found that this method
agreed well with Voigt profile fits performed simultaneously for the
different transitions.  This is a clear example that the two methods
agree with each other.}  The integral over velocity of the optical
depths that are corrected for saturation yields $\log N({\rm
Si~II})=13.95^{+0.05}_{-0.03}$.

\subsubsection{N~II}\label{N_II}

From our determinations of $N$(O~I) and $N$(Si~II) discussed above, we
conclude that [Si~II/O~I]$\equiv$ $\log N({\rm Si~II})-\log N({\rm O~I})
-\log ({\rm Si/O})_\odot=+0.63$ (henceforth, we will adopt a notation
[$A_i/B_j$], i.e., an extension of the standard bracket notation [$A/B$]
for total element abundances, to denote the difference between the
logarithm of the ratio of element $A$ in ionization state $i$ to element
$B$ in state $j$ compared to the logarithm of the ratio of the elements'
solar abundances $\log (A/B)_\odot$)\footnote{For solar abundances, we
have adopted the following values on a logarithmic scale with H set to
12.00: C~=~8.39  (Allende Prieto, Lambert, \& Asplund 2002), N~=~7.93 
(Holweger 2001), O~=~8.66  (Asplund et al. 2004), Si~=~7.51  (Asplund
2000), S~=~7.20  (Grevesse \& Sauval 2002), Ar~=~6.18  (Asplund et al.
2004), and Fe~=~7.46  (Asplund 2000).  There is a brief report by
Asplund, Grevesse \& Sauval  (Asplund, Grevesse, \& Sauval 2004) that
the solar abundance of N may be 0.15~dex lower than the value determined
by Holweger (2001).}.  This apparently enormous enhancement of Si over
O, compared to their solar abundance ratio, suggests that most of the
Si~II resides in a region containing fully (or mostly) ionized hydrogen,
where there should be very little O~I.  Such a region will also hold
most of the N~II.  For this reason, the velocity structure of the N~II
profile is expected to be very similar to that of Si~II.

There is an unfortunate coincidence between the location of the $\lambda
1084$ N~II line and that of a probable O~VI $\lambda 1032$ line at
$z=0.13541$ (see Paper~I). We have measured the strength of the $\lambda
1038$ O~VI line for this system and obtained marginal detections
($W_{\rm r}=14.0\pm 12.6\,$m\AA\ in the STIS spectrum and $W_{\rm
r}=16.8\pm 10.8\,$m\AA\ in the {\it FUSE\/} spectrum).  Combining the
two and assuming the $\lambda 1032$ line is twice as strong as the
$\lambda 1038$ one that we measured, we find an expected $W_{\rm
r}(1032) = 31.2\pm 16.4\,$m\AA.  If we assume that the equivalent widths
of the N~II and O~VI lines simply add to each other (which may not be
entirely correct if the lines are actually very deep and overlapping),
we find that the difference between the two, $W_{\rm r} = 55.0\pm
17.7\,$m\AA, represents our best determination of the strength of the
Lyman limit system's N~II absorption strength after allowing for the
possible contamination by the O~VI absorption feature from the other
system.

Our nominal value for the strength of the $\lambda 1084$ line expressed
in terms of $W_\lambda/\lambda$ is about equal to that of the Si~II line
at 1304$\,$\AA.  Hence, if we assume that N~II saturates in the same
fashion as Si~II (see above), $\log N({\rm N~II})=\log N({\rm Si~II})
+\log (f\lambda)_{1304}-\log (f\lambda)_{1084}=13.92$.  Likewise, this
line's $1\sigma$ upper limit for $W_{\rm r}$ translated into
$W_\lambda/\lambda$ is about the same as that for the $\lambda 1190$
Si~II line, which leads to $\log N({\rm N~II})=14.41$ by the same
argument.  The lower limit for $W_{\rm r}$ is substantially weaker than
the $\lambda 1304$ Si~II line; if we make the conservative assumption
that it is unsaturated and $N$ scales linearly with $W_{\rm r}$ (see
\S\ref{weak}), we derive $\log N({\rm N~II})=13.51$.  (A similar line of
reasoning was applied to derive the $2\sigma$ error limits shown later
in Figure~\ref{no_nh}.)

\subsubsection{Nearly Saturated Doublets: C~IV and
Si~IV}\label{doublets}

The doublets of C~IV and Si~IV both exhibit moderately strong
saturation: the ratio of the line strengths for each species is $\gtrsim
1.3$, which is less than the expected 2.0 for these lines if they were
unsaturated.  We used the standard doublet ratio method  (Str\"omgren
1948) to derive the column densities.  This method was chosen so that we
could implement a simple, yet reasonably conservative method of
estimating the errors in the column densities.  In each case, for
$1\sigma$ upper and lower bounds we repeated the analysis using $W_{\rm
r}({\rm weak})+1\sigma$ and  $W_{\rm r}({\rm strong})-1\sigma$ for the
former and $W_{\rm r}({\rm weak})-1\sigma$ and $W_{\rm r}({\rm
strong})+1\sigma$ for the latter.  We feel that the conservatism of
adopting the worst combinations of equivalent widths for the limits is
balanced by the shortcomings of the not so certain assumption that a
doublet ratio analysis gives precise answers.  Both species exhibit a
best value for $b$ equal to $9\,{\rm km~s}^{-1}$, but they are offset
from the lower ions by $+7\,{\rm km~s}^{-1}$.

In addition to the main feature of C IV, there is a broad, shallow
component centered at about $v = -50 \,{\rm km~s}^{-1}$.  Traces of
neutral hydrogen associated with this material may explain the presence
of absorption to the left of the Ly$\alpha$ Voigt profile depicted in
Fig.~\ref{lalpha}.

\subsubsection{S~III, C~III and O~VI}\label{S_III}

The two S~III lines ($\lambda 1012$ and $\lambda 1190$) have intrinsic
line strengths that differ by only 0.2$\,$dex, and the relative error in
the equivalent width of the former of the two is large.  For these
reasons, we felt that an application of the doublet ratio analysis was
not appropriate.  Instead, we noted that the radial velocity of S~III is
similar to those of C~IV and Si~IV (see Fig.~\ref{vstack}), so we
derived $N$(S~III) from $W_{\rm r}$ of the $\lambda 1190$ line recorded
by STIS under the assumption that $b=9\,{\rm km~s}^{-1}$ from the C~IV
and Si~IV doublet ratio analyses.

The single available line of C~III at 977.02$\,$\AA\ is very strong and
is almost certain to be so badly saturated that we can only derive a
lower limit that is probably far below the true value.  As with S~III,
we assumed that the velocity structure of C~III approximates that of
C~IV and Si~IV (i.e., $b=9\,{\rm km~s}^{-1}$), but our extreme lower
limit was based on $\tau_0=4$ (which, with the assumed value for $b$,
would only produce a line with $W_{\rm r}=77\,$m\AA\ if the velocity
profile were a pure Gaussian).

In Paper~I we identified a feature that was likely to be caused by O~VI
$\lambda 1032$ absorption, but with a velocity displacement of
$+110\,{\rm km~s}^{-1}$ with respect to the low ionization species.  We
were not able to confirm this identification by observing the $\lambda
1037$ feature because it is coincident with the Galactic Fe~II line at
1121.97$\,$\AA.  Another unfortunate coincidence is that the redshifted
O~VI $\lambda 1032$ line has a wavelength nearly the same as that of the
Lyman 0$-$0~P(3) feature from Galactic H$_2$.  However, we rejected this
H$_2$ line as the principal source of this absorption because it was
slightly stronger than the Lyman 1$-$0~P(3) line whose transition
strength should be 3.3 times as large.  Our improved S/N for the new
{\it FUSE\/} spectrum warrants a reinvestigation of this issue.  This
time, we find that the 0$-$0~P(3) line has $W_{\rm r}=50.7\pm
7.0\,$m\AA\ (previously reported as $78\pm 16\,$m\AA) and the 1$-$0~P(3)
line gives $W_{\rm r}=71.3\pm 7.4\,$m\AA\ (previously reported as $63\pm
11\,$m\AA).  This inversion of the relative strengths of the two H$_2$
lines now makes it more reasonable to assert that the line is actually
from Galactic H$_2$ instead of O~VI at a redshift near that of the Lyman
limit system.  Moreover, the new spectrum shows the line to be narrow,
again favoring H$_2$ as the origin rather than O~VI.  Finally, the Fe~II
line that would have blocked (or enhanced) the O~VI $\lambda 1037$ line
now seems to have about the right strength in relation to other Fe~II
lines with similar transition probabilities.  Had there been an O~VI
feature at the same wavelength, one would expect this line to appear
anomalously strong relative to the other Fe~II lines.  Thus, we withdraw
our previous interpretation that the feature that appears at
1115.85$\,$\AA\ in the {\it FUSE\/} spectrum arises from O~VI.

\subsubsection{Fe~II}\label{Fe_II}

A doublet ratio analysis based on the $\lambda 1145$ and $\lambda 2600$
lines indicates the saturation of the former is not large ($\tau_0=0.80$
for the nominal column density, 0.44 for the lower limit, and 1.49 for
the upper limit, when derived using the prescription given in
\S\ref{doublets}).   The $\lambda 1143$ line of Fe~II is too weak for a
useful determination of $N$(Fe~II), but its strength is consistent with
what we would expect.  Unfortunately, the $\lambda 1608$ line of Fe~II
is just beyond our STIS wavelength coverage.

\subsubsection{Very Weak or Undetected Lines: C~II\/$^*$, N~I, S~II and
Ar~I}\label{weak}

The lines of N~I and S~II are so weak that it is safe to assume that the
derived values of $N$ are directly proportional to the equivalent
widths.  Hence $N = 1.13\times 10^{20}(W_\lambda/\lambda)/
(f\lambda)\,{\rm cm}^{-2}$ for these features.  There is some small gain
in considering all of the useful lines of a given case instead of just
the strongest one.  Thus, for these two species we evaluated the column
densities $N_i$ and their respective errors $\sigma(N_i)$ for the
individual line measurements and then averaged the $N_i$ with weights
proportional to $\sigma(N_i)^{-2}$, yielding a combined result with an
overall error equal to $(\sum \sigma(N_i)^{-2})^{-0.5}$.

Neither C~II$^*$ nor Ar~I lines are visible in the spectra.  For Ar~I,
the $\lambda 1066$ transition is so much weaker than the $\lambda 1048$
one that it is of little value to consider it when the stronger line is
not seen.  The measurements of $W_{\rm r}$ at the locations of the
C~II$^*$ 1037$\,$\AA\ and 1336$\,$\AA\ lines and the 1048$\,$\AA\ line
of Ar~I happen to give formal numbers that are negative, but the
magnitudes of the accompanying errors are larger than or comparable to
them.  For these cases, we define  $1\sigma$ upper bounds for $W_{\rm
r}$ by using the method of Marshall  (1992) for treating apparent
nondetections of quantities that are not allowed to be negative; we then
assume that the column density limit scales with the equivalent width
according to the direct proportionality given above.

\subsubsection{C II}\label{C_II}

Features of C~II were recorded in both the {\it FUSE\/} and STIS
wavelength bands.  Unfortunately, both lines are very heavily saturated. 
Repeating the argument we made in connection with N~II (see
\S\ref{N_II}), we argue that C~II probably has a velocity profile that
is not very much different from the one for Si~II.  We note that both
the $\lambda 1036$ and $\lambda 1335$ lines show a larger value of
$W_\lambda/\lambda$ than that of the strongest Si~II line at
1193.23$\,$\AA.  Hence, a conservative lower limit arises if we assume
the lines have equal strength: $\log N({\rm C~II})\gg \log N({\rm
Si~II}) + \log (f\lambda)_{1193} - \log (f\lambda)_{1036} = 14.70$.

\subsubsection{Molecular Hydrogen}\label{h2}

If H$_2$ molecules are present, they are most likely to appear in the
$J=1$ or $J=3$ rotational levels, since these states have large
statistical weights and only moderate excitation energies. 
Table~\ref{h2lines_eqw} shows our equivalent width measurements for the
strongest transitions out of these levels that were clear of other
features identified in Paper~I (including locations where lines might be
present but were below the detection threshold).  In no case were we
able to claim that any individual line could be seen well above the
noise.  While this may be so, we could increase our sensitivity to small
amounts of H$_2$ by evaluating a weighted average of all of the formal
equivalent width measurements arising from the strongest lines from
either of the two levels.  These weighted averages were computed in the
same manner as for the multiple weak lines of N~I and S~II (see
\S\ref{weak}).

\placetable{h2lines_eqw}
\begin{deluxetable}{
l    
r    
r    
l    
r    
}
\tabletypesize{\small}
\tablecolumns{5}
\tablewidth{0pt}
\tablecaption{Equivalent Widths of H$_2$ Lines\label{h2lines_eqw}}
\tablehead{
\colhead{Transition} & \colhead{Observed} & \colhead{Transition} &
\colhead{} & \colhead{$W_{\rm r} \pm 1\sigma$ Error}\\
\colhead{Name\tablenotemark{a}} & \colhead{$\lambda$
(\AA)\tablenotemark{b}} &
\colhead{$\lambda$ (\AA)} & \colhead{$\log f\lambda$\tablenotemark{c}} &
\colhead{(m\AA)\tablenotemark{d}}\\
}
\startdata
\cutinhead{$J=1$}
W 4$-$0 R(1) &1004.920 & 929.687 &1.172&$  -5.7\pm   8.2$\\
W 3$-$0 R(1) &1022.970 & 946.386 &1.090&   2.3$\pm   6.1$\\
L 9$-$0 R(1) &1072.290 & 992.013 &1.252&$  -8.4\pm   9.3$\\
L 8$-$0 R(1) &1083.574 &1002.453 &1.256&   9.1$\pm  20.9$\\
W 0$-$0 R(1) &1090.108 &1008.497 &1.326&$ -16.6\pm  16.6$\\
W 0$-$0 Q(1) &1091.484 &1009.770 &1.384&$ -16.7\pm  18.1$\\
L 7$-$0 R(1) &1095.446 &1013.436 &1.307&$  -6.7\pm   9.2$\\
\cutinhead{$J=3$}
W 3$-$0 P(3) &1028.684 & 951.672 &1.092&$  -2.6\pm   5.9$\\
W 1$-$0 P(3) &1071.604 & 991.379 &1.075&   8.3$\pm   7.6$\\
W 0$-$0 Q(3) &1094.630 &1012.681 &1.386&   5.0$\pm   9.6$\\
L 7$-$0 P(3) &1102.003 &1019.502 &1.050&   5.2$\pm   8.5$\\
L 6$-$0 P(3) &1114.639 &1031.192 &1.055&$  -1.5\pm   7.1$\\
L 5$-$0 P(3) &1127.945 &1043.502 &1.060&  15.4$\pm  11.5$\\
L 4$-$0 R(3) &1139.268 &1053.977 &1.137&  24.0$\pm  12.8$\\
L 3$-$0 R(3) &1153.856 &1067.473 &1.028&   2.4$\pm   9.1$\\
\enddata
\tablenotetext{a}{{L} = Lyman band; W = Werner band.}
\tablenotetext{b}{Not measured values, but computed using the laboratory
wavelength and our best general fit to the absorption system's redshift,
$z_{\rm abs}=0.080923$.}
\tablenotetext{c}{Transition $f$-values are from Abgrall \& Roueff 
(1989).}
\tablenotetext{d}{Equivalent width in the rest frame of the Lyman limit
system.}
\end{deluxetable}

For our weighted averages of the column densities and their errors, we
obtained $N({\rm H}_2, J=1)=-4.3\pm 3.6\times 10^{14}{\rm cm}^{-2}$ and
$N({\rm H}_2, J=3)=4.2\pm 3.0\times 10^{14}{\rm cm}^{-2}$.  Thus, we
were not able to detect H$_2$ in the $J=1$ rotational level, but we
obtained a very marginal ($1.4\sigma$) detection in the $J=3$ state.  We
then arrive at a limit $N({\rm H}_2,J=1)+N({\rm H}_2,J=3)=-0.1\pm
4.6\times 10^{14}{\rm cm}^{-2}$, which translates into a $2\sigma$ upper
limit of $1.06\times 10^{15}{\rm cm}^{-2}$ if once again we invoke the
method of assigning upper limits outlined by Marshall  (1992).  To
arrive at a total column density for H$_2$, we must make a special
assumption about the distribution in different rotational states.  If
the rotational temperature $T_{\rm rot}$ of the H$_2$ were as high as
about 1000$\,$K, about the largest value seen in the general
interstellar medium of our Galaxy  (Spitzer, Cochran, \& Hirshfeld 1974;
Savage et al. 1977; Jenkins et al. 2000b) and elsewhere  (Levshakov et
al. 2002), the occupations of the $J=1$ and $J=3$ levels would be about
equal to each other and together they would comprise 62\% of the total
over all levels.  Thus, for $T_{\rm rot}\approx 1000\,$K, we expect that
our upper limit for H$_2$ in all $J$ levels would be about $1.7\times
10^{15}{\rm cm}^{-2}$.  It follows that the average fraction of hydrogen
in molecular form $f({\rm H}_2)\equiv 2N({\rm H}_2)/[2N({\rm
H}_2)+N({\rm H~I})]<3.6\times 10^{-3}$.  If $T_{\rm rot}$ is much lower,
say around $50-100\,$K that is typically found in denser regions of our
Galaxy  (Savage et al. 1977; Rachford et al. 2002), the limit for
$N({\rm H}_2,J=1)$ places a more stringent upper limit on the total
molecular column density, $N({\rm H}_2,{\rm all}~J)<9.4\times
10^{14}{\rm cm}^{-2}$.  There is little danger that we are overlooking
molecules in the $J=0$ state, because an extraordinarily low $T_{\rm
rot}<19\,$K would be needed to make $N({\rm H}_2,J=0)>N({\rm H}_2,J=1)$.

\subsection{Temperature of the Neutral Gas}\label{temp}

Having derived $N$(H~I) for the Lyman limit system (\S\ref{H_I}), we can
deduce its velocity dispersion from the convergence of the Lyman series
lines at wavelengths just above the Lyman limit, even if it was observed
with an instrumental resolution that is far broader than the widths of
the lines  (Jenkins 1990; Hurwitz \& Bowyer 1995).  We use this
technique to determine the widths of weak Lyman series lines, and from
this we can determine the temperature of the H~I-bearing material.

\begin{figure}
\plotone{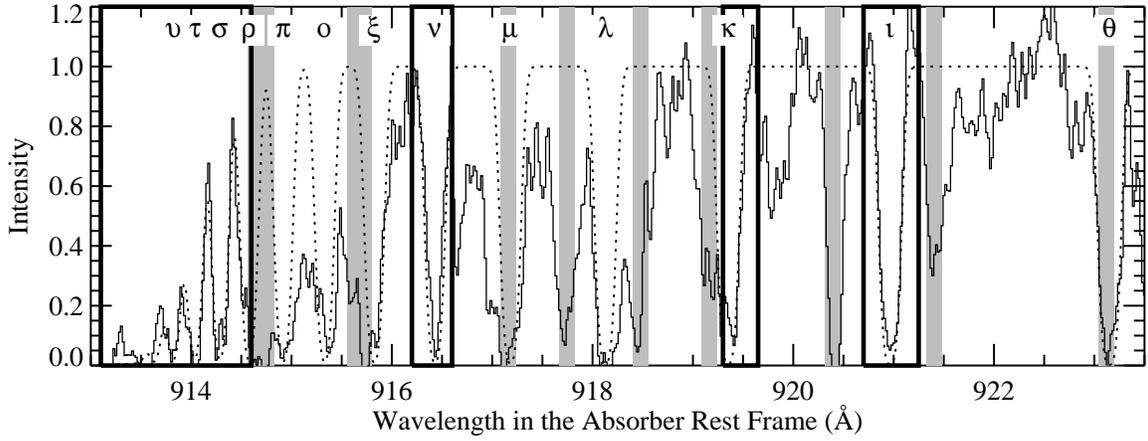}
\caption{{\it Histogram-style line:\/} Combined {\it FUSE\/} night-time
observations of intensities at wavelengths just below the Lyman limit
for the absorption system at $z=0.080923$ in front of PHL~1811.  In this
presentation, the spectrum has been binned to a resolution $\Delta
v=6\,{\rm km~s}^{-1}$ for easier viewing (best-fit calculations were
done at full resolution).  Shaded vertical bars indicate positions of
easily identified, but irrelevant features, and the boxes enclose
regions where the values of $\chi^2$ between the observed spectrum and
the synthetic trial spectrum ({\it dotted curve\/}) were evaluated. 
Greek letters indicate wavelength positions of the members of the Lyman
series lines.\label{lylimit_best_fit}} 
\end{figure}

Figure~\ref{lylimit_best_fit} shows the FUSE spectrum (data taken only
during orbital night to avoid telluric O~I $\lambda 989$ emission) at
wavelengths covering the high members of the Lyman series absorptions. 
The large number of redshifted systems toward PHL~1811, coupled with the
profusion of H$_2$ lines from our Galaxy, creates a large number of
features that interfere with the Lyman series lines of the Lyman limit
system.  Those that we identified in Paper~I (plus the O~I Galactic
lines near 989$\,$\AA) are underneath the vertical gray bars in the
figure.  Hence these regions were not used in the analysis.  Wavelength
intervals that contain Lyman series lines that seem to be free of
interference are within the tall rectangles.  Within such regions, we
evaluated the goodness of fit (as measured by $\chi^2$) of a
reconstructed spectrum to the observed one for different trial values of
the velocity dispersion $b$ and instrumental resolution.  The best fit
occurred for $b=16.4\,{\rm km~s}^{-1}$ if $\log N({\rm H~I})$ equals the
preferred value of 17.98 (\S\ref{H_I}).  If $\log N({\rm H~I})=17.88$,
i.e., the preferred value minus a $2\sigma$ deviation, the best fit
moves only to $b=17.2\,{\rm km~s}^{-1}$, with a formal upper bound of
$17.7\,{\rm km~s}^{-1}$ at the point where $\chi^2-\chi^2_{\rm
min}=1.0$.  Similarly, for $\log N({\rm H~I})=18.08$ we found that $b$
could ultimately be as low as $15.3\,{\rm km~s}^{-1}$.

\placefigure{lylimit_best_fit}

We found the best solution for the instrumental resolution to be
$29\,{\rm km~s}^{-1}$ (FWHM for a Gaussian profile), which is on the
high side of the range $22-30\,{\rm km~s}^{-1}$ found by other {\it
FUSE\/} investigators  (Heckman et al. 2001; H\'ebrard et al. 2002;
Lebouteiller et al. 2004; Sembach et al. 2004; Williger et al. 2004).  A
slight reduction in our resolution may have arisen from the difficulty
in registering the offsets of individual low S/N observing sessions with
respect to each other.

\begin{figure}
\plotone{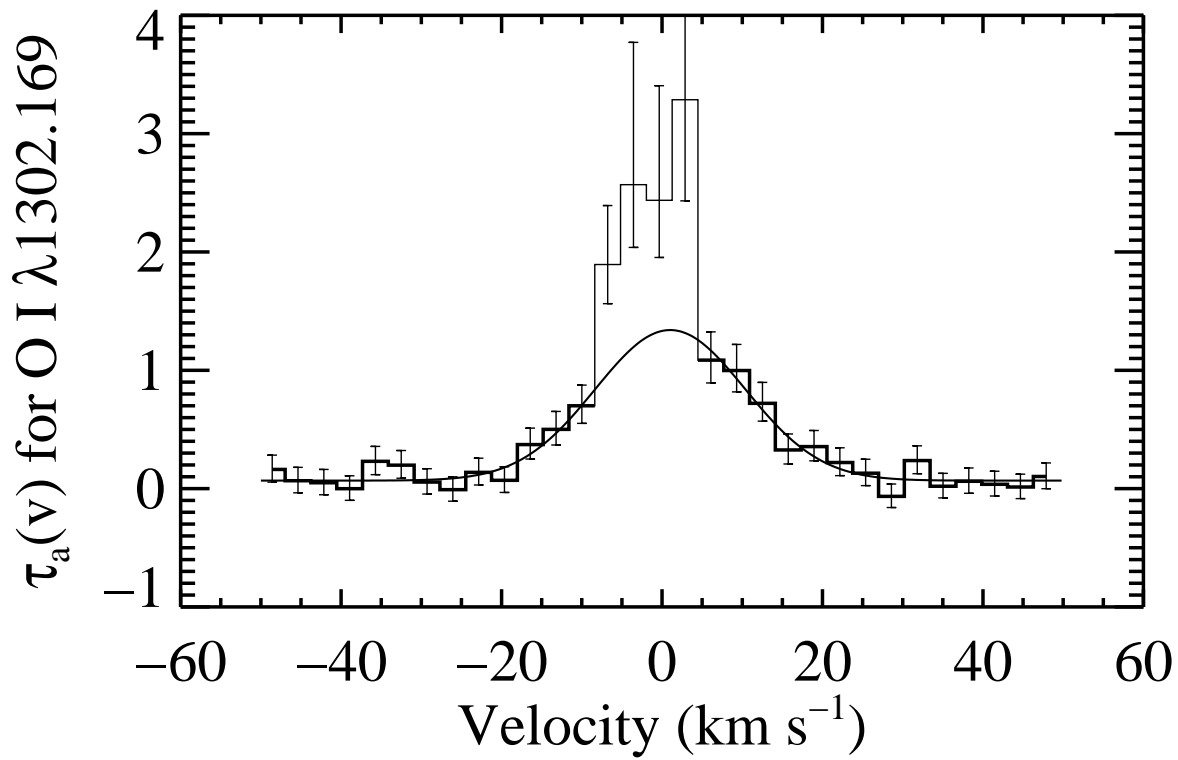}
\caption{The apparent optical depth $\tau_{\rm a}(v)\equiv \ln[I_{\rm
cont.}/I(v)]$ as a function of velocity for the $\lambda 1302$ feature
of O~I in the Lyman limit system at $z=0.080923$ in front of PHL~1811. 
The histogram-style plot shows the observations and their errors.  The
smooth profile is the best-fit Gaussian function that defines the
velocity dispersion of O~I, exclusive of the central portion that is
indicated with a thin line.\label{oi_profile}}
\end{figure}

We can use the $\lambda 1302$ O~I absorption observed by STIS to
indicate the contribution by turbulent broadening. 
Figure~\ref{oi_profile} shows the conversion of this feature into
apparent optical depths $\tau_{\rm a}(v)$.  Here, we can see that the
lower portion of the velocity profile approximates a Gaussian
distribution with $b=13.2\pm 1.4\,{\rm km~s}^{-1}$, which may be reduced
to $12.6\,{\rm km~s}^{-1}$ when a compensation is made for instrumental
smearing.  The behavior of the upper portion of the O~I profile has no
effect on our temperature derivation because the overall strength of the
O~I feature is much greater than those of the H~I Lyman series
absorptions that defined $b_{\rm H~I}$.  (In effect, this portion of the
profile should be completely saturated in the H~I lines we analyzed.) 
If we solve the equations
\begin{mathletters}
\begin{equation}
b^2_{\rm O~I}={kT\over 8m_{\rm p}}+b^2_{\rm turb}
\end{equation}
\begin{equation}
b^2_{\rm H~I}={2kT\over m_{\rm p}}+b^2_{\rm turb}
\end{equation}
\end{mathletters}
we find that
\begin{equation}\label{temp_eval}
T={8m_{\rm p}(b^2_{\rm H~I}-b^2_{\rm O~I})\over
15k}=7070^{+3860}_{-4680}\,{\rm K}
\end{equation}
for the preferred values of $b_{\rm H~I}$ and $b_{\rm O~I}$, with the
error limits defined from their worst extremes working in opposite
directions.

\placefigure{oi_profile}

In principle, it is possible that in this analysis we are being deceived
by a small amount of hydrogen that could have a much larger turbulent
velocity dispersion.  For the ratio of H~I to O~I in this absorption
system, Lyman series lines near Ly$\sigma$ (at 914.3$\,$\AA), which
probably have the most influence in determining $b$(H~I), have optical
depths about 10 times that of the O~I $\lambda 1302$ feature.  In
essence, our analysis rests on the assumption that the Gaussian profile
shown in Fig.~\ref{oi_profile} remains the same down to $\tau_{\rm
a}\sim 0.1$, which may not be true.  However, it is reassuring that the
observed Ly$\iota$ profile is not wider than the theoretical one, and
the fit to the right-hand edge of Ly$\alpha$ shown in Fig.~\ref{lalpha}
gives a value $b=16.4\,{\rm km~s}^{-1}$ with a formal uncertainty in the
fit of $0.4\,{\rm km~s}^{-1}$.  This line is 340 times as strong as
Ly$\iota$, which gives us some reassurance that the broadening is indeed
purely thermal.  Thus, except for hydrogen at negative velocities (shown
by the shaded portion of the profile in Fig.~\ref{lalpha}), the Gaussian
shape seems to continue to much lower $\tau_{\rm a}$ levels in the
hydrogen lines than we can see in the faintest wings of the equivalent
profile in O~I.

\section{Interpretation of the Spectra}\label{spectra_interpretation}

\subsection{Electron Density}\label{elec_dens}

From our upper limit for $N$(C~II$^*$), we can derive a limit for the
excitation rate of this upper fine-structure level and, in turn, the
electron density in the C~II-bearing material (which is probably mostly
ionized, so we can ignore collisional excitation by hydrogen atoms). 
The electron density is given by the relation\footnote{This formula
applies only to the case where $T\gg \Delta E/k = 95\,$K and $N({\rm
C~II^*})\ll N({\rm C~II})$.  For the more general formula that applies
to the case where these restrictions are violated, see, e.g., Eq.~5 of
Jenkins, Gry \& Dupin  (2000).  This reference also gives the sources of
the atomic data that were incorporated into this equation.}
\begin{equation}\label{ne_formula}
n(e)=0.189T^{0.5}\left[{N({\rm C~II^*})\over N({\rm C~II})}\right]
\end{equation}
Since we have only a lower limit for $N$(C~II), and one which is
probably well below the real value, we can instead use $N({\rm
Si~II})({\rm C/Si})_\odot$ (or the same with S~II, since $[{\rm
S~II/Si~II}]\sim 0$) as a proxy for $N$(C~II).\footnote{From the
standpoint of possible element mixtures that could arise from a less
chemically evolved system than the present-day gas in our Galaxy, Fe
would be a better match to C  (Dessauges-Zavadsky et al. 2003a;
Matteucci \& Chiappini 2003), even though these two elements are
produced chiefly from different nucleosynthetic sources.  However, Fe
has the disadvantage of possibly being depleted onto dust grains in the
medium that we are examining.  Had we used Fe instead of Si as the proxy
for C, we would have derived limits for $n_e$ that were a factor 2 less
stringent than those given in Eq.~\ref{ne_result} (see
Table~\protect\ref{rel_abund}).  In our environment where partial
photoionization takes place, we expect that the ion fractions of Si, S
and C are very similar.  For instance, in the CLOUDY models discussed in
\protect\S\ref{cloudy}, Si and S have fractions in the singly ionized
state within 0.1$\,$dex of that of C and nearly identical derivatives
thereof with respect to $\log U$ for either $\log U = -4.4$ or $-3.6$. 
The difference in depletion of Si and C onto grains is likely be very
small, since we find that [S~II/Si~II]$\sim 0$, and S usually does not
deplete appreciably in our Galaxy  (Savage \& Sembach 1996).} 
Numerically, we find that for the temperature and limits thereof
expressed in Eq.~\ref{temp_eval} that Eq.~\ref{ne_formula} yields the
values
\begin{eqnarray}\label{ne_result}
n(e) & \leq & 0.032\,{\rm cm}^{-3}~{\rm for}~T=2390\,{\rm K}\nonumber\\
& \leq & 0.056\,{\rm cm}^{-3}~{\rm for}~T=7070\,{\rm K}\nonumber\\
& \leq & 0.069\,{\rm cm}^{-3}~{\rm for}~T=10930\,{\rm K}~.
\end{eqnarray}

\subsection{General Patterns of Element
Abundances}\label{general_patterns}
\begin{deluxetable}{
c    
c    
}
\tablecolumns{2}
\tablewidth{200pt}
\tablecaption{Relative Abundances\tablenotemark{a} of Specific Atoms and
Ions\label{rel_abund}}
\tablehead{
\colhead{Ratio} & \colhead{Value}\\
}
\startdata
\cutinhead{Species relative to H~I}
{[C~II/H~I]}&$\gg 0.33$\\
{[N~I/H~I]}&$-1.47^{+0.30}_{-\infty}$\\
{[O~I/H~I]}&$-0.17\pm 0.07$\\
{[Si~II/H~I]}&$0.46^{+0.07}_{-0.06}$\\
{[S~II/H~I]}&$0.52^{+0.17}_{-0.25}$\\
{[Ar~I/H~I]}&$<0.44$\\
{[Fe~II/H~I]}&$0.15^{+0.15}_{-0.13}$\\
\cutinhead{Species relative to Si~II}
{[C~II/Si~II]}&$\gg -0.13$\\
{[N~II/Si~II]}&$-0.45^{+0.49}_{-0.41}$\\
{[S~II/Si~II]}&$0.06^{+0.16}_{-0.25}$\\
{[Fe~II/Si~II]}&$-0.31^{+0.14}_{-0.13}$\\
\enddata
\tablenotetext{a}{See the text at the beginning of \S\protect\ref{N_II}
for the definition of an abundance ratio.  Error limits represent $\pm
1\sigma$ deviations, but do not include possible systematic errors that
could arise from incorrect $f-$values or solar reference abundances.  In
this table, no ionization corrections have been applied. 
Table~\protect\ref{cloudy_abund} shows how ionization corrections can
affect the inferred abundances.}
\end{deluxetable}

The top portion of Table~\ref{rel_abund} summarizes the abundances of
various elements with respect to H~I, compared to their respective solar
abundance ratios.  There are striking differences from one element to
the next.  At the most superficial level, we note that the singly
ionized forms of C, Si, S and Fe appear to have supersolar abundances,
while those of the neutral atoms N and O are subsolar. The ions all have
ionization potentials more than a few eV higher than that of H, while
the neutral atoms N and O do not.  Thus, the pattern we note here is
probably caused by conditions that favor having a substantial amount of
the hydrogen in a fully or partially ionized condition, and the hydrogen
ions far outnumber the neutral atoms that coexist with the N~I and O~I.

\placetable{rel_abund}

\begin{figure}
\plotone{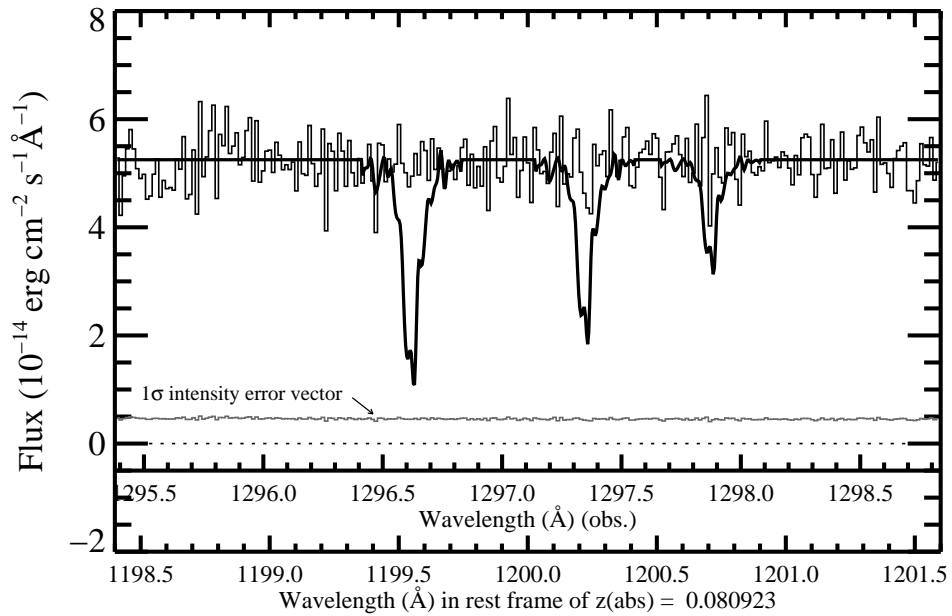}
\caption{The STIS E140M spectrum of PHL~1811 covering the expected
wavelength of the 1200$\,$\AA\ triplet for the Lyman limit absorption
system at $z=0.080923$ ({\it Histogram style trace\/}).  Overplotted are
three replications of the 1302$\,$\AA\ O~I feature, but with their
optical depths rescaled to show how N~I features would appear if N~I and
O~I had the same velocity profiles and relative abundances that agreed
with the solar ratio.\label{ni_spectrum}}
\end{figure}

In the region containing H~I, there is a spectacular deficiency of
nitrogen with respect to oxygen: [N~I/O~I]~$\lesssim -1.3$.  We
illustrate the strength of this abundance disparity for these two
neutral atoms in Figure~\ref{ni_spectrum}, where the observed spectrum
covering the 1200$\,$\AA\ triplet of N~I has three replicas of the
1302$\,$\AA\ feature overplotted on it, each with an apparent optical
depth that is rescaled to agree with what the respective underlying N~I
line would look like if [N~I/O~I] were equal to zero.  Later, in
\S\S\ref{collisional}--\ref{more_detailed}, we will examine the
secureness of our conclusion that a low value of [N~I/O~I] indicates a
true underabundance of nitrogen and that it is not simply a consequence
of drastically different responses to various ionizing processes.

\placefigure{ni_spectrum}

To examine the abundance patterns of the ions, we use Si~II as a
standard, since its abundance has the smallest errors.  The bottom
portion of Table~\ref{rel_abund} indicates that Fe and N appear to be
underabundant, while the other two elements, C and S, have abundances
that are consistent with the solar ratios.  The mild underabundance of
Fe may be caused by either its depletion onto dust grains, since Fe
generally depletes more rapidly than the other elements  (Savage \&
Sembach 1996; Jenkins 2004), or by a smaller contribution from Type~Ia
supernovae compared to the mix of sources that contributed to the
chemical enrichment of our Galaxy.

It is important to note that the atoms within the region are shielded
from a uniform, external ionizing radiation field by a column of neutral
hydrogen that approaches only $5\times 10^{17}{\rm cm}^{-2}$.  Moreover,
if the region is porous, or is in a thin sheet inclined to the line of
sight, or contains internal sources of ionization (i.e., recently formed
stars), the shielding could be considerably smaller.  Hence, we cannot
simply assume that virtually all atoms or ions in a region containing
H~I must be concentrated within the lowest stage that has an ionization
potential greater than that of hydrogen.  It is therefore clear that
meaningful interpretations of element abundances must allow for
alterations that might arise from elements being distributed in
different ionization levels, many of which are unseen.  In the three
subsections that follow, we consider the consequences of this problem. 
We start with the possibility that collisional ionization could play a
role, and then we examine the effects of photoionization from several
different perspectives.

\subsection{Collisional Ionization}\label{collisional}

For the temperature range of the H~I-bearing material that we derived in
\S\ref{temp}, there should be negligible ionization of atoms and first
ions to higher stages of ionization if the gas is in equilibrium 
(Sutherland \& Dopita 1993).  While this may seem to rule out
collisional ionization as an important factor, we must not overlook the
possibility that the gas might have been very hot at some earlier time
and has cooled radiatively.  The time scale for such cooling is shorter
than the recombination time, so the higher ionization from an earlier
time may effectively be ``frozen in'' at the current epoch, or at least
partially so.  For a minimum electron density $n(e)=10^{-3}{\rm
cm}^{-3}$ that we derive in \S\ref{uniform_slab} below and a
representative rate of recombination for singly-ionized atoms
$\alpha(T)=10^{-12}{\rm cm}^3{\rm s}^{-1}$ at $T\sim 10^4\,$K, the
recombination time scale is short: $\tau=[(n(e)\alpha({\rm
H},T)]^{-1}=3\times 10^7{\rm yr}$.  It seems implausible that we are
viewing the gas at just the right moment within the probable total
lifetime of the system.

\subsection{Photoionization in a Uniform Slab Illuminated by the
Intergalactic Field}\label{uniform_slab}

\subsubsection{General Considerations}\label{general_considerations}

A simple picture to consider is a uniform, infinite slab that is
immersed in a bath of ionizing radiation.  In this regime, we can
implement the CLOUDY ionization code (v94.0, Ferland et al. 1998) and
compare the calculations with our observations.  We employ CLOUDY
following the procedures described in Tripp et al.  (2003); we assume
that the gas is photoionized by the UV background from QSOs according to
the calculations of Haardt \& Madau
 (1996), and we set the intensity of the flux at 1~Rydberg to $J_{\nu} =
1 \times 10^{-23}$ ergs s$^{-1}$ cm$^{-2}$ Hz$^{-1}$ sr$^{-1}$, a value
in accord with current observational constraints  (Shull et al. 1999;
Dav\'e \& Tripp 2001; Weymann et al. 2001).
\begin{figure}
\epsscale{0.7}
\plotone{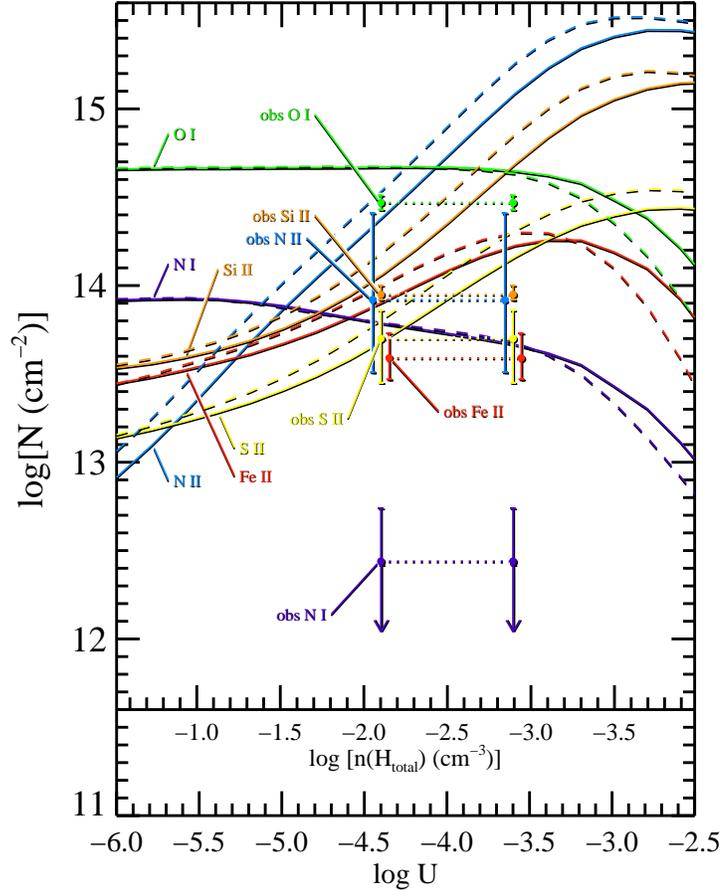}
\caption{The trends for the logarithms of predicted column densities
from a CLOUDY model calculation for different atoms and ions,
each depicted by a different color, as a
function of the logarithm of the ionization parameter $U$, as seen
through a slab perpendicular to the line of sight with a thickness that
matches the observed $N({\rm H~I})=10^{17.98}{\rm cm}^{-2}$ (solid
curves) and irradiated on both sides by an intergalactic radiation field
specified by Haardt \& Madau  (1996).  All curves are based on solar
abundance ratios of the various elements with respect to hydrogen.  The
dashed curves show the same information, except that here the slab is
assumed to be inclined to the line of sight by $60\arcdeg$, making the
physical thickness of the slab smaller by a factor of 2, but with
perceived column densities that are increased by the same factor to
compensate.  An extra scale for the $x$ axis shows the values of $\log
n(H_{\rm total})$ (i.e., the sum of the ionized and neutral hydrogen
local densities).  The pairs of points with error bars show the observed
column densities for comparison, located at assigned values of $\log
U=-4.4$ and $-3.6$ that seem to give a good agreement with the curves
using different criteria (\S\S\protect\ref{lUm4.4} and
\protect\ref{lUm3.6}).}\label{cloudy}
\end{figure}

Figure~\ref{cloudy} shows how the expected column densities of different
species vary as a function of the ionization parameter $U$, which equals
the ratio of density of hydrogen ionizing photons to the total density
of hydrogen (in both neutral and ionized forms).  To make the
comparisons of the observations and theory easy to interpret in terms of
the solar abundances, the intrinsic abundances of the elements in the
CLOUDY calculation are set to be equal to the solar abundances.  The
solid curves depict the outcome for a slab perpendicular to the line of
sight, so that the observed column density $N({\rm H~I})=10^{17.98}{\rm
cm}^{-2}$ equals its true thickness.  Of course, there is a good chance
that the slab is not perpendicular to our viewing direction.  In fact,
with random possible orientations the median inclination would be
$60\arcdeg$, and under this circumstance the true thickness of the disk
(and hence the shielding of material well inside the slab) would be
reduced by a factor of 2.  The effects arising from this inclination are
small, and they are shown by the dashed lines in the figure.  The range
of temperatures calculated by CLOUDY start from 940$\,$K at $\log
U=-6.0$ and increase steadily to 8300$\,$K at $\log U=-2.5$.

\placefigure{cloudy}

The results of our observations of singly-ionized or neutral species in
the Lyman limit system are shown in Fig.~\ref{cloudy} by the vertical
locations of pairs of points with error bars (joined by dashed lines). 
Figure~\ref{vstack} indicates that the multiply ionized atoms C~IV,
Si~IV and S~III (shown in the lowest three panels in the figure) are
slightly offset in velocity from the others.  Hence, either they arise
from a location that is distinctly different from places where the lower
excitation species are found, or perhaps they reside in a special
boundary region where there is some velocity shear with respect to the
region with lower excitation.  For this reason, we feel that it is
probably unwise to use these stages as a guide for determining the
conditions that affect the lower ions.  Indeed, we find that the CLOUDY
models cannot simultaneously match the column densities of the low and
high ions at a single ionization parameter.  This corroborates our
conclusion that the absorber is a multiphase entity.

In the following two subsections, we explore the outcomes for two
different values of $\log U$, a free parameter that we can adjust to
give the best level of self consistency for the element abundances.

\subsubsection{An Initial Choice for the Ionization
Parameter}\label{lUm4.4}
\begin{deluxetable}{
c    
c    
c    
c    
c    
c    
c    
}
\tablecolumns{7}
\tablewidth{0pt}
\tablecaption{Relative Element Abundances from the CLOUDY
Model\tablenotemark{a}\label{cloudy_abund}}
\tablehead{
\colhead{} & \multicolumn{2}{c}{$\log U=-4.4$} & \colhead{} &
\multicolumn{2}{c}{$\log U=-3.6$}\\
\colhead{} & \multicolumn{2}{c}{at the Indicated Inclinations} &
\colhead{} & \multicolumn{2}{c}{at the Indicated Inclinations}\\
\cline{2-3} \cline{5-6}\\
\colhead{Ratio} & \colhead{$0\arcdeg$} & \colhead{$60\arcdeg$}
& \colhead{~~} & \colhead{$0\arcdeg$} & \colhead{$60\arcdeg$} &
Error\tablenotemark{b}\\
}
\startdata
{[C/H]}&$\gg -0.16$&$\gg -0.32$&&$\gg -0.75$&$\gg -0.93$&\nodata\\
{[N/H]}$_{\rm N~II}$\tablenotemark{c}&$-0.41$&$-0.56$&&$-1.11$&$-1.27$& 
+0.49, $-0.41$\\
{[N/H]}$_{\rm N~I}$\tablenotemark{d}&$-1.35$&$-1.35$&&$-1.24$&
$-1.24$&+0.30, $-\infty$\\
{[O/H]}&$-0.20$&$-0.20$&&$-0.18$&$-0.18$&$\pm 0.07$\\
{[Si/H]}&$-0.11$&$-0.22$&&$-0.77$&$-0.88$&+0.07, $-0.06$\\
{[S/H]}&0.01&$-0.10$&&$-0.47$&$-0.60$&+0.17, $-0.25$\\
{[Fe/H]}&$-0.34$&$-0.42$&&$-0.66$&$-0.72$&+0.15, $-0.13$\\
\enddata
\tablenotetext{a}{Assumes that a uniform slab is illuminated by the
intergalactic radiation field calculated by Haardt \& Madau  (1996) and
that $N({\rm H~I})=10^{17.98}{\rm cm}^{-2}$ for two assumed values of
$\log U$.  Slightly different results are obtained for two inclinations,
as indicated by the solid and dashed curves in
Fig.~\protect\ref{cloudy}.}
\tablenotetext{b}{The errors expressed in this column apply to all
entries in a given row.  The limits indicate $1\sigma$ uncertainties in
the observed ratios of ions or atoms and do not include systematic
errors arising from uncertainties in the atomic parameters, solar
reference abundances, or assumptions about the model.}
\tablenotetext{c}{From the determination of $N$(N~II), which emphasizes
the material associated with the more highly ionized gas.}
\tablenotetext{d}{From the determination of $N$(N~I), which emphasizes
the gas associated with the more neutral gas.}
\end{deluxetable}

Since the nucleosynthetic origins of O, Si and S are very similar 
(Wheeler, Sneden, \& Truran 1989; Thuan, Izotov, \& Lipovetsky 1995;
Chen et al. 2002) and the depletions of these elements onto grains is
not appreciable in regions of low density  (Savage \& Sembach 1996;
Jenkins 2004), we may start with a provisional assumption that there are
not any serious deviations from the solar abundance ratios between these
three elements.  If we do so, we find that a CLOUDY model with $\log U
\approx -4.4$ gives a generally acceptable fit to the observations for
most of the low ions, but with a glaring exception for N~I.  From the
flatness of the predictions for O~I and the tight error limits
associated with our observation of $N({\rm O~I})$, together with the
knowledge that O usually is only mildly depleted in the ISM of our
Galaxy  (Cartledge et al. 2001, 2004; Andr\'e et al. 2003), we may state
that the most secure measure of an overall metallicity is given by
$[{\rm O/H}]=-0.20\pm 0.07$, which represents the distance the
measurement is below the model prediction for a solar abundance ratio. 
The abundances of other elements depend more sensitively on the assumed
value for $\log U$.  The left-hand portion of Table~\ref{cloudy_abund}
shows the element abundances for $\log U = -4.4$; considering that O is
slightly subsolar, Si and S seem to have close to their expected
abundances, but Fe appears to be depleted by approximately $-0.2\,$dex. 
The temperature calculated by CLOUDY at this value of $\log U$ is
5900$\,$K, a value consistent with our analysis discussed in
\S\ref{temp}.

\placetable{cloudy_abund}

\subsubsection{Alternative Choice for the Ionization
Parameter}\label{lUm3.6}

While our choice of setting $\log U$ to $-4.4$ gives relative abundances
of O, Si, S and Fe that are generally consistent with abundance ratios
for stars and the ISM of our Galaxy and other systems, the model
indicates a very strong deficiency of N.  Beyond this, however, we find
an untidy mismatch between the observations and predictions for nitrogen
in the neutral and singly-ionized forms.  We can obtain a better
concordance for these two forms by raising the value of $\log U$, but
this comes at the expense of making other $\alpha$-process elements
discordant with their solar abundance ratios.  If we raise $\log U$ by
0.8$\,$dex to the value $-3.6$, the two values for the nitrogen
abundance match each other.  The right-hand part of
Table~\ref{cloudy_abund} shows the abundance ratios of all elements with
this higher value of $\log U$.  The disparity in the results for the two
forms of N vanishes, but N is still very deficient relative to the other
elements.  It now becomes harder to define an overall metallicity, since
the outcome depends on which element one choses as the standard.  The
predicted temperature of the gas $T=7000\,$K at this new value of $\log
U$ is slightly higher than before, but it is still completely consistent
with what we derived in \S\ref{temp}.

We believe that it is unlikely that $\log U$ could be much higher than
about $-3.5$ because the disparity between the inferred abundances of
different $\alpha$-process elements becomes unacceptably large.  If the
intergalactic field is the only radiation field present, this finding
sets a lower limit on the electron density $n(e)>10^{-3}{\rm cm}^{-3}$. 
Any local sources of additional ionizing photons, a possibility that we
will consider in the following subsection, should raise this limit.

For either choice of the inclination angle and $\log U<-3.5$, nearly all
of the silicon is singly ionized.  If we accept our finding that $-0.11
< [{\rm Si/H}] < -0.88$, depending on our choice for the inclination and $\log
U$, our measured $\log N({\rm Si~II})=13.95$ implies that the amount of
hydrogen in neutral and ionized forms amounts to $18.55 < \log N({\rm
H}_{\rm total}) < 19.34$.  If the silicon in the system is depleted onto
grains (at most, by no more than 0.4$\,$dex -- see
\S\ref{more_detailed}), this hydrogen column density could be somewhat
larger.

\subsection{More Detailed Photoionization Modeling for N and
O}\label{more_detailed}

The striking difference between the abundances of N~I and O~I warrants
further study.  Up to now, we have ignored the possibility that the
photoionization of the gas might be enhanced over that provided by the
extragalactic radiation field.  In this case, one could imagine that we
are being misled by the effects arising from an intense, internal
radiation field provided by embedded hot stars.  The configuration could
be such that there would be insignificant shielding of the gas, or much
of it, by any layers of neutral hydrogen or helium.

At first glance, we might suppose that any additional ionizing flux
should be very weak, since there is empirical evidence that the rate of
star formation per unit area generally scales in proportion to the mass
surface density to the 1.4 power  (Kennicutt 1998).  In our Lyman limit
system, our estimate for the surface density $N({\rm H}_{\rm
total})=10^{18.54}-10^{19.34}{\rm cm}^{-2}=0.04-0.23\,{\rm M}_\odot~{\rm
pc}^{-2}$ derived from $N$(Si~II) is considerably lower, for instance,
than the indications from H~I in the local region of our Galaxy, $N({\rm
H~I})=10^{20.8}{\rm cm}^{-2}=6.4\,{\rm M}_\odot~{\rm pc}^{-2}$  (Dickey
\& Lockman 1990) (and might be lower still if the Lyman limit system is
flat and inclined to the line of sight).  Nevertheless, we cannot
exclude the possibility that we are viewing through a small gap in a
disk system that is, in general, considerably thicker than our
observations indicate.  If this system subtends a small angle in the
sky, we probably would not be able to see its starlight because we are
blinded by the light from the background quasar.

For this new picture with internal sources of ionization, it might seem
that we are confronted with a hopeless task of grasping the enormous
range of possible conditions that govern the relative ionizations of
various elements and converging upon a limited solution space.  However,
the problem becomes tractable when we focus on the ratios of species
that are only mildly susceptible to the strength of the ionization, and
then use other combinations of observed species that are very sensitive
to ionization effects to serve as indicators of the general severity of
ionizing field.  We will use this approach in the analysis that follows.

Our goal is now to try to explain the result shown in
Table~\ref{rel_abund} that [N~I/H~I] is less than [O~I/H~I] by about
1.3$\,$dex or more.  Fortunately, under most circumstances the neutral
fractions of these three elements are coupled to each other by strong
charge exchange reactions  (Field \& Steigman 1971; Steigman, Werner, \&
Geldon 1971).  However, at some point when the ionizing field becomes
very strong and/or the densities are low, the coupling will start to
break down.  We can see this effect for the predictions shown in
Fig.~\ref{cloudy}.  The column densities of N~I and O~I for a fixed
value of $N$(H~I) are stable for $\log U<-5.0$, but above this value
$N$(N~I) starts on a slow downward drift.  The behavior of $N$(O~I)
remains flat up to $\log U=-3.5$, beyond which there is a steep decline. 

There is good astrophysical evidence to support the notion that
ionization effects can lower [N~I/O~I].  In the local interstellar
medium within about 100$\,$pc of the Sun, $[{\rm N~I/O~I}]\approx -0.2$
instead of the expected value of zero, and this is consistent with the
calculated effects of photoionization in a low density medium
illuminated by local sources of EUV radiation  (Jenkins et al. 2000a;
Lehner et al. 2003).  It is important to establish whether or not the
low value of [N~I/O~I] in the Lyman limit system under investigation
here is a reflection of a true abundance anomaly or, alternatively,
simply a consequence an even more severe shift in ionization caused by a
strong EUV radiation bath (or that it arises from a combination of the
two).

In the Appendix of this paper, we develop the equations that govern the
ionization equilibria of H, He and a trace element (either N or O in
this study).  Using these equations (Eqs.~\ref{n0} to \ref{n(e)}), we
have determined the expected ratios for the neutral fraction of N to
that of O, $f_0({\rm N},T)/f_0({\rm O},T)$, in the presence of
unshielded radiation fields that are identical to the fluxes computed by
Sternberg, Hoffmann \& Pauldrach  (2003) for 3 different stellar
temperatures $T_*$ and surface gravities $g$, ones that correspond to
spectral types O9.5~I, O7.5~I and O5~I  (Vacca, Garmany, \& Shull 1996). 
Our ionization calculations were performed for 3 values of the gas
kinetic temperature $T$, equal to the lower limit, best value and upper
limit derived in \S\ref{temp}.  For all cases, we evaluated the results
for progressively stronger overall intensities (or lower densities).  In
order to use this information, however, we must define a practical limit
for the ratio of the radiation density divided by the particle density.

It is clear from the simple application of the CLOUDY calculation shown
in Fig.~\ref{cloudy} that the expected amount of Si~II responds rapidly
to changes in $U$.  We can make good use of this behavior, since we have
high quality observations of this element for two stages of ionization,
Si~II and Si~IV.  (Unfortunately, the Si~III $\lambda 1206$ line of the
Lyman limit system partly overlaps the Galactic Si~II $\lambda 1304$
line.)  Thus, to gauge the strength of the ionization we supplemented
our calculations for N and O with evaluations of the expected behavior
of Si$^+$ from the same equations (but with ionization levels one higher
than those indicated by Eqs.~\ref{n0}$-$\ref{n+}).

\begin{figure}
\plotfiddle{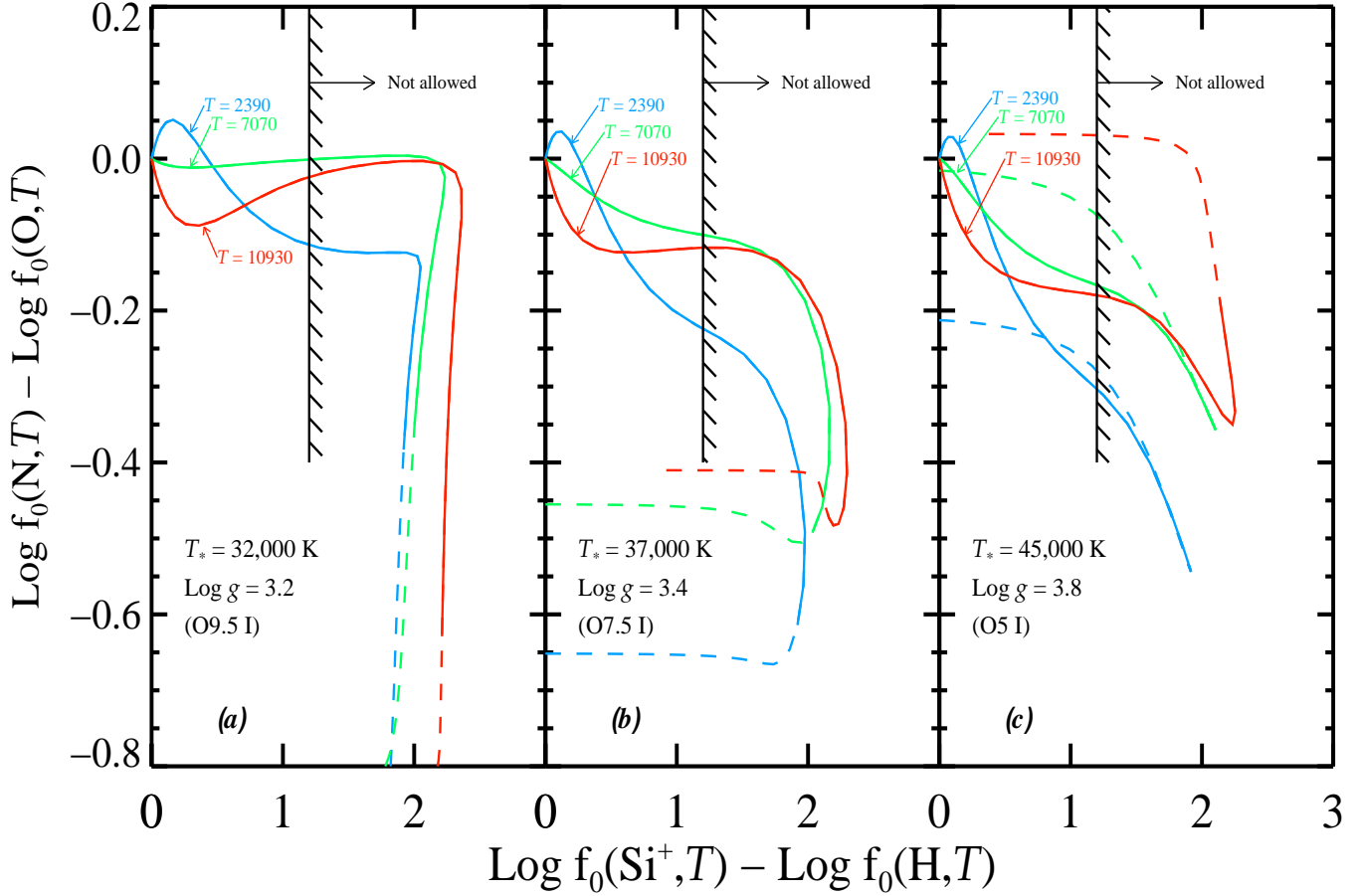}{1in}{0}{600}{400}{-100}{0}
\caption{\small {Logarithmic presentations of the expected neutral
fraction of N divided by that of O, as a function of the fraction of Si
in the singly ionized state divided by the neutral fraction of H.  The
gas is assumed to be exposed to ionizing radiation fields characterized
by the fluxes from supergiant stars at 3 different temperatures $T_*$ 
(Sternberg, Hoffmann, \& Pauldrach 2003), as indicated by the label in
each panel.  Self shielding of the gas by neutral hydrogen or helium is
assumed to be negligible.  As the curves progress away from the origin,
the density of the ionizing radiation divided by that of the atoms
increases.  The three curves in each panel trace the relationships for
the preferred value of temperature derived in \S\ref{temp} (green curve) plus
reasonable deviations on either side of this temperature defined by the
error limits (blue and red curves).  Values of
 $\log [f_0({\rm Si^+},T)/f_0({\rm H},T)]$ to
the right of the vertical feathered line are not allowed (see text). 
The dashed extensions of the curves represent portions of the curves
where the predicted value of ${\rm Si^{+3}/Si^+}$ exceeds 0.7, which
represents our largest value of $N$(Si~IV) divided by our smallest
$N$(Si~II).}}\label{myioniz}
\end{figure}

Figure~\ref{myioniz} shows how $\log f_0({\rm N},T)-\log f_0({\rm O},T)$
behaves as a function of our ionization strength indicator, $\log
f_0({\rm Si^+},T)-\log f_0({\rm H},T)$.  At the point where the curves
bend sharply downward in the first two figure panels ($a$ and $b$, for
$T_*=32,000$ and 37,000$\,$K, respectively), the value of Si$^+$/H is no
longer useful as an ionization index, but somewhat beyond this point
Si$^{+3}$/Si$^+$ emerges as an indicator.  This is important, because at
extremely large ratios of radiation density to gas density, Si$^+$/H
starts to return downward (as shown by the doubling back of the curves
in the middle panel $b$).  However, we can exclude these extreme values
because they would produce higher values of Si$^{+3}$/Si$^+$ than the
observed ratio of $N$(Si~IV) (plus a $1\sigma$ error) to $N$(Si~II)
(minus a $1\sigma$ error) that should not exceed 0.70.  The dashed
portions of the curves indicate the regions where this violation occurs. 
For the highest stellar temperature shown in the right-hand panel $c$,
the  Si$^{+3}$/Si$^+$ constraint is not useful, since the curves double
back toward the same general values of $\log f_0({\rm N},T)-\log
f_0({\rm O},T)$.

\placefigure{myioniz}

We must now use our observations to provide an estimate for the maximum
permissible value for $f_0({\rm Si^+},T)/f_0({\rm H},T)$ in our gas
system.  From Table~\ref{rel_abund} we reported that $[{\rm
Si~II/H~I}]=0.46^{+0.07}_{-0.06}$.  Were it not for ionization effects,
we would conclude from our result $[{\rm
S~II/Si~II}]=0.06^{+0.16}_{-0.25}$ that Si is not more depleted than S
by more than about 0.22$\,$dex\footnote{It is generally believed that S
has very little depletion in the Galactic interstellar medium, although
we caution that contributions from H~II regions might mask the effects
of real depletions in the H~I regions.}.  If we allow for the
possibility that S might be slightly depleted and that there are small
differences in the response of S and Si to ionization effects (as
indicated in Fig.~\ref{cloudy}), we can adopt a conservative position
that the depletion of Si onto grains could be as high as 0.4$\,$dex.  If
this is true, and we assume that [Si/O] is not appreciably different
from zero, we can state that ionization conditions that would be
expected to create values of $\log f_0({\rm Si^+}) - \log f_0({\rm H}) >
1.20$ should be excluded by our observations.\footnote{The value 1.20 is
obtained from the sum of the following logarithmic factors: (1)
[Si~II/H~I]~=~0.46, (2) the estimated error in [Si~II/H~I]~=~0.07 (3)
$-$[O/H]~=~0.19, (4) the estimated error in [O/H]~=~0.08, and (5) the
estimate for the strongest possible depletion of Si of 0.40.  We are not
considering here the possibility that S, which was used to create a limit
for the depletion of Si, could be below its solar abundance relative to
O, i.e., [S/O] $<0$, as suggested by the outcome for calculations with
only the extragalactic irradiation and $\log U=-3.6$
(\S\protect\ref{lUm3.6}).}  This limitation is shown by the vertical
feathered line in each panel of Fig.~\ref{myioniz}.  Ultimately, we find
that to the left of this line $\log f_0({\rm N},T)-\log f_0({\rm O},T) >
-0.12$, $-0.22$ and $-0.30$ for $T_*=32,000$, 37,000 and 45,000$\,$K,
respectively.  In turn, this means that under the most extreme
conditions ionization effects should not erode the apparent elemental
deficiency of N relative to O by more than 0.30$\,$dex.

Our ionization calculations also indicate that for $\log f_0({\rm Si^+})
- \log f_0({\rm H}) < 1.20$, virtually all of the Si is still singly
ionized, that is, in the allowed region all of the changes in the ratio
of Si~II to H~I are caused by changes in the ionization of H.  This is a
useful finding, since it shows that our earlier determination of
$N$(H$_{\rm total}$) based on $N$(Si~II) is still valid.

\section{Reduction and Analysis of the ACS Image}\label{ACS}

\subsection{Basic Reduction}\label{basic_reduction}

To better understand the origin of the Lyman limit absorption system, we
dedicated one orbit of our allocated HST time to obtaining an image of
the field using the Advanced Camera for Surveys (ACS) instrument. These
data were taken with the Wide Field Camera (WFC) on 05 May 2003 and were
archived as a dataset with the root name J8D90501. We used the F625W
filter, which is the equivalent of the {\it Sloan Digital Sky Survey\/}
$r$-band filter  (Pavlovsky et al. 2003). 

To aid with the removal of cosmic rays and hot pixels, four separate
exposures were taken at four pointings using the ACS-WFC-DITHER-BOX
pattern and no CR-SPLIT. The spacing between sub-exposures was 0.265
arcsecs, and each was 520 sec long. The four sub-images were shifted and
combined using the `multidrizzle' task available in the {\tt dither}
package of the STSDAS and PyRAF\footnote{STSDAS and PyRAF are products
of the Space Telescope Science Institute, which is operated by AURA for
NASA.} software suite.  Fluxes can be obtained from the final co-added 
image by multiplying the recorded ADU sec$^{-1}$ pix$^{-1}$ by the
inverse sensitivity constant (PHOTOFLAM): we used the most current value
available at the STScI web site, $1.195\times10^{-19}$~\flux\ for the
WFC data. These fluxes can in turn be converted to magnitudes and
referenced to the ST magnitude zero-point of 21.10. Data from the HRC
were not used in our analysis, since no bright galaxies were detected. 

\begin{figure}
\centerline{\psfig
{figure=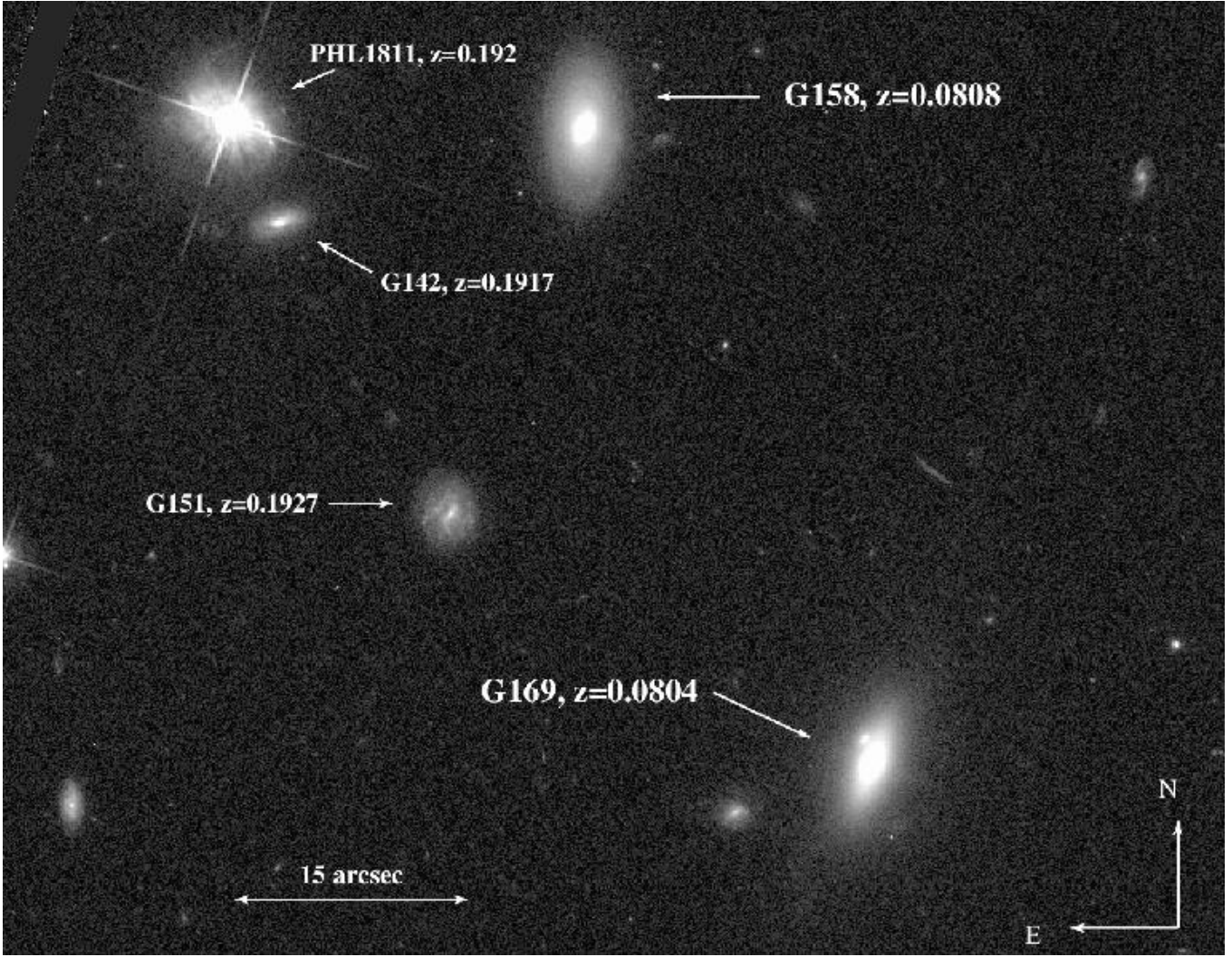,height=14cm,angle=0}}
\caption{Section of a 2080$\,$s image of the field of PHL~1811 taken
with ACS using the F625W filter. Galaxies with redshifts presented in
Paper~I are identified in the image using the same nomenclature used in
that paper.  Details about the galaxies G158 and G169 are given in
Table~\protect\ref{tab_acs}. An image that covers a much larger field
surrounding the quasar is shown in Paper~I.\label{fig_acs1}}
\end{figure}

Part of the image is reproduced in Fig.~\ref{fig_acs1}. We choose to
show the region of the data that contains the QSO and the two galaxies
at $z\simeq 0.08$ (although we label all the galaxies covered in this
region that had redshifts measured in Paper~I). The most striking result
from the image is that the $z\simeq 0.08$ galaxies appear, at first
sight, to be lenticular in shape. The galaxies clearly have a bright
bulge at their centers, but both show signs of disks with no spiral
pattern, indicative of S0 galaxies.  This is particularly interesting,
since the origin of S0 galaxies is widely debated.

\placefigure{fig_acs1}

\subsection{Surface Photometry}\label{surface_phot}

\begin{figure}
\centerline{\psfig
{figure=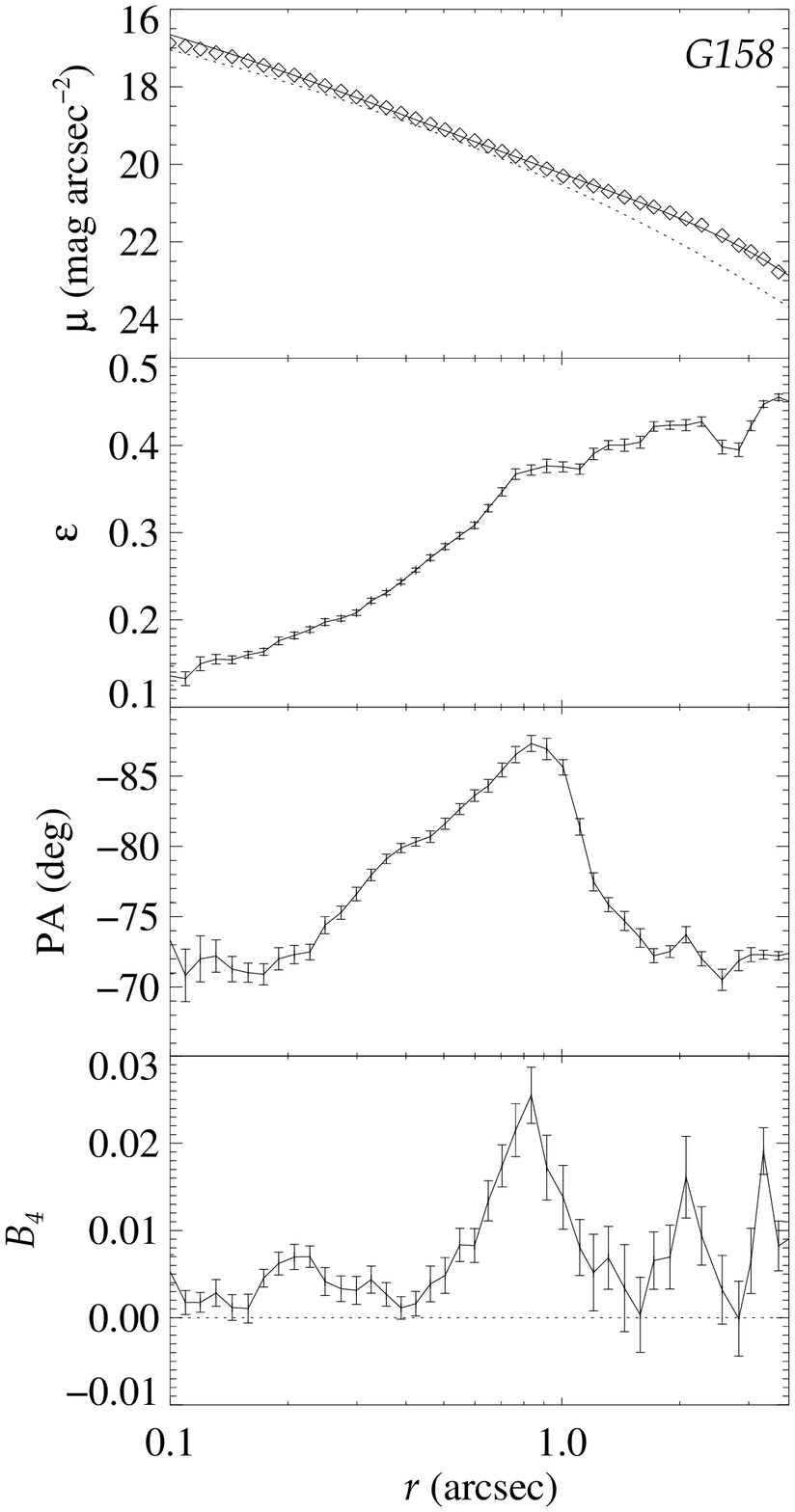,height=15cm,angle=0}
\psfig
{figure=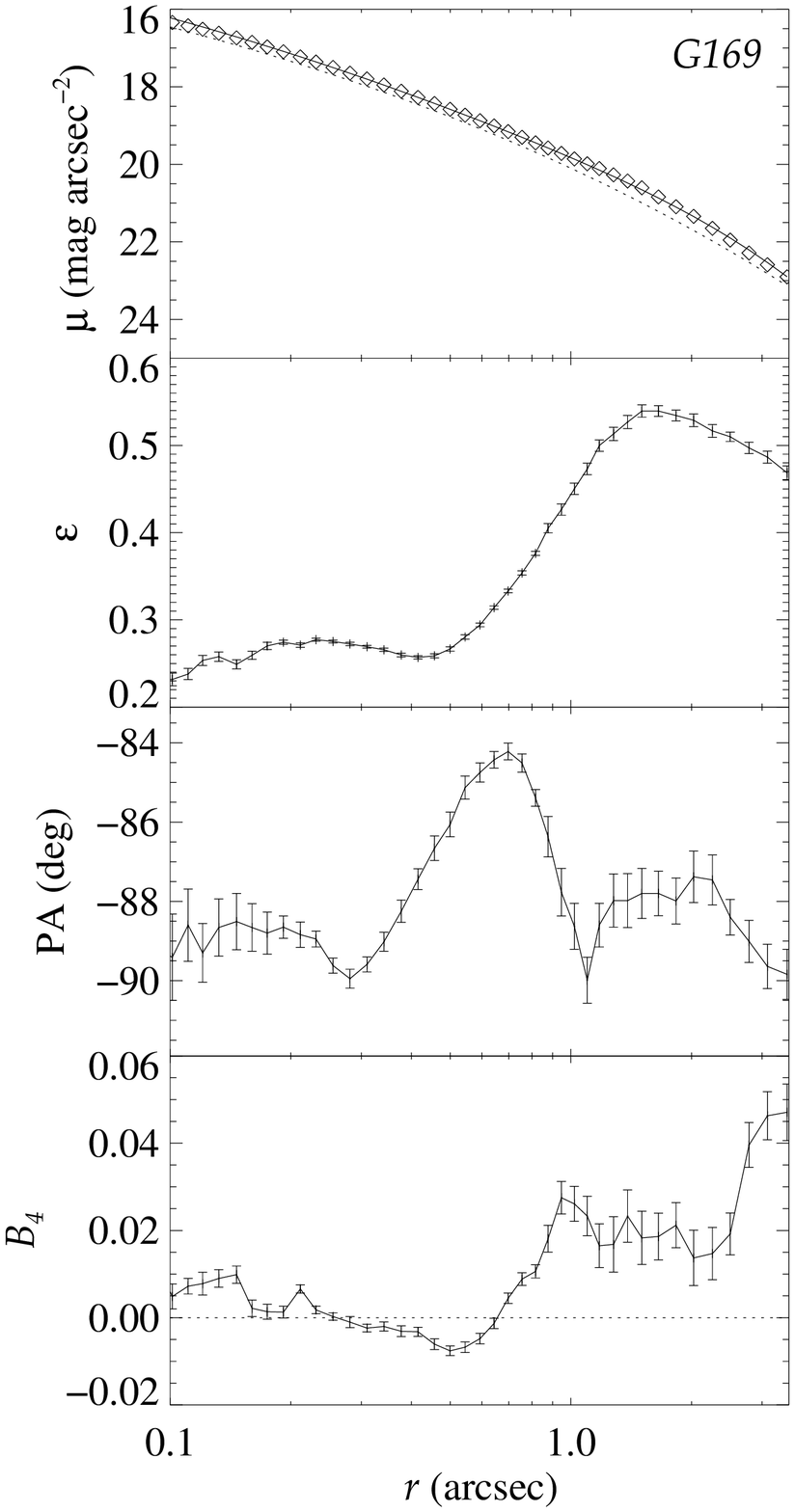,height=15cm,angle=0}} \caption{Results from
ellipse fitting to the surface brightness of the two galaxies near
$z=0.08$. All quantities are plotted against the equivalent radius,
$r=a\sqrt{1-\epsilon}$, where $a$ is the length of the semi-major axis
and $\epsilon$ is the ellipticity.  Top panel:  surface brightness
profile, with the best fit of a de Vaucouleurs plus exponential disk fit
shown as a solid line. To highlight the disk component, a dotted line
shows a fit using only a de Vaucouleurs profile to the inner part of the
profile.  Second panel: ellipticity $\epsilon$ of the isophotes. Third
panel: position angle PA of the isophotes; Fourth panel: Fourier
coefficient $B_4$ of the isophotes.\label{fig_ellipses}}
\end{figure}

To better quantify the morphology of both G158 and G169, we investigated
the surface brightness profiles of the galaxies using the {\tt isophote}
package available with STSDAS [ (Jedrzejewski 1987); see also
Milvang-Jensen \& J\o rgensen  (1999) and references therein]. Given a
particular value of the semimajor axis $a$ measured from the center of a
galaxy, ellipses can be fitted to isophotes of the intensity of light
from the galaxy, resulting in a measure of the center of the ellipse,
the surface brightness $\mu$ at the value of $a$, the position angle PA
of the ellipse, and its ellipticity $\epsilon$. These values can be
plotted against $a$, or, more commonly, against the equivalent radius,
$r=a\sqrt{1-\epsilon}$. We show the results of ellipse fitting for G158
and G169 in Figure~\ref{fig_ellipses}, and discuss the results from each
in \S~\ref{s_g158} and \ref{s_g169}.

\placefigure{fig_ellipses}

One method used to quantify the deviations of a perfect elliptical fit
to an isophote is to express the deviations as a Fourier sum.  The
details of this technique are well documented and are not reproduced
here [see the above references for further details; for a discussion of
the errors inherent in the fitting process, see, e.g., Rauscher  (1995)
or Busco  (1996)]. Of particular interest is the $B_4$ coefficient (the
$\cos4\theta$ term in the Fourier expansion): when $B_4>0$, the
isophotes are pointed, or 'disky', whereas when $B_4<0$ the light is
distributed in a more rectangular shape. Thus $B_4$ can serve as a
useful discriminator of galaxy type: although $B_4=0$ for regular
elliptical galaxies, and $B_4<0$ for more boxy ellipticals, $B_4>0$ can
indicate the presence of an S0 galaxy. We examine the values of $B_4$
for both G158 and G169 below.

A second way to examine the nature of the galaxies is to measure the
disk and bulge components seen in the surface brightness profiles.  To
that end, we fitted theoretical profiles to the observed values of $\mu$
by minimizing the value of $\chi^2$ between the fit and the data. We
adopted the simplest profile: a de Vaucouleurs $r^{1/4}$ profile for the
bulge component added to an exponential profile for the disk component.
When the intensity profile of a galaxy is converted to surface
brightness, these profiles become:

\begin{eqnarray}
\mu & = & \mu_e + 8.325 [ (\frac{r}{r_e})^{1/4} -1 ]\\
\mu & = & \mu_0 + 1.086\:\frac{r}{r_d}
\end{eqnarray}

\noindent
where $r_e$ and $r_d$ are the scale lengths of the bulge and disk
components respectively, and $\mu_e$ and $\mu_0$ are the surface
brightness values at those radii.

Our profile fits to the surface brightness profiles of G158 and G169 are
shown in the top panels of Fig.~\ref{fig_ellipses} and discussed in
\S\S~\ref{s_g158} and \ref{s_g169}. We only fitted points beyond twice
the FHWM of the core of the ACS Point Spread Function, or $\approx 0.18$
arcsec (a distance that corresponds to 3.6 ACS WFC pixels, since each
pixel is 0.05\arcsec~pix$^{-1}$). At smaller radii, the surface
brightness turns over due to the finite resolution. We note that other,
more complicated bulge profiles could have been used for the fit.
However, the simple $r^{1/4}$ profile is sufficient for our purposes of
identifying the bulge component in our galaxies. 

The conversion to magnitudes also included three corrections. We first
subtracted 0.13 mags to account for Galactic extinction: we assumed $A_r
= 2.5\:E(B-V)$  (Fitzpatrick 1999) and calculated $E(B-V)=0.05$ from the
dust maps of Schlegel et al.  (1998)\footnote{Instructions on how to
calculate $E(B-V)$ interactively for a given Galactic longitude and
latitude can be found at http://astron.berkeley.edu/dust/dust.html}. 
Second, we subtracted a $k$-correction of $2.5\log(1+z) = 0.08$ mags.
This term was expected to be small for all galaxy types at such a low
redshift  (Fukugita, Shimasaku, \& Ichikawa 1995). Finally, we
subtracted a value of $10\log(1+z) = 0.34$ mags to account for
cosmological dimming of the surface brightness. No correction was made
for extinction internal to the galaxies, since we have no information on
the magnitude or type of the extinction. 

The parameters deduced from the surface brightness profiles are given in
Table~\ref{tab_acs}. From these values, we can derive the disk-to-bulge
($D/B$) ratio. If $I_e$ and $I_d$ are the intensities corresponding to
$\mu_e$ and $\mu_0$, then the bulge fraction ($B/T$) is given in the
usual way by  

\begin{equation}
B/T = \frac{r_e^2 I_e}{r_e^2 I_e + 0.28 r_d^2 I_d}
\end{equation} 

\noindent
with $D/B$ = ($T/B$)$-$1  (Binney \& Merrifield 1998). The values of
$D/B$ are given in column 8 of Table~\ref{tab_acs} and are used in
\S\S~\ref{s_g158} and \ref{s_g169} to further indicate the morphological
types of G158 and G169.

\placetable{tab_acs}

Also included in Table~\ref{tab_acs} are the measured and absolute
magnitudes of the galaxies, $m_r$ and $M_r$, respectively.  Apparent
magnitudes were initially measured using {\tt sextractor}  (Bertin \&
Arnouts 1996); the values were compared with the integrated counts
available from the ellipse fitting routine, and found to be identical
with the {\tt sextractor} values when integrated to the last reliable
isophote.  The absolute magnitudes were derived from $m_r$ after
correcting for Galactic extinction and applying the $k$-correction given
above.

\subsection{Unsharp Masking}\label{unsharp_masking}

\begin{figure}[h!]
\centerline{\psfig
{figure=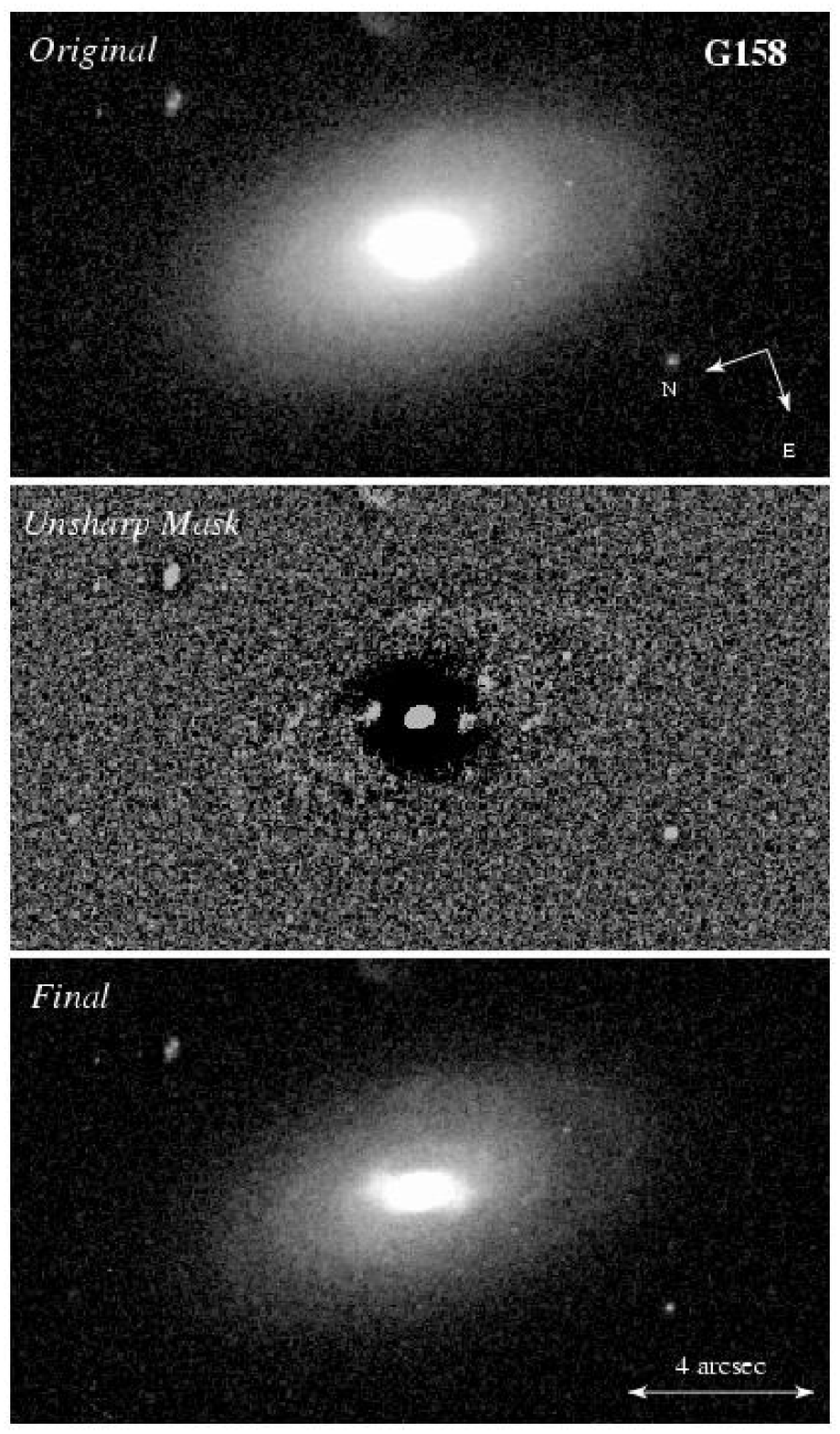,height=15cm,angle=0}
\psfig
{figure=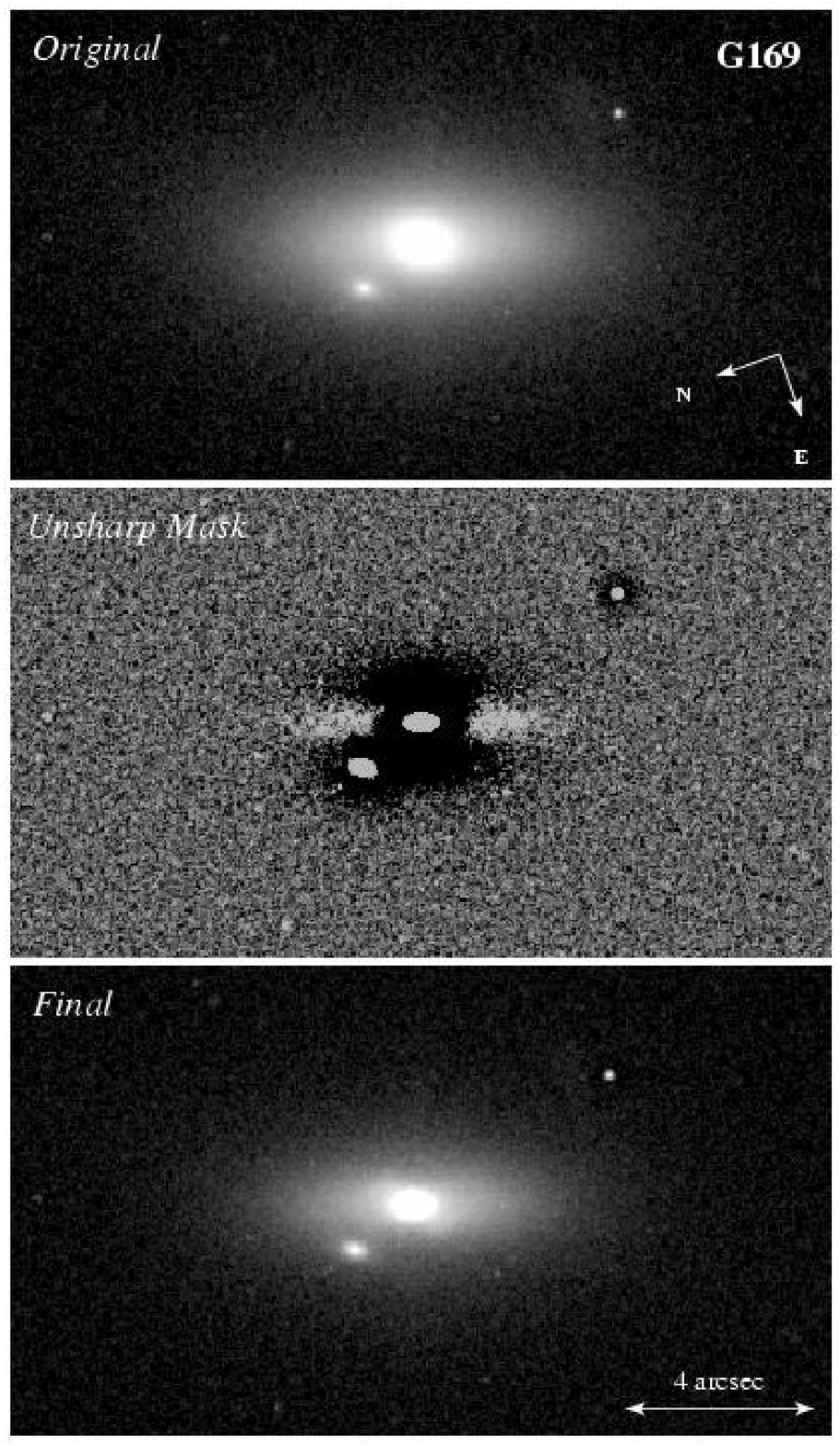,height=15cm,angle=0}}
\caption{Gray scale representations of G158 and G169 reproduced from the
ACS F625W image. Top panels show the galaxies as detected from the final
drizzled image, plotted on a logarithmic scale. The middle panels show
the unsharp mask of the galaxies, produced by smoothing the original
image with a Gaussian of width $\sigma=5$ pixels, then subtracting it
from the original. The final image is the sum of the unsharp mask and
the original image, plotted with the same logarithmic scaling as used
for the top panel.\label{fig_sharp1}}
\end{figure}

Finally, we have used the technique of unsharp masking  (Malin, Quinn,
\& Graham 1983) to highlight some of the small-scale structure in G158
and G169. Unsharp masking suppresses large-scale, low-frequency
variations, and emphasizes small-scale brightness variations. Example of
how useful this technique can be for galaxies observed with HST can be
found in, e.g., Erwin \& Sparke  (2003). To produce our unsharp masks,
we smoothed the original data by convolving it with a Gaussian function
of width $\sigma=5$ pixels, then subtracted this from the original
image. To produce a 'final' image, the unsharp mask can be added back to
the original data. The results for G158 and G169 are shown in
Fig.~\ref{fig_sharp1}. For each galaxy, we show the original image (top
panel), the unsharp mask (middle panel), and the 'final' image (bottom
panel). Our interpretations of the unsharp masks for the galaxies are
given later in \S\ref{s_g158} and \S\ref{s_g169}.
\begin{deluxetable}{lccccccccccccccc}
\setlength{\tabcolsep}{0.02in}
\tabletypesize{\tiny}
\tablecolumns{16}
\tablewidth{0pc} 
\tablecaption{Properties of galaxies close to the PHL~1811 sight line at
$z=0.08$ \label{tab_acs}}
\tablehead{
\colhead{} &
\colhead{} &
\colhead{} &
\colhead{} &
\colhead{} & 
\multicolumn{5}{c}{Surface Brightness Parameters} &
\colhead{} & 
\colhead{} &
\colhead{} &
\colhead{} &
\colhead{} &
\colhead{} \\
\cline{5-10}
\multicolumn{2}{c}{Designations} &
\colhead{} &
\colhead{$\rho$} &
\colhead{} & 
\colhead{} &
\colhead{$r_e$} &
\colhead{} &
\colhead{$r_d$} &
\colhead{} &
\colhead{} & 
\colhead{} &
\colhead{} &
\colhead{} &
\colhead{} &
\colhead{} \\
\cline{1-2}
\colhead{Paper~I} &
\colhead{2MASS} &
\colhead{$z$} &
\colhead{(\h )} &
\colhead{} & 
\colhead{$\mu_e$} &
\colhead{($''$)} &
\colhead{$\mu_0$} &
\colhead{($''$)} &
\colhead{$D/B$} &
\colhead{} & 
\colhead{$m_r$} &
\colhead{$m_J$} &
\colhead{$m_H$} &
\colhead{$m_{K_s}$} & 
\colhead{$M_r - 5\log h_{70}$} \\
\colhead{(1)} &
\colhead{(2)} &
\colhead{(3)} &
\colhead{(4)} &
\colhead{} & 
\colhead{(5)} &
\colhead{(6)} &
\colhead{(7)} &
\colhead{(8)} &
\colhead{(9)} & 
\colhead{} & 
\colhead{(10)} &
\colhead{(11)} &
\colhead{(12)} &
\colhead{(13)} &
\colhead{(14)}
}
\startdata
G158 &J21545996$-$0922249& 0.0808 & 34 && 19.48  & 0.51  & 20.24  & 1.47 
& 1.2  && 17.48 & 15.68 & 14.95 & 14.63 & $-20.5$   \\
G169 &J21545870$-$0923061\tablenotemark{a}& 0.0804 & 87 && 19.51  & 0.73 
& 20.11  & 1.00  & 0.3  && 17.22 & 15.15 & 14.38 & 13.96 & $-20.8$   \\
\enddata
\tablecomments{Column entries:\\
Col.~(1): Galaxy designation, as given
  in Paper~I; \\
Col.~(2): Designation in the 2MASS point-source catalog;\\
Col.~(3): Galaxy redshift, as reported in Paper~I;\\
Col.~(4): Separation between QSO and galaxy on the plane
  of the sky, assuming $q_0=0$;\\ 
Col.~(5): Surface brightness at the
  effective radius $r_e$ as measured in the F625W band, in mag
  arcsec$^{-2}$, for the component of the surface brightness profile
  fit given by a de Vaucouleurs profile;\\ 
Col.~(6): Effective radius,
  in arcsec, of the same de Vaucouleurs profile;\\ 
Col.~(7): Surface brightness at the disk length $r_d$  as measured in
  the F625W band, in mag arcsec$^{-2}$, for the component of the
  surface brightness profile fit given by an exponential brightness
  distribution. NB., Both $\mu_e$ are $\mu_0$ are corrected for Milky
  Way extinction (0.13 mags), cosmological expansion [10$\log$(1+$z$) =
  0.34 mags], and $k$-correction [$2.5\log$(1+$z$)= 0.08 mags];\\
Col.~(8): Disk length, in arcsec, for the same exponential brightness
  distribution;\\
Col.~(9): Disk-to-bulge ratio from Surface Brightness profile
  fits---see text;\\
Col.~(10): Total magnitude of galaxy measured through the ACS F625W
filter;\\
Cols.~(11$-$13): $J$, $H$ and $K_s$ profile-fit magnitudes listed in the
2MASS point-source catalog;\\
Col.~(14): Absolute magnitude of galaxy, from $m_r$, after
correction for Milky Way extinction and $k$-correction.
}
\tablenotetext{a}{Also listed in the 2MASS extended source catalog, with
the designation 2MASX~J21545868$-$0923057.}
\end{deluxetable}
\clearpage

\placefigure{fig_sharp1}

\subsection{PSF Subtraction}\label{psf_subtraction}

A persistent problem when searching for galaxies that might be
responsible for quasar absorption lines is that the absorbing galaxy may
be so small and well centered along a sightline that it cannot be
detected against the glare of the bright background source.
To help overcome this problem, one can subtract a properly scaled
version of the instrument's point spread function (PSF) in an attempt to
remove the  extended pattern of light from the image of the background
object.  Any believable features detected after the PSF cancellation
arise either from the host galaxy of the QSO, or from an object that
could, in principle, be at the redshift of the intervening absorption
line system.

The high spatial resolution of the ACS offers a powerful tool for
searching for objects close to a QSO sight line. One disadvantage,
however, is that the PSF is a complicated two-dimensional function whose
shape varies with time and field position (for a variety of
reasons---see, e.g. Krist  (2003)). The use of the ACS coronagraph has
proved to be particularly effective in suppressing the central intensity
from bright sources  (Clampin et al. 2003), including QSOs  (Martel et
al. 2003). Unfortunately, we had insufficient time available to execute
a coronagraphic exposure of PHL~1811 with the suitable PSF control
sample.  Instead, we made a normal exposure with the intent of using a
nearby star in the field to act as a model for the PSF.

\begin{figure}[h!]
\plotone{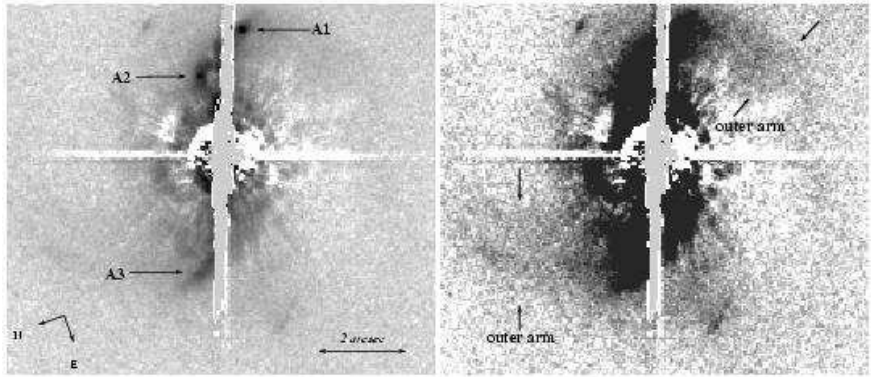}
{\caption{Results of subtracting a PSF of a nearby star from  the QSO.
Both panels show gray scale representations of the same image, plotted
on a logarithmic scale, but the right-hand panel has a more extreme
stretch to show the possible presence of outer spiral arms, which are
likely to arise from the host galaxy.  At the redshift of the QSO, the
2\arcsec\ bar shown at the bottom of the left-hand panel corresponds to
6.4\h .  The labels A1, A2 and A3 are used to indicate bright nebulous
regions in the image that are discussed in the text.\label{fig_qsopsf}}}
\end{figure}

The core of PHL~1811 in our image is saturated, with the excess charge
bleeding down the columns that were centered on the QSO.  The radial
structure of the PSF is clearly visible, since the brightness of the QSO
ensures that the wings of the PSF are captured with moderate S/N. To
provide a PSF with a similar charge saturation, column bleed, and S/N in
the wings (in order to provide better subtraction in that region), we
elected to use the one star\footnote{The star chosen to give the model
PSF was GSC2 S32133137 located $76\arcsec$ from PHL~1811.  It has
magnitudes $F=13.49$ and $J=14.47$ compared to $F=14.07$ and $J=14.08$
for PHL~1811.} in the field with a brightness similar to that of
PHL~1811. To obtain the best possible subtraction, we fitted the stellar
PSF to the QSO PSF by minimizing the $\chi^2$ statistic, allowing a
shift in the stellar PSF image to vary in both the $x$ and $y$
direction, as well as a multiplicative brightness scale factor between
QSO and stellar PSF, and a rotation of the PSF about its center (which
overcomes the effect of remapping to correct for the field distortion).
The results are shown in Fig.~\ref{fig_qsopsf} and are discussed below
in \S\ref{s_qso}. 

\placefigure{fig_qsopsf}

\section{Interpretation of the ACS Image}\label{ACS_interpretation}

\subsection{G158 \label{s_g158}}

G158 is the galaxy closest to the PHL~1811 sight line with a redshift
that is similar to that of the Lyman limit system.  At the redshift
$z=0.0808$ reported in Paper~I, the separation of $22\farcs 9$ from the
line of sight corresponds to a lateral distance of 34~\h . We measured
an $r$-band magnitude of G158 equal to 17.35 using the methods described
in \S\ref{basic_reduction}, after correcting for Galactic extinction
(\S\ref{surface_phot}) but with no $k$-correction.  We can compare the
resulting absolute magnitude $M_r-5\log h_{70}=-20.43$ with those of
other galaxies at a similar redshift.  Blanton et al.  (2003) have
derived the $r$-band luminosity function for galaxies at $z\simeq 0.1$
in the SDSS.  They find that an $L^*_r$ galaxy at $z=0.1$ has an
absolute magnitude of $M^*_r - 5 \log h_{70} = -21.21$. Since the
difference in the $k$-correction between a galaxy at $z=0.08$ and
$z=0.1$ is negligible, G158 is therefore a $0.5 L^*_r$ galaxy.

The left hand panel of Fig.~\ref{fig_ellipses} shows the results of our
surface photometry analysis of G158. The surface brightness profile
shows a clear deviation from the $r^{1/4}$ law beginning at $r\approx
0.9$\arcsec . The deviation is easily accounted for by fitting an
exponential disk model. The presence of a disk with no obvious spiral
arm patterns suggests that the galaxy is of type S0. It is interesting
to note, however, that the unsharp mask image presented in the middle
panel of Fig.~\ref{fig_sharp1} reveals faint rings in the galactic disk,
which may be residuals of spiral arm structures, which were once more
prevalent in the ealier history of the galaxy.  However, since these
features would not normally be apparent in unprocessed images, it seems
appropriate to retain the classification of G158 as an S0 galaxy.  The
disky nature of the galaxy is also suggested by the peak in a positive
value of $B_4$ at a similar radius. The $D/B$ ratio is 1.2, which lies
in the middle of the range expected for S0 galaxies, $0.1<D/B<2$
(Burstein 1979; Kent 1985; Bothun \& Gregg 1990). 

Also intriguing is the peak in the PA. Such maxima, combined with peaks
in ellipticity, are often associated with a galactic bar. We see little
evidence in our data for a well defined maximum in $\epsilon$ to
accompany the peak in PA, but there are examples in nearby galaxies
where a bar (or some other feature) does not have an obvious ellipticity
peak  (Erwin \& Sparke 2003). If we look at the unsharp mask in
Fig~\ref{fig_sharp1}, we can see small peaks of emission at the edge of
the dark ring that surrounds the nucleus of the galaxy. Since the end of
bars in early type galaxies have sharp cutoffs in their luminosity
profile  (Ohta, Hamabe, \& Wakamatsu 1990) galactic bars often show up
well in unsharp mask images (which emphasize small-scale, high frequency
structures). The features seen in Fig.~\ref{fig_sharp1} do not extend
all the way to the center of the galaxy, but are suggestive of the
existence of a bar. Similar features can be seen arising from the bars
in the unsharp masks of NGC~2880, NGC~2962, NGC~6654 presented by Erwin
\& Sparke  (2003). When the unsharp mask image is combined with the
original to give the `final' image shown at the bottom of
Fig~\ref{fig_sharp1}, the bar shows up convincingly. For these reasons,
we extend our classification of G158 to that of an SB0 galaxy. 

\subsection{G169 \label{s_g169}}

G169 is the other galaxy in the field that we were able to identify at a
redshift near that of the Lyman limit system.  Again from Paper~I we
find $z=0.0804$, a separation of $58\farcs 9$ from the line of sight,
and an impact parameter of 87~\h .  Our current $r$-band magnitude
measurement of G169 is 17.09 when corrected for Galactic extinction (and
no $k$-correction), which corresponds to an absolute luminosity of
$M^*_r - 5\log h_{70} = -20.69$.  Again using the value of $M^*_r$ from
Blanton et al.  (2003), we find that G169 is a $0.6 L^*_r$ galaxy,
similar to that of G158. 

The right-hand panel of Fig.~\ref{fig_ellipses} shows the results of our
surface photometry analysis of G169. The surface brightness profile is
close to that expected for an $r^{1/4}$ law, but there is a small
deviation at $r\approx 1$\arcsec, which can be accounted for using a
disk exponential profile. The $B_4$ coefficient also shows positive
values at 1\arcsec\ and beyond. Although the high inclination of the
galaxy makes it hard to see any spiral structure which may exist, no
spiral features are obviously present. This fact, combined with the
detection of a weak disk, again leads us to classify G169 as an S0
galaxy. The $D/B$ ratio of G169, 0.3, is consistent with that of S0
galaxies, which as noted above, have $0.1<D/B<2$.

Both the PA and ellipticity profiles show peaks, again suggesting the
presence of a bar. However, the maxima do not coincide at the same
radius. The unsharp mask also shows emission at the edges of the dark
ring surrounding the core of the galaxy, but we interpret these as being
due to the disk of the galaxy. Hence the case for the existence of a bar
is much weaker for G169 than for G158. 

We also note the presence of a possible companion north of the center of
G169, 1.56\arcsec\ from the center of the galaxy. Although this object
could be unassociated with G169 and merely lie fortuitously along the
line of sight, it may also be a satellite of G169. Such satellites may
not be uncommon: Madore, Freeman \& Bothun  (2004) find that isolated
elliptical galaxies often have $+1.0\pm 0.5$ companions within a
projected radius of 70~\h .  We estimate that the satellite has an
observed magnitude of $r\simeq 21.3$, which would correspond to $M_r =
-16.9$ (after correction for Galactic extinction and $k$-correction).

A broad, tail-like structure extending from G169 toward the south is
barely visible in the ACS picture, but it is more easily seen in the
image taken from the ground that was reported in Paper~I.  This
structure may be indicating that some tidal interaction may have
occurred, but the tail does not point toward the direction of our sight
line to PHL~1811.

\subsection{QSO \label{s_qso}}

The results from removing a stellar PSF from the center of PHL~1811 are
shown in Fig.~\ref{fig_qsopsf}. We show the results twice at two
different stretches to emphasize different features. The PSF subtraction
is far from perfect, which is to be expected primarily because of
differences in the structure of the PSF as it varies with position
across the WFC field of view, and also because the star has a somewhat
different spectral energy distribution than the QSO, as indicated by the
photographic magnitudes given in the footnote in
\S\ref{psf_subtraction}. Some of the radial 'spokes' which are a common
feature of ACS PSFs have been removed successfully, but some remain, and
the inner diffraction rings can still be seen.

We do not believe that any feature seen within $\sim 1\farcs 5$ of the
quasar should be considered real.  However, we identify certain
structures farther out which do appear to be genuine.  At the top of the
left-hand panel of Fig.~\ref{fig_qsopsf} we identify at least two
nebulous cores (which we label A1 and A2) that lie in a meandering
`snake' of emission, which winds from the top right quadrant to the top
left quadrant and down towards the center of the QSO.  We also detect an
arc of emission, labeled A3, which appears from under the saturated
column in the bottom left quadrant. None of these features are artifacts
of the PSF subtraction, since A1 and A3 are just visible in the original
image before the PSF was subtracted (see Fig.~\ref{fig_acs1}). Nor are
they inherent features of the ACS PSF wings, since they are not aligned
with the usual radial features or the diffraction rings seen in ACS
PSFs; no other stellar PSFs in the field show similar images; and
synthetic PSFs generated by the software program {\tt
TinyTim}\footnote{http://www.stsci.edu/software/tinytim/tinytim.html} do
not show such features. 

The right-hand panel shows the results of our PSF subtraction at a much
harder stretch. The features labeled in the left-hand panel are clearly
burned out, but the snake in which A1 and A2 are embedded is seen to be
much wider. Even more significant is the apparent presence of outer
spiral arms. The arm in the lower left quadrant is connected to the arc
A3; and while part of the emission in the inner upper right quadrant is
lost because of poor PSF subtraction, there is good evidence for a
second spiral arm at the outer edge of the quadrant. The emission can be
traced out to 3.8 and 4.8\arcsec\ for the arm in the upper right and
lower left quadrants, respectively.

The results from the PSF subtraction therefore appears to indicate the
presence of a spiral galaxy directly in front of, or surrounding, the
QSO.  This could be responsible for the Lyman limit system at
$z=0.080$.\footnote{One can go even further and offer the conjecture
that if this foreground galaxy is well centered on the QSO, it may
gravitationally lens the QSO and increase its apparent magnitude.  
Against this hypothesis, we should note that PHL~1811 appears as a
single image, which requires a rather unusual (but not impossible)
configuration of the lens and object.}  Alternatively, the object could
be the host galaxy of the QSO at $z=0.192$. The detection of a QSO
shining through the center of a faint galaxy is not unprecedented: we
detected such an alignment of the low surface brightness galaxy
SBS$\,$1543+593 in front of the quasar HS~1543+5921 for example  (Bowen,
Tripp, \& Jenkins 2001).  Nevertheless, it is still striking from
Fig.~\ref{fig_qsopsf} that the central AGN appears to lie exactly at the
center of the disk structure, which suggests that we are seeing the host
of the QSO. 

\section{Large Scale Structure of Galaxies near PHL~1811}\label{LSS}

\begin{figure}
\vspace*{-1.25cm}\centerline{\psfig
{figure=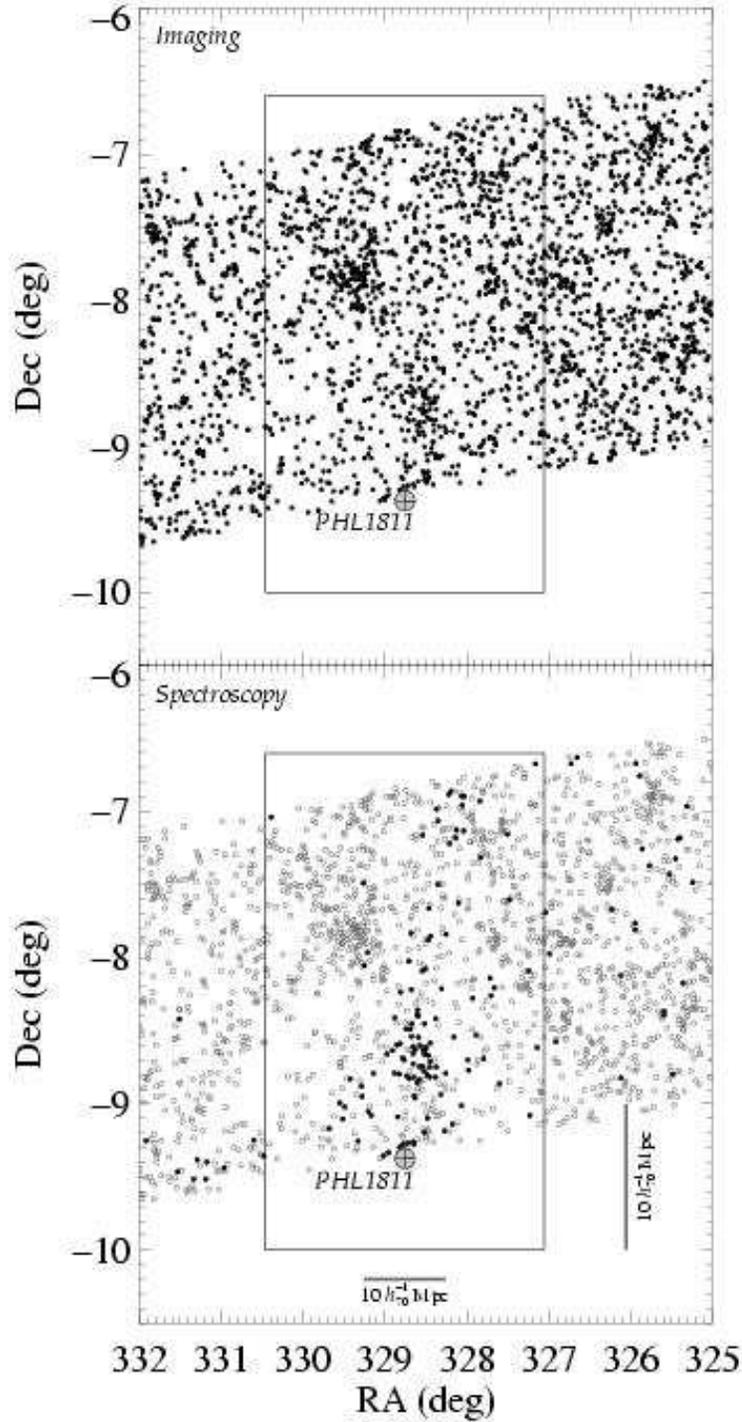,height=20cm,angle=0}}
\caption{{\it Top panel:} the distribution of objects cataloged by DR3
of the SDSS near the sight line of PHL~1811,  for objects identified as
galaxies on the available CCD data, and with Petrosian $r$-band
magnitudes $< 17.7$. {\it Bottom panel:} positions of objects with $r <
17.7$ confirmed spectroscopically as galaxies, and with $z>0.001$. The
galaxies indicated with filled circles lie in the interval $0.070 < z <
0.080$. The overplotted rectangle indicates the survey area used in
Fig.~\protect\ref{fig_wedges}.  The lengths of the bars in the lower
panel indicate comoving distances of $10\,h_{70}^{-1}{\rm Mpc}$ at
$z=0.08$.\label{fig_dr3cover} }
\end{figure}

To understand the distribution of galaxies close to the sight line of
PHL~1811, we searched the third data release (DR3) of the SDSS 
(Abazajian et al. 2004) within four degrees of the position of PHL~1811
using the SDSS
SkyServer\footnote{http://cas.sdss.org/dr3/en/get/SkyQA.asp?}. We used
the {\tt galaxy} ``view'', which is a subset of the {\tt PhotoPrimary}
imaging catalog containing objects classified as galaxies. This
classification is based on identifying resolved objects, and includes no
spectrographic information. We selected galaxies with Petrosian
magnitudes $r < 17.7$, which is the magnitude at which the SDSS aims to
be complete spectroscopically  (Strauss et al. 2002).  The results are
shown in the top panel of Fig.~\ref{fig_dr3cover}. The SDSS apparently
covers a strip of the sky $\sim 2$\degr\ wide, running north of the
PHL~1811 sight line, just missing the QSO itself. To collate the
available redshift information, we performed the same search using the
{\tt SpecObj} view, which contains a cleaned catalog of available
spectroscopic information taken from the {\tt SpecObjAll} table of the
DR3.  The galaxies identified spectroscopically are shown in the bottom
panel of Fig.~\ref{fig_dr3cover}.

\placefigure{fig_dr3cover}

\begin{figure}
\vspace*{-0.95cm}\centerline{\psfig
{figure=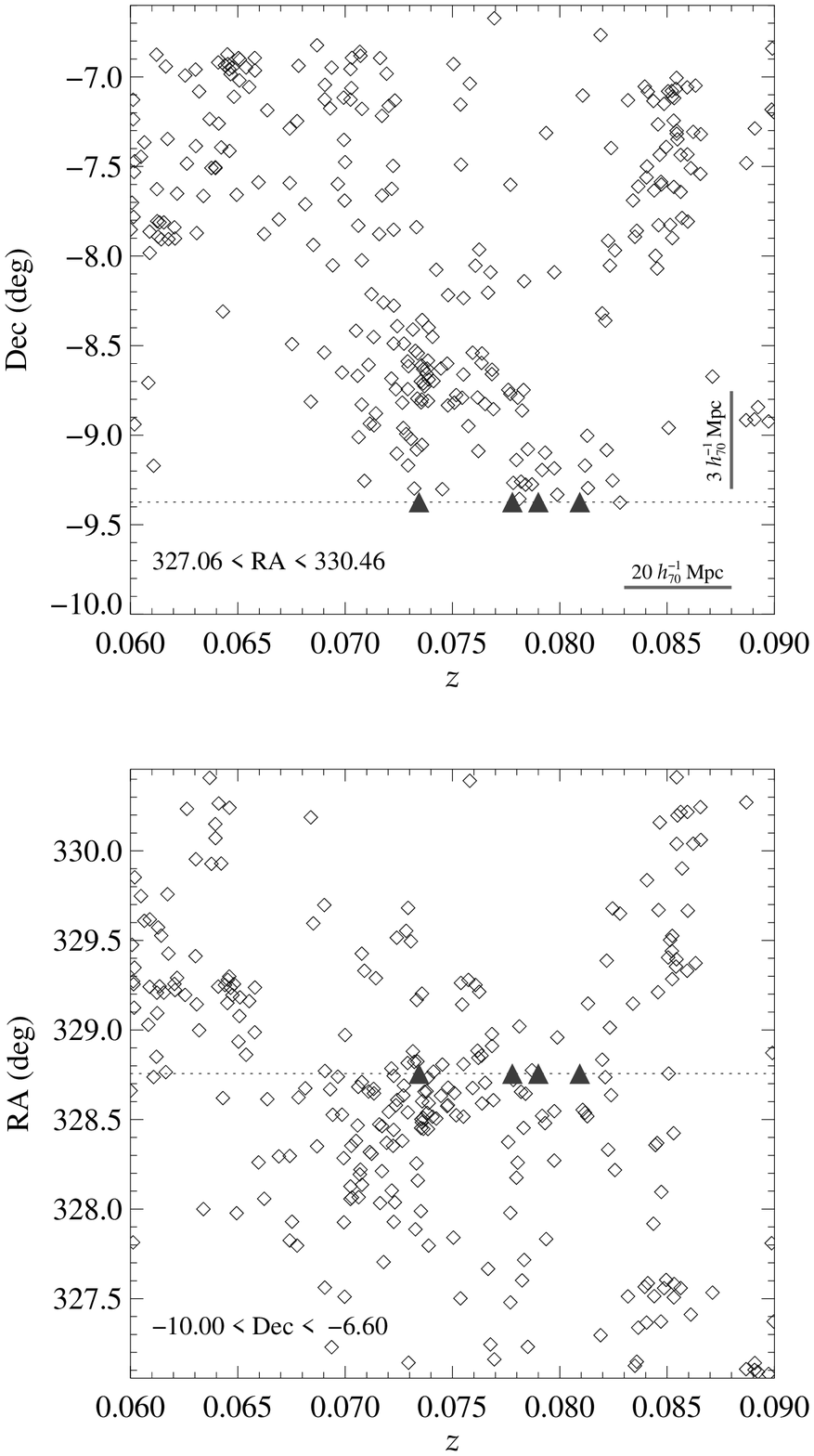,height=21cm,angle=0}}
\caption{Distribution in redshift of $r<17.7$  SDSS galaxies with: {\it
top panel}---declination, summed over a 3.4\degr\ right ascension 
interval;  {\it bottom panel}---right ascension, summed over a 3.4\degr\
declination interval. The sight line to PHL~1811 is shown as a dotted
line, and the redshifts of the absorption systems are indicated by
triangles. The lack of galaxies below Dec $\simeq -9.5$ in the top panel
is due to the limited coverage of DR3.\label{fig_wedges}}
\end{figure}

In the bottom panel of Fig.~\ref{fig_dr3cover} we have highlighted the
galaxies found in the redshift interval $0.070<z<0.080$, the interval
that covers the Lyman limit system discussed in this paper, as well as
the other metal-line absorption systems detected at slightly lower
redshifts at $z= 0.07344, 0.07779$ and 0.07900. The figure shows the
galaxies in this redshift range are largely confined to a narrow strip
in right ascension (RA) centered at 328.5\degr\ and that spans at least
2\degr\ in declination (Dec), with a significant conglomeration of
galaxies at [RA,Dec] = [328.5,$-$8.7]\degr . We explore this
distribution further in Fig.~\ref{fig_wedges}: the top panel shows the
redshift distribution of galaxies with Dec, including all galaxies
within a 3.4\degr\ interval over a range $327.06 <$ RA $< 330.46$.  It
is striking that the galaxies appear to follow a line from the top left
of the figure to the bottom right, indicative of a filament, or sheet,
of galaxies. The cut shows that the two S0 galaxies at $z=0.080$
discussed in \S\ref{s_g158} and \S\ref{s_g169} (not plotted here, since
they were not covered by the SDSS) lie at the edge of the filament,
assuming that the filament would continue to run to the south of the QSO
line of sight (obviously, more data are required to confirm such an
assumption).

\placefigure{fig_wedges}

A comparison between the objects cataloged in the {\tt galaxy} catalog
and those in {\tt SpecObj} over the enclosed area of the sky shown in
Fig.~\ref{fig_dr3cover} suggests that the spectroscopic observations are
72~\%\ complete over the magnitude range $14<r<17.7$, with the fraction
of incompleteness roughly the same at all magnitudes over that range. 
Given this high level level of completeness, and that the redshift
interval shown is small, the clustering evident in Fig.~\ref{fig_wedges}
is likely to be real, and not a result of incompleteness in the SDSS
redshift survey.

The bottom panel of Fig.~\ref{fig_wedges} shows a similar distribution
of galaxies with redshift but with a cut made orthogonally to the view
in the top panel. In the lower diagram, we sum galaxies over the
interval $-10.0 <$ Dec $<-6.60$ and plot their RA as a function of $z$.
This distribution in RA is less well defined than the cut in Dec, but
shows that the galaxies seen at the top left of the upper panel tend to
be confined to the upper left of the lower panel. This may suggest that
the galaxies are distributed in a more filamentary structure, rather
than as a sheet.

Bregman, Dupke \& Miller  (2004) have noted that the direction toward
PHL~1811 is not far from a line in the sky that extends from the
clusters Abell~2402 (50\arcmin\ away at $z=0.0809$) and Abell~2410
(109\arcmin; $z=0.0809$) to Abell~2376 (136\arcmin; $z=0.0896$) and
Abell~2377 (139\arcmin; $z=0.0808$), all of which lie within the
Aquarius~B Supercluster.  The fundamental theme of their paper is that
lines joining clusters are probably more likely than usual to be near
over-dense accumulations of the warm-hot intergalactic medium (WHIM), a
conclusion suggested by the outcomes of hydrodynamical simulations of
structure formation in the universe  (Cen \& Ostriker 1999; Dav\'e et
al. 2001).  The redshifts of these clusters are remarkably close to each
other and to that of the Lyman limit system under study here. 
Unfortunately, the clusters and the line joining them are at a
declination just below the southern limit of the SDSS, so we are not
able to check whether or not there is an enhancement in the density of
galaxies near the line.

\section{Discussion}\label{disc}

\subsection{Chemical Evolution}\label{chemev}

The relative proportions of different elements in either stars or an
interstellar medium convey important information on the history and
nature of the nucleosynthetic sources in a particular environment.  This
is a rich field of study, but one that presents major theoretical
challenges to interpret the many factors that can influence the
enrichment process, ranging across such topics as the star formation
rates and how they vary with time, the distribution of stellar masses
(which may also change with time), the inflow of pristine material from
outside the system, and the loss of gas into the intergalactic medium,
either through SN-driven winds or dynamical processes [see Matteucci 
(2001) for a coverage of many of the fundamentals these subjects].  In
this section, we will explore some implications that arise from our
findings on relative element abundances in the Lyman limit system at
$z=0.08$ in front of PHL~1811.  In turn, this information provides
insights on the properties of probable contributors to the gas that we
will discuss in \S\ref{source}. 

We return to our conclusions on the abundance ratios of the elements
that arose from the CLOUDY calculations for a slab inclined by
$60\arcdeg$ to the line of sight, as presented for two values of $\log
U$ in Table~\ref{cloudy_abund}.  These two conditions represent our most
plausible choices for the state of the gas that is exposed to the
intergalactic radiation field, but without any appreciable influence
from possible internal sources of ionization.  Both values of $\log U$
have certain ways of deviating from our usual expectations however.  For
the lower value of the ionization parameter, i.e., $\log U=-4.4$, the
$\alpha$-process elements O, Si and S have relative abundances
consistent with their solar abundance ratios, while for N we obtain
different abundances derived from $N$(N~I) and $N$(N~II), which may
reflect real differences in a poorly mixed system with varying levels of
ionization.  The opposite is true for the other choice for $\log U$,
i.e., the one equal to $-3.6$: the two nitrogen abundances are
consistent with a single (low) abundance of N, but there are mild
disparities in the abundances of the $\alpha$-process elements compared
to their solar values.  For either choice of $\log U$, several important
facts emerge: (1) the abundance of O is slightly sub-solar, (2) the
abundance of N, even if it varies somewhat, is far below that of O,
compared to the solar abundance ratio and (3) the value of [Fe/O] is
somewhat below zero.  As is usually the case when Fe is compared to O or
some other $\alpha$-process element, it is not certain if a deficiency
of Fe is caused by the formation of dust grains or a relative lack of
production by Type~Ia supernovae.  

For the moment, we focus on the result for nitrogen when $\log U=-4.4$
in the CLOUDY model.  In this case, the two forms of N that we observe,
neutral and singly-ionized, present an intriguing dilemma since they
give mutually inconsistent results for the abundance of this element. 
This disagreement is not strong, however, since the $+1\sigma$ limit for
[N/H] based on N~I is only slightly below the $-1\sigma$ limit based on
N~II.  These error limits do not include additional deviations that
might arise from uncertain assumptions in the CLOUDY model or atomic
parameters.  Nevertheless, if we accept at face value the possibility
that nitrogen is very deficient with respect to other elements in the
H~I-bearing material but is closer to normal in the more fully ionized
gas, we must adopt the notion that we are viewing a chemically
inhomogeneous gas system.  One possibility is that the gas complex is
not simply an intergalactic cloud, but is a system of gas and stars that
has some internal H~II regions that have been specially enriched by some
bursts of star formation.  This is not a new concept; it is one that has
been proposed for the blue compact dwarf galaxies Mrk~59, I~Zw~36,
I~Zw~18 and NGC~1705, based on the observed differences of abundances in
their H~I and H~II regions  (Thuan, Lecavelier des Etangs, \& Izotov
2002; Aloisi et al. 2003; Lebouteiller et al. 2004; Lee \& Skillman
2004; Aloisi et al. 2005, in preparation).\footnote{However, the
assertions about abundance differences for I~Zw~18 have been questioned
recently by Lecavelier des Etangs et al  (2004).}  While we do not see
stars or H~II regions in our ACS image, it is possible that our ability
to do so was compromised by the glare of the quasar
(\S\ref{psf_subtraction}), or that perhaps the spiral galaxy discussed
in \S\ref{s_qso} is indeed the Lyman limit system and not the quasar's
host galaxy.  Of course, this dichotomy departs from our simple
construct of a chemically homogeneous cloud whose ionization is
maintained by an external radiation field, and for the more fully
ionized gas we may be approaching a regime that more closely
approximates the locally enhanced ionization conditions discussed in
\S\ref{more_detailed}.  Even in the light of this more complex picture,
we can still maintain that our determination of the range of $N({\rm
H}_{\rm total})$ (\S\ref{uniform_slab}) should remain intact because it
encompasses our worst extremes for [Si/H], and in all circumstances most
of the Si is in the form of Si~II.

We now switch to considerations that arise when $\log U=-3.6$.  Here,
the two forms of N give a consistent result in the simple CLOUDY model,
[N/H]~=~$-1.25$, but while [O/H] still remains at a level near $-0.2$,
the abundances of the other two $\alpha$-process elements decrease to
lower amounts, ranging from $-0.88$ for [Si/H] to $-0.60$ for [S/H].  As
we stated earlier, these elements are generally considered to have a
common origin.  Even so, there is some evidence from the analyses of
stars in the thin disk of our Galaxy that as metal enrichment progresses
(as measured by [O/H]), there are gradual increases in [Mg/O] and [Si/O] 
(Bensby, Feltzing, \& Lundstr\"om 2004).  In essence, the abundances of
Si and O do not necessarily increase exactly in lock step with each
other (and this may apply to S as well).  On the theoretical side,   the
composition of supernova ejecta in the models of Woosley \& Weaver 
(1995) show that supernovae with initial masses of about 35$\,$M$_\odot$
have strong yields of O with much less production of Si and S, while the
converse is true for progenitor stars in the range $15-25\,{\rm
M}_\odot$.\footnote{Figure~6 in the review article by McWilliam  (1997)
very effectively communicates this outcome at a glance.}  Thus, our
findings may reflect upon a stronger than usual emphasis of high mass
stars in the mix of progenitors for the Type~II supernovae that created
the elements.  In turn, we could interpret the excess of O relative to
other $\alpha$-process elements to arise from either an initial mass
function (IMF) that is skewed more toward higher masses, relative to the
IMF of our Galaxy, or by the fact that our sample is dominated by gas
that was enriched very soon after a burst of star formation where only
the highest mass stars had a chance to evolve to the supernova stage. 

Irrespective of the value of $\log U$ (between the two extremes we have
chosen), there is a clear deficiency of nitrogen.  The study of this
element relative to the $\alpha$-process elements in different
astrophysical sites is useful because [N/$\alpha$] increases as a system
ages and intermediate-mass stars (and stars from later episodes of star
formation) have a chance to inject additional nitrogen into the
interstellar gas. Deviations in the trend of [N/$\alpha$] versus a
general measure of metallicity, such as [$\alpha$/H], can give insights
on the star formation history within a gas system or its stellar mass
function.  For instance, Pilyugin, Thuan \& V\'ilchez  (2003) have
argued that galaxies that had most of their star formation at an early
time have higher N/O trends vs. O/H than those for which most of the
stars formed more recently.

However, nitrogen nucleosynthesis is a complicated topic, one that is
not yet fully understood. Fundamentally, carbon and oxygen must already
be present in the region of the star where fusion occurs in order for
nitrogen to be synthesized. Consequently, two nitrogen production
channels have been recognized: (1) ``primary'' nitrogen production, in
which carbon and oxygen are first produced in the star and then are
mixed into a region where nitrogen is subsequently created from the
mixed-in C/O, and (2) ``secondary'' nitrogen production, in which the
star forms from gas that was already enriched with C and O from previous
generations of stars. Nitrogen nucleosynthesis trends vs. time depend on
various factors, including the initial metallicity of the interstellar
gas, the masses of stars that predominantly contribute to the primary
and secondary N production, and whether or not stars of various masses
rotate.  Izotov \& Thuan  (1999) have suggested that massive stars play
an important role in primary nitrogen synthesis. In low-metallicity blue
compact galaxies, they argue that the oxygen and nitrogen come from  the
same massive stars, and there is no delay between the time of the oxygen
and (primary) nitrogen enrichment. Stellar rotation can enable massive
stars to synthesize primary N, but Meynet \& Maeder  (2002a, b) conclude
that stellar rotation will actually cause intermediate-mass stars to be
the main source of primary nitrogen, in which case there should be a lag
between the oxygen enrichment and the nitrogen injection.  If rotation
is not important, then primary N synthesis can occur when C and O from a
He-burning core diffuse into a hydrogen-burning shell, the so-called
``hot bottom burning;'' this should also occur mainly in
intermediate-mass stars  (Marigo 2001). At any rate, a nitrogen
underabundance indicates that a system is young because either the
enriching stars are early-generation stars formed in initially
low-metallicity gas, or there has not been sufficient time for
longer-lived intermediate-mass stars to contribute N to the interstellar
gas (or both).

\begin{figure}
\epsscale{1.0}
\plotone{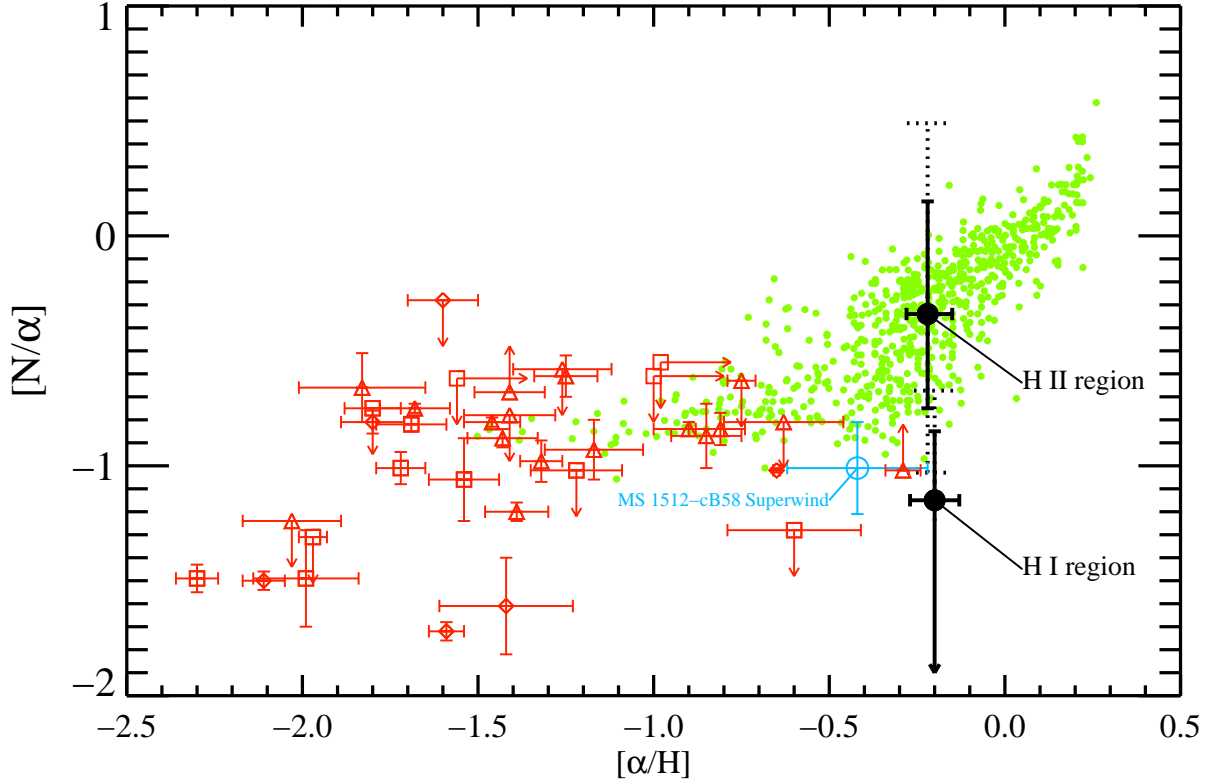}
\caption{Observed trends in the ratios of nitrogen to $\alpha$-process
elements as a function of the ratios of an $\alpha$-process element to
hydrogen.  {\it Large black dots with error bars:\/} Our determinations for
the regions containing mostly ionized gas (upper point) and gas
containing some neutral hydrogen (lower point) for the CLOUDY model
results for $\log U=-4.4$ listed in Table~\protect\ref{cloudy_abund} for
a slab of gas at an inclination of $60\arcdeg$, with $1\sigma$ errors
({\it solid error bars\/}) and $2\sigma$ errors ({\it dotted error
bars\/}) based on column density uncertainties only (i.e., {\it not\/}
including errors in the ionization model).  Silicon and oxygen were used
as the $\alpha$-process reference elements for the Lyman limit system's
H~II region and the H~I-bearing gas, respectively.  For comparison,
abundance measurements in other systems are shown: The result for the
superwind from the gravitationally lensed Lyman break galaxy
MS~1512$-$cB58  (Pettini et al. 2002b) is shown in blue, the small,
green dots represent the measurements of abundances for H~II regions in
various types of galaxies summarized by Pilyugin et al  (2003), and the
remaining points with error bars and limit arrows (red) depict the results for
damped Lyman alpha absorption (DLA) systems summarized by Centuri\'on et
al  (2003) (but see the footnote to the text in \S\ref{chemev}), where,
in decreasing order of preference, the $\alpha$-process element is
represented by O ({\it diamond markers\/}), S ({\it triangle markers\/})
or Si ({\it square markers\/}).}\label{no_nh}
\end{figure}

In Figure~\ref{no_nh} we show in the [N/$\alpha$] vs. [$\alpha$/H]
diagram the locations of our two measurements of N assuming $\log
U=-4.4$, one that applies to the H~I-bearing gas where we have plotted
[N/H]$_{\rm H~I}-$[O/H] vs. [O/H] with the values
($-1.15^{+0.31}_{-\infty}$, $-0.20\pm0.07$; errors are the expected
$1\sigma$ uncertainties) and the other to the mostly ionized material
indicated by [N/H]$_{\rm H~II}-$[Si/H] vs. [Si/H] with the values
($-0.34^{+0.49}_{-0.41}$, $-0.22^{+0.07}_{-0.06}$).  For $\log U=-3.6$
the definition of an $\alpha$ abundance depends on which of the elements
O, S or Si is adopted as a standard.  For instance, if we use Si as the
standard, we find that $[{\rm N}/\alpha]=-0.37^{+0.31}_{-0.41}$, which
is nearly identical to our determination for the H~II region for the
case with $\log U=-4.4$.  Conversely, if we adopt O for the $\alpha$
element, we obtain $[{\rm N}/\alpha]=-1.07^{+0.31}_{-0.41}$, which is
slightly greater than our derivation for the H~I region.  (To avoid
clutter, these results for $\log U=-3.6$ are not shown in the figure.)

To compare our results with findings elsewhere, we show (1) the results
for the relative abundances in a high speed outflow from a
gravitationally lensed Lyman break galaxy MS~1512$-$cB58 studied by
Pettini et al.  (2002b), (2) emission-line studies of H~II regions in
galaxies of various types  (Pilyugin, Thuan, \& V\'ilchez 2003), and (3)
the compilation of N and $\alpha$-process elements in damped Lyman alpha
(DLA) systems by Centuri\'on et al  (2003).\footnote{Exceptions: In all
cases, we adjusted the abundance ratios so that they were referenced to
the solar abundances that we adopted (see the footnote for
\S\protect\ref{N_II}).  Si was substituted for O for the $z=2.456$ DLA
towards Q1409+0950 and the $z=2.076$ system toward Q2206$-$199 reported
by Pettini et al  (2002a) because we felt that the O~I line was too
saturated for a reliable result;  the detection of N~I at $z=2.0762$
toward Q2206$-$199 by Molaro et al.  (2003) replaces the upper limit
given by Pettini et al.  (2002a); the ionization-corrected values of
Prochaska et al.  (2002b) replace the values obtained by previous
authors for the $z=2.6256$ system toward GB 1759+7539.  Additions:  We
used ionization-corrected determinations of Dessauges-Zavadsky et al. 
(2003b) for [S/H] and an upper limit for [N/H] for the sub-DLA at
$z=2.668$ appearing in the spectrum of Q1409+095.  Values of $N$(S~II)
and $N$(N~I) were taken from Dessauges-Zavadsky  (2004) for the
$z=1.776$ DLA toward Q1331+17 and the $z=2.066$ system toward
Q2231$-$00.  Finally, we included a determination [O/H] and [N/H] for
the sub-DLA toward PG1216+069 carried out by Tripp et al  (2004).} For
these systems, the availability (or suitability) of various
$\alpha$-process elements varied from one case to the next.  Generally,
the element of choice was O, since it is generally undepleted onto dust
grains, and its ionization relative to H is well behaved. 
Unfortunately, the absorption features from O were often badly
saturated.  Therefore, other elements had to replace O, such as S
(second choice: generally undepleted but sometimes requires ionization
corrections in particular systems) and Si (third choice, since it can
require ionization corrections, can sometimes be depleted, and might
have some contributions from low-mass stars).  The many weaknesses
behind the use of Si might make one question its use, but empirically
Centuri\'on et al  (2003)) and Prochaska et al.  (2002a) claim that
[O/Si] seems not to deviate far from zero in the systems that they have
observed.  (Of course, this principle is violated here if $\log U=-3.6$
applies to the gas in our investigation.)  In Fig.~\ref{no_nh}, it is
clear that our result for N in the mostly ionized gas appears to follow
the general trend exhibited by other systems, while the N associated
with H~I shows an anomalously low value of [N/$\alpha$], one that seems
appropriate for DLAs with [$\alpha$/H]~$<-1$ rather than systems that
have a near solar abundance of the $\alpha$-process elements.

\placefigure{no_nh}

Figure~\ref{no_nh} also shows that while H~II regions in galaxies
exhibit a well defined plateau at [N/$\alpha$]~$=-0.9$ for
[$\alpha$/H]~$<-1.0$, for the DLA systems with low metallicity there is
a large scatter, and even a suggested bimodality, in the values of
[N/$\alpha$].  This phenomenon has captured the attention of Pettini et
al  (2002a), Henry \& Prochaska  (2003) and Prochaska et al.  (2002a),
who have explored a number of possible explanations.  One is that the
variations are a natural consequence of viewing systems associated with
star systems with different ages.  For this proposal, it is difficult to
understand why, if DLAs form at an approximately constant rate, there
are so many them detected with ages less than about the 250$\,$Myr
interval that elapses before the production of secondary nitrogen starts
to elevate [N/$\alpha$] above the level created by high mass stars 
(Henry, Edmunds, \& K\"oppen 2000).  However, stellar rotation could
significantly increase the time lag between the oxygen and nitrogen
enrichment  (Meynet \& Pettini 2003).  Another possibility is that there
are differences in the ratios of the N and O effective yields that are
driven by metallicity differences between Population~III and
Population~II stars, or that such metallicity differences govern the
time delay for the production of secondary nitrogen.  Finally, Henry \&
Prochaska  (2003) and Prochaska et al.  (2002a) have experimented with
models where the the initial mass function may be more top-heavy
(flatter) than usual, or truncated below some mass threshold, and these
models were successful in explaining the low N abundances.

For the H~I-bearing gas in our Lyman limit system, we have the added
challenge of explaining the low N abundance even when O seems to have
nearly a solar abundance.  A hint may come from findings of similar
conditions for the gas in a superwind arising from the distant,
gravitationally lensed Lyman break galaxy MS~1512$-$cB58 observed by
Pettini et al.  (2002b) (see Fig.~\ref{no_nh}).  Here, very rapid star
formation could cause a rapid growth in the metallicity and, at the same
time, the violent winds created by an abnormally large number of
supernovae may eject gas so quickly that the secondary nitrogen has not
had a chance to build up in the system.  It is possible that a similar
process occurred with the gas associated with our Lyman limit system. 
It is not clear how the gas became stabilized into a kinematically well
defined system with a low velocity dispersion, but perhaps this outcome
could arise from an outflow that creates an outward moving shell that
subsequently fragments into small clouds as it cools  (Theuns, Mo, \&
Schaye 2001).

The abundances of iron-group elements relative to $\alpha$-process
elements represent another approach for determining the time scale for
the chemical evolution of a system.  Consider first the case where $\log
U=-4.4$.  If we adopt the view that the fully ionized and partially
ionized media possibly have a different mix of nucleosynthesis sources,
we can only focus on the implications of our iron abundance for the
fully ionized medium since it dominates over the material associated
with H~I (Fe~II should be found in both).  We found that [Fe/Si] is only
moderately depressed ($-0.3\,$dex), and given that some Fe may be
depleted onto dust grains, this result supports our finding from
$N$(N~II) that nitrogen in the fully ionized gas is not very far from
the usual trends for H~II regions of other galaxies (shown in
Fig.~\ref{no_nh}), and thus the contrast that we found against the
neutral material may be real.  If the applicable value of $\log U$ is
$-3.6$, the abundance of Fe relative to either Si or S does not deviate
appreciably from its respective solar ratios, but, like Si and S, it is
strongly deficient relative to O.  An injection of excess O by
abnormally large contributions from the explosions of high mass stars
could occur without upsetting the balance between Si, S and Fe, but the
deficiency of N with respect to these three elements indicates that the
chemical enrichment of the gas differs in still other respects from that
of our Galaxy.

\subsection{Possible Sources of the Gas}\label{source}

It may seem odd that we have identified a gas system with a moderately
large column density and an oxygen abundance only slightly below solar
that is not obviously part of some galactic system.  As we discussed in
Paper~I, the detection of an absorbing galaxy $34\,h^{-1}_{70}\,$kpc
from a QSO sightline is consistent with the separations found for the
Mg~II absorbing galaxy population as a whole.  Our ACS image reveals no
obvious interloping galaxy between G158 and the PHL~1811 sightline that
might be responsible for the absorption instead of G158 itself.  Without
further observations that define the redshift of the extended emission
centered on the image of the quasar, it is difficult for us to state
categorically that there is or is not a face-on spiral galaxy, exactly
centered on the quasar, that has a redshift $z=0.0809$.  It should be
possible to obtain a spectrum of this galaxy from the ground during
conditions of good seeing, since some of the emission is $\gtrsim
2\arcsec$ from the quasar (see Fig.~\ref{fig_qsopsf}). 

The maximum characteristic size of the Lyman limit system's absorbing
region is given by our estimate for $N({\rm H}_{\rm
total})=3.5-20\times10^{18}{\rm cm}^{-2}$ divided by our minimum value
for $n(e)$ ($10^{-3}{\rm cm}^{-3}$), both of which were derived in
\S\ref{uniform_slab} under the assumption that our simple CLOUDY
calculations are correct.  The outcome is $1.4-8\,$kpc, which could be
somewhat larger if $N({\rm H}_{\rm total})$ is larger than our estimate
because some of the Si is depleted onto grains in the system.  This size
is smaller than the separation of the line of sight from either of the
two nearby S0 galaxies, indicating that it is probably not a large-scale
halo belonging to one or both of the galaxies.

Again, if the gas does not have embedded sources of radiation that would
invalidate the uniform slab model presented in \S\ref{uniform_slab} (and
if the temperature of the H~II region is not much different from the
H~I-bearing gas in the model for $\log U=-4.4$), we derive a range for
the thermal pressure $4<p/k<140\,{\rm cm}^{-3}\,$K. A warm-hot
intergalactic medium (WHIM) that could be associated either with the
line of galaxies shown in Figs.~\ref{fig_dr3cover} and \ref{fig_wedges},
or with a hypothetical bridge of material between several Abell clusters 
(Bregman, Dupke, \& Miller 2004), could have $\rho/\bar\rho\sim 20$ 
(Dav\'e et al. 2001) and thus might consolidate and pressure-confine the
material in our Lyman limit system. If this is correct, we do not
require the presence of a gravitational well created by stars or dark
matter to hold the gas together.

The presence of two S0 galaxies that have transverse distances less than
100$\,$kpc from the line of sight raises the intriguing possibility that
the Lyman limit system represents material either torn or ejected away
from one or both of the galaxies by tidal disruption  (Morris \& van den
Bergh 1994), ram-pressure stripping  (Quilis, Moore, \& Bower 2000), or
a superwind  (Bond et al. 2001; Pettini et al. 2002b).  We had mentioned
earlier (\S\ref{s_g169}) the possible presence of a tidal tail in G169. 
Also, the image of G169 shows a possible companion, and tidal stripping
or induced star formation and supernova activity inside it may be a
possible source of the gas  (Wang 1993).  A different picture is
presented by G158, where the regularity of the disk pattern that is
evident in the unsharp mask picture (Fig.~\ref{fig_sharp1}) argues
against any significant tidal disturbance.

In \S\ref{temp} we noted that the O~I $\lambda 1302$ line profile
appears to have a narrow core superimposed on a broader component (see
Fig.~\ref{oi_profile}).  This narrow core does not affect the
temperature constraint that we derived, but it likely does indicate that
some additional colder gas is present in this multiphase system.  It is
possible that if the cloud has been propelled through a medium of lower
density, we might be witnessing the effects of some ablation of the gas,
as has been proposed for some Milky Way high-velocity clouds  (Tripp et
al. 2003; Ganguly et al. 2004) .  In this case, the broader O~I
component could arise in gas that has started to ablate, while the
narrow O~I feature arises from gas that has remained intact within the
core that has not yet been subjected to the influence of the stripping
process.  The existence of small amounts of neutral hydrogen and C~IV at
large negative velocities (see Figs.~\ref{vstack} and \ref{lalpha}) may
represent a continuation of this process, since small amounts of
material could be shredded and accelerated from the main bulk of gas.

If the galaxies are moving rapidly enough with respect to a dense WHIM,
we might be witnessing the outcome of a stripping process similar to
that studied by Quilis, Moore \& Bower  (2000), who suggested it as an
explanation for the transformation of a spiral galaxy into an S0 galaxy. 
They proposed that the time scale for this process can be less than
100$\,$Myr, but in order to explain the low nitrogen abundances, the
onset of stripping would need to have occurred at a special time in the
galaxy's history, which is not very appealing.  By contrast, the
hypothesis that we are seeing some remnants of a superwind allows for a
natural reason for the low abundance of N: a sudden burst of star
formation would create Type~II supernovae that could expel the gas  (Mac
Low \& Ferrara 1999; Martin 2004; Veilleux 2004), and this process could
occur before an appreciable buildup of secondary nitrogen could arise
from intermediate mass stars.  Another clue that this could be happening
soon after such a burst is the possible excess of O over Si and S (but
only if $\log U=-3.6$), since there may not have been enough time for
any but the most massive stars to evolve to the point that they explode. 
While this is possible, an alternative explanation for this excess
could be that supernovae arising from the most massive stars have
higher explosion energies (Iwamoto et al. 1998; Mazzali et al. 2002;
Hamuy 2003) and thus are more likely to
expel their material from a galaxy.  Although
the models of Mac Low \& Ferrara  (1999) indicate that significant
ejection should operate only for galaxies with masses considerably
smaller than the S0 galaxies that we are considering ($10^{10} -
10^{11}{\rm M}_\odot$), we propose that even if the material is only
expelled into the galaxy's halo and is unable escape on its own, the
combination of low density and a less deep gravitational potential could
make it especially prone to being removed by ram pressure stripping from
the intragroup medium, much more so than low-metallicity gas residing
within the disk.  If the combined wind and stripping activity were
vigorous enough, it is conceivable that the gas from a spiral may be
driven away, star formation would then diminish, and an S0 galaxy with
perhaps some faint remnants of spiral structure would remain.

\section{Summary}\label{summary}

In the direction of the bright quasar PHL~1811, we encountered a low-$z$
intergalactic gas system that presented an especially favorable prospect
for studying its composition, physical conditions and galaxy
environment.  We exploited this opportunity by implementing an {\it
HST\/} observing program that recorded a spectrum of the quasar with the
STIS E140M medium-resolution echelle spectrograph (E140M) and obtained
an $r$-band image of the nearby environment with the WFC on ACS.

In analyzing the absorption lines, we were particularly fortunate in
being able to measure accurately the column densities of H~I and O~I. 
Also, we were able to provide a sensitive limit on the maximum amount of
N~I that could be present.  There are two important properties of these
three elements that make them easy to interpret.  First, charge exchange
reactions make their neutral fractions strongly coupled to each other,
and this vastly reduces the uncertainties arising from ionization
corrections.  Second, their relative abundances are not strongly
influenced by depletions caused by the condensation of atoms into solid
form on dust grains.  As for the utility of the relative abundances of
H, O and N, we argue that the first two indicate the general amount of
chemical enrichment of the system, while the second two indicate the
time scale for the chemical enrichment, since elevated levels of
nitrogen, compared to other $\alpha$-process elements, arise from the
production of secondary nitrogen, which is regulated by the evolution
time for intermediate and low mass stars.  Our large value for the
abundance of O, i.e., $[{\rm O/H}]=-0.19\pm 0.08$, indicates that the
gas has evolved chemically almost to the level of that of our Galaxy. 
However, if the stellar IMF was not markedly unusual, it must have done
so over a very short time ($\lesssim 0.25\,$Gyr) in order to explain the
low nitrogen abundance that we derived, $[{\rm
N/O}]=-1.15^{+0.31}_{-\infty}$ for $\log U=-4.4$,
$-1.07^{+0.31}_{-0.41}$ for $\log U=-3.6$, or simply $[{\rm N/O}]<-0.59$
at the $2\sigma$ confidence level.

Relative to O~I and N~I, the large abundances of singly-ionized forms of
the elements C, S, and Si indicate that there is a substantial amount of
material associated with gas where the hydrogen is fully ionized (or
nearly so).  The relative proportions of these ions agree with solar
abundance ratios, except for a mild underabundance of Fe, which might
arise from either depletions onto dust grains, a relative lack of
contributions of nucleosynthetic products created by Type~Ia supernovae,
or the fact that Fe reverts to the doubly ionized form more rapidly than
Si or S when the ionization parameter is increased.  In contrast to the
very low value of [N/O] indicated by N~I and O~I, the N~II in the more
ionized material gives an abundance that seems to be only mildly
deficient: $[{\rm N/Si}]=-0.34^{+0.49}_{-0.41}$ if $\log U=-4.4$.

Strongly ionized material, represented by C~IV, Si~IV and S~III, is seen
at velocities slightly displaced from the other features toward more
positive velocities -- see Fig.~\ref{vstack}.  These features are not
accompanied by a matching absorption from O~VI.

The fraction of hydrogen atoms in molecular form $f({\rm H}_2)$ is of
potential value for gaining a better understanding of the dust content
and photodissociation rate within a gas complex  (Browning, Tumlinson,
\& Shull 2003).  For DLAs in general, there is a large dispersion in
$f({\rm H}_2)$, which reinforces the picture that the systems have
strong differences in physical conditions from one case to the next 
(Curran et al. 2004).  For the Lyman limit system in front of PHL~1811,
we were unable to detect features arising from H$_2$, which lead to a
limit $f({\rm H}_2)<3.6\times 10^{-3}$.  This limit is not very
stringent because $N$(H~I) is not very large.  By comparison, a {\it
FUSE\/} survey of intermediate velocity clouds in the halo of our Galaxy
by Richter et al  (2002) yielded a number of detections of H$_2$ with
$-5.3 < \log f({\rm H}_2) < -3.3$ for $19.3 < \log N({\rm H~I}) < 20.3$. 
However, our H$_2$ fraction is lower than that found by Richter, Sembach
\& Howk  (2003) for a small, high-density cloud in the halo, assuming
that their value of $N$(H~I) is approximately correct (based on an
observation $N$(O~I) and the assumption that the gas has an
approximately solar abundance ratio of O to H).

From the convergence of high Lyman series lines in the {\it FUSE\/}
spectrum, we could derive a value for the velocity dispersion of H~I. 
We used the profile of O~I to indicate how much of this spread was due
to turbulent motions, so that the pure thermal Doppler broadening of H~I
could be estimated.  We derived a temperature $T = 
7070^{+3860}_{-4680}\,$K for the H~I-bearing gas using this technique. 
A narrow core in the O~I $\lambda 1302$ absorption profile indicates
that some colder gas is probably also present.

By comparing our upper limit for the column density of C~II$^*$ to the
measurement of $N$(Si~II) (and assuming a solar abundance ratio), we
derived upper limits for the electron density $n(e)$ that ranged from
$0.032\,{\rm cm}^{-3}$ at $T=2390\,$K to $0.069\,{\rm cm}^{-3}$ at
$T=10930\,$K.  These upper limits combined with a lower limit
$n(e)>10^{-3}\,{\rm cm}^{-3}$ from a CLOUDY model with the maximum
permissible ionization parameter define our acceptable range for $n(e)$.

A major drawback of observing systems with $N({\rm H~I}) \lesssim
10^{19.5}{\rm cm}^{-2}$ is that many elements can have major fractions
of their atoms in unseen ionization stages  (Viegas 1995; Jenkins 2004),
leading to erroneous conclusions about abundances.  Indeed, even systems
with much higher column densities may not be immune to such effects 
(Howk \& Sembach 1999; Izotov, Schaerer, \& Charbonnel 2001; Vladilo et
al. 2003) [although some contrary views are presented by Vladilo et al. 
(2001)].  A very important indicator of whether or not one is being
misled by partial ionization effects is Ar~I  (Sofia \& Jenkins 1998;
Vladilo et al. 2003), but unfortunately our upper limit for $N$(Ar~I) is
so high that it is not useful for this purpose.  We therefore felt a
strong need to investigate the consequences of various possible
ionization scenarios, including (1) collisional ionization, (2)
photoionization caused by the penetration of an external, intergalactic
radiation field into the region, and (3) photoionization from embedded
early-type stars that are not visible to us.  We claim that option (1)
is unlikely to be an important factor.  Option (2) was investigated
using CLOUDY calculations for two values of $\log U$.  At the lower
level of ionization ($\log U=-4.4$) we found that, except for N, various
elements were mildly below their solar abundances relative to H by a
nearly uniform amount.  The two forms of N gave inconsistent results
however.  That inconsistency could be resolved by raising $\log U$ to
$-3.6$, but this came at the expense of having various $\alpha$-process
elements showing some mutually discordant results when compared to their
solar abundances.  The results for both values of $\log U$ are
summarized in Table~\ref{cloudy_abund}.  Option (3), for which one could
devise a wide range of possible conditions, was investigated only from
the limited perspective of how much damage it could create for our
conclusions on the relative ionization fractions of N and O, using
independent information from the ionization of Si to constrain the
severity of the ionization.  From this study we found that the apparent
deficiency of N with respect to O in the H~I-bearing region could be
exaggerated by, at most, only 0.3$\,$dex.  Thus, regardless of which
ionization model we adopted, the fact that [N/H] is lower than [C, O,
Si, S and Fe/H] seems inescapable.

In Paper~I we identified two galaxies not far from the line of sight to
PHL~1811 that had redshifts nearly the same as that of the Lyman limit
system.  The ACS $r$-band image that we obtained now allows us to learn
more about the properties of these galaxies.  From various surface
photometry metrics given in Table~\ref{tab_acs}, we conclude that both
galaxies are of type S0, and unsharp masking of the image reveals faint
spiral arm features and a bar in one of them (G158).  One or both of
these galaxies may have contributed the gas that we see in the Lyman
limit absorption system.  As for the exact mechanism of transport,
several possibilities may be considered.  First, the gas may have been
expelled in a superwind created by Type~II supernovae following a burst
of star formation, a process which would provide a natural explanation
for the low nitrogen abundance if previous episodes on chemical
enrichment were far less advanced.  Another possibility is that some
stars and gas have been torn away by a tidal interaction.  For instance,
G169 shows a very faint tail-like structure that might support this
interpretation.  Finally, ram-pressure stripping may have removed most
of the gas in the disk of what was once a spiral galaxy, and we are
seeing remnants of this material.

A large-scale structure of galaxies in the vicinity of the Lyman limit
system (revealed by data from the SDSS) may have been influential in
enhancing our chances of coming across such a system, as well as the
other systems at nearly the same redshift reported in Paper~I.  Also,
the band of galaxies seen in the SDSS data, or perhaps a possible unseen
bridge of material between some Abell clusters at the same redshift, may
signal the presence of an accompanying excess of warm-hot intergalactic
matter that might help to confine the gas in the Lyman limit system.   

Finally, our ACS image revealed the presence a diffuse, $S$-shaped
emitting structure that may possibly be a face-on spiral galaxy with two
arms having large pitch angles.  With the evidence at hand, we are
unable to determine conclusively whether this structure is at the
redshift of the Lyman limit system or that of the quasar.

\acknowledgements

We thank an anonymous referee for useful ideas that broadened our
approach to the problems discussed in this paper.  Support for our {\it
HST\/} guest observer program (nr. 9418) was provided by the National
Aeronautics and Space Administration (NASA) through a grant to Princeton
University from the Space Telescope Science Institute, which is operated
by the Association of Universities for Research in Astronomy,
Incorporated, under NASA contract NAS5-26555.  Additional support was
provided by the NASA LTSA grant NNG04GG73G to the University of
Massachusetts and by subcontract 2440-60014 to Princeton University
under the NASA prime contract NAS5-32985 to Johns Hopkins University. 
The {\it FUSE\/} spectrum of PHL~1811 was obtained for the Guaranteed
Time Team by the NASA-CNES {\it FUSE\/} mission operated by Johns
Hopkins University. This publication makes use of data products from (1)
the Two Micron All Sky Survey, which is a joint project of the
University of Massachusetts and the Infrared Processing and Analysis
Center/California Institute of Technology, funded by NASA and the
National Science Foundation (NSF) and (2) the Sloan Digital Sky Survey
(SDSS), which is supported by the Alfred P. Sloan Foundation, NASA, NSF,
the U. S. Department of Energy, the Japanese Monbukagokusho, and the Max
Planck Society.  The SDSS project is managed by the Astrophysical
Research Consortium (ARC) for the participating institutions that
include the University of Chicago, Fermilab, the Institute for Advanced
Study, the Japan Participation Group, the Johns Hopkins University, Los
Alamos National Laboratory, the Max-Planck-Institute for Astronomy
(MPIA), the Max-Planck-Institute for Astrophysics (MPA), New Mexico
State University, University of Pittsburgh, Princeton University, the
United States Naval Observatory, and the University of Washington.

\appendix
\section{Photoionization Equilibrium Equations}\label{photo_eqns}

In a medium where the hydrogen is partly ionized with atomic and proton
densities $n({\rm H})$ and $n({\rm H}^+)$, respectively, the densities
of an element X in its 3 lowest levels of ionization X, X$^+$ and
X$^{++}$ are governed by the equilibrium equations\footnote{The
development here follows that given by Eqs. 12--19 of Sofia \& Jenkins 
(1998), except that we have added the He charge exchange reaction as an
additional channel for reducing the ionization of element X in its
doubly charged form.  We have corrected Eqs.~\protect\ref{equi1} and
\protect\ref{n0} here to include a missing $n({\rm H^+})$ term, which
was a typographical oversight in the earlier presentation.}
\begin{equation}\label{equi1}
\big[\Gamma({\rm X})+C({\rm X}^+,T)n({\rm H^+})\big]n({\rm
X})=\big[\alpha({\rm X},T)n(e)+C^\prime({\rm X}^+,T)n({\rm
H})\big]n({\rm X}^+)
\end{equation}
\begin{equation}\label{equi2}
\Gamma({\rm X}^+)n({\rm X}^+)=\big[\alpha({\rm X}^+,T)n(e)+C^\prime({\rm
X}^{++},T)n({\rm H})+D^\prime({\rm X}^{++},T)n({\rm He})\big]n({\rm
X}^{++})
\end{equation}
where $\Gamma({\rm X}^y)$ is the photoionization rate of element X in
its state $y$ (neutral, +, or ++) and $\alpha({\rm X}^y,T)$ is the
recombination rate of the $y+1$ state with free electrons as a function
of temperature $T$.  The charge exchange rate constants $C^\prime({\rm
X}^+,T)$ and $C^\prime({\rm X}^{++},T)$ refer to the reactions ${\rm
X}^++{\rm H}\rightarrow {\rm X}+{\rm H}^+$ and ${\rm X}^{++}+{\rm
H}\rightarrow {\rm X}^++{\rm H}^+$, respectively.  The rate constant
$C({\rm X}^+,T)$ for the reverse endothermic reaction ${\rm X}+{\rm
H}^+\rightarrow {\rm X}^++{\rm H}$ can be obtained from $C^\prime({\rm
X}^+,T)$ by applying the principle of detailed balancing.  Reactions
with neutral He can be important when the hydrogen is much more strongly
ionized than He, so we include the rate constant $D^\prime({\rm
X}^{++},T)$ that applies to ${\rm X}^{++}+{\rm He}\rightarrow {\rm
H}^++{\rm He}^+$.  The simultaneous solution to these two equations
yields the fractional abundances in the three ionization levels
\begin{mathletters}
\begin{eqnarray}\label{n0}
f_0({\rm X},T) & \equiv & {n({\rm X})\over n({\rm X})+n({\rm
X}^+)+n({\rm X}^{++})}\nonumber\\
&=&\left( 1\, +\, {\big[\Gamma({\rm X})+C({\rm X}^+,T)n({\rm H^+})\big]
\big[\Gamma({\rm X}^+)+Y\big]\over \big[\alpha({\rm X},T)n(e) +
C^\prime({\rm X}^+,T)n({\rm H})\big]Y}\right)^{-1}~,
\end{eqnarray}
with
\begin{equation}
Y=\alpha({\rm X}^+,T)n(e)+C^\prime({\rm X}^{++},T)n({\rm
H})+D^\prime({\rm X}^{++},T)n({\rm He})
\end{equation}
\end{mathletters}
\begin{equation}\label{n++}
f_{++}({\rm X},T) \equiv {n({\rm X}^{++})\over n({\rm X})+n({\rm
X}^+)+n({\rm X}^{++})}={1-f_0({\rm X})\over 1+Y/\Gamma({\rm X}^+)}~,
\end{equation}
and
\begin{equation}\label{n+}
f_+({\rm X},T)\equiv {n({\rm X}^+)\over n({\rm X})+n({\rm X}^+)+n({\rm
X}^{++})}=1-f_0({\rm X},T)-f_{++}({\rm X},T)~.
\end{equation}
In applying Eqs.~\ref{n0}--\ref{n+} to N and O in the analysis that
supported the conclusions in \S\ref{more_detailed}, we used the
photoionization cross sections $\Gamma$ from the analytic approximations
of Verner, et al.  (1996), the recombination coefficients $\alpha$ from
the fit equations of Shull \& Van Steenberg  (1982), and the charge
exchange rates $C^\prime$ from Kingdon \& Ferland's  (1996) fits to the
calculations of Butler \& Dalgarno  (1979) for N and Chambaud et al. 
(1980) for O.  We adopted the rate constants $D^\prime({\rm N}^{++},T)$
from Butler \& Dalgarno  (1980) and $D^\prime({\rm O}^{++},T)$ from
Butler, Heil \& Dalgarno  (1980).  Of course, before the ionization
fractions of N or O can be derived, we must determine the ionization
balance of helium in the presence of hydrogen by solving the same
equations (substituting He for X and eliminating the $D^\prime$ term)
along with the equation for the hydrogen ionization balance,
\begin{mathletters}
\begin{equation}
{n({\rm H}^+)\over n({\rm H})}={\Gamma({\rm H})+Z\over \alpha({\rm
H},T)n(e)-Z}
\end{equation}
with
\begin{equation}
Z=0.1n({\rm H})\big[C^\prime({\rm He}^+,T)f_+({\rm He},T)+C^\prime({\rm
He}^{++},T)f_{++}({\rm He},T)\big]~,
\end{equation}
\end{mathletters}
and the constraints
\begin{equation}\label{He_total}
n({\rm He})+n({\rm He}^+)+n({\rm He}^{++})=0.1n({\rm H})\left[1+{n({\rm
H}^+)\over n({\rm H})}\right]
\end{equation}
and
\begin{equation}\label{n(e)}
n(e)=n({\rm H}^+)+n({\rm He}^+)+2n({\rm He}^{++})~.
\end{equation}
%

%

\clearpage

\clearpage

\clearpage

\clearpage
%
\end{document}